# Pretreatment DCE-MRI Resolves Response Quality Within Pathologic Endpoints in Neoadjuvant Breast Cancer

*A four-tier structural framework in 1,200 neoadjuvant patients across four institutions, with fifth-institution validation*

Dattatreya Kantha[1], Murray H. Loew[1]

[1]Medical Imaging & Image Analysis Laboratory, Department of Biomedical Engineering, George Washington University, Science and Engineering Hall, 800 22nd St., NW, Suite 5000, Washington, DC 20052, USA
Corresponding author: Murray H. Loew, Ph.D., P.E. loew@gwu.edu
First author contact: Dattatreya Kantha, M.S. dattatreya.kantha@gwu.edu

## Abstract

Pathologic complete response (pCR) is a strong neoadjuvant endpoint, yet 5-15% of complete responders recur and clinical/genomic variables do not reliably identify them. We tested whether pretreatment dynamic contrast-enhanced MRI entropy - intratumoural enhancement heterogeneity - resolves response quality hidden within pCR and residual cancer burden (RCB). Across four cohorts (1,200 patients), a prespecified entropy threshold defined favourable and adverse structural states. Crossing structure with pathology yielded a four-tier framework spanning 4.1-fold recurrence in I-SPY1 and 7.7-fold at response extremes. In I-SPY2, 55 of 219 complete responders (25.1%) were structurally adverse pretreatment. In an external HER2-positive responder synthesis (I-SPY1 pathology-confirmed pCR plus UCSF best-response proxy; n = 33, 10 events), adverse structure was associated with higher recurrence risk (HR = 2.87, 95% CI 1.38-5.96) capturing 7 of 10 recurrences, enriching rather than determining risk. In a HER2-positive RCB-0 subset, recurrence was 12.5% with favourable and 80.0% with adverse structure; Firth Cox regression preserved the association (HR = 8.13, 95% CI 1.71-49.21; n = 21, 6 events). In Duke (n = 908; 76 events), favourable structure remained independently associated with lower distant-recurrence risk (adjusted HR = 0.61, 95% CI 0.41-0.91). RNA linked favourable structure to a directionally reproduced immune-architecture programme among non-overlapping patients within I-SPY2; EMT-pathway enrichment was favourable-side, while the adverse tier contained a broadly immune-depleted substate. Yet full-cohort RNA models weakly discriminated structural state and did not recover continuous entropy. Pretreatment MRI therefore does not replace pCR or RCB; it reveals response-quality differences that these endpoints compress and identifies a recurrence-enriched group for prospective validation.

## 1. Introduction

Pathologic complete response (pCR) is among the most reassuring outcomes in neoadjuvant breast cancer and the basis of regulatory accelerated-approval pathways[3]. Adaptive platform trials graduate experimental arms by pathologic response — I-SPY2[13] and, more recently, I-SPY2.2[4,5] — and in practice pCR drives de-escalation of adjuvant chemotherapy, surveillance intensity, and therapy-omission decisions that directly affect patients. Yet pCR is, by construction, a binary post-treatment label: it confirms that invasive disease was not found at surgery, but not whether the complete response it certifies is biologically durable, immune-supported, or free of later recurrence. The same binary label can mark biologically different complete responses. Multicohort analyses show that 5–15% of patients who achieve pCR nonetheless recur within ten years[2,8,9,45] — including recurrence documented within a high-dose radiation boost volume after confirmed pCR[48,49] — and the largest systematic search for prognostic factors within pCR, a pooled analysis of 2,066 pCR patients across five trials[7], found that molecular subtype, Ki-67, and standard clinicopathologic variables were not significantly associated with recurrence risk inside the label. The conventional clinical and genomic toolkit cannot see inside the pCR label.

The clinical stakes of this blind spot are asymmetric. Patients with residual disease after neoadjuvant therapy now have validated adjuvant escalation — for example, T-DM1 for HER2-positive residual disease[46] — whereas a complete responder who is nonetheless destined to recur has no residual-disease-triggered signal to act upon, because the pathologic category that defines her reports nothing to escalate. Residual recurrence risk is therefore concentrated precisely where the current clinical system offers no further resolution, and identifying that subgroup has been recognised as a clinical priority[6]. This patient-level question is distinct from the trial-level surrogacy debate: the CTNeoBC pooled analysis of nearly 12,000 patients found pCR prognostic for individuals yet an unreliable trial-level surrogate for survival[8,32], which is why the FDA positions pCR as a basis for accelerated — not definitive — approval[3,31], and later analyses and reviews reinforced the disconnect between pCR-rate gains and survival improvements[36,37]. These analyses established that pCR is informative but incomplete; they did not identify the biological source of that incompleteness.

We use the term response quality for this missing dimension — not how much disease remains, but whether the tumour's baseline state predisposes the response to be durable or fragile. Dynamic contrast-enhanced MRI records treatment-response information[44] that pathologic assessment does not: the distributional heterogeneity of tumour enhancement. Critically, this signal is read before therapy — a predictor that helps interpret the pathologic response observed later, not a substitute for that pathology. In a companion study[1], we established that baseline structural entropy — a measure of intratumoural enhancement-distribution heterogeneity on pretreatment DCE-MRI — dominates the organising axis of longitudinal treatment-response geometry, and that this axis is non-redundant with receptor subtype, PAM50 classification, MammaPrint[30] genomic risk, and baseline tumour burden. A complementary question is whether this non-redundant imaging axis carries measurable same-patient molecular context, and whether that context is sufficient to recover the imaging phenotype itself. Association and reconstruction are distinct questions: a molecular programme may covary with an imaging phenotype without containing sufficient patient-level information to reproduce that phenotype. The present study asks what that pretreatment structural axis means for patients once pathology has spoken.

In DCE-MRI, the segmented tumour contains many voxels, each with a signal-enhancement ratio describing its contrast-enhancement behaviour. Enhancement entropy summarises the distribution of these voxel-level values: low entropy means the values are concentrated within a relatively narrow range, whereas high entropy indicates a broader mixture of enhancement behaviours within the tumour. Entropy is a first-order distribution statistic and therefore does not encode the spatial arrangement of individual voxels. We refer to the resulting continuous phenotype as a structural axis because it reproducibly organises treatment trajectories, biological states, and recurrence patterns.

Our central proposition is that crossing pretreatment structure with pathologic response creates a four-tier response-quality framework that resolves heterogeneity compressed within each pathologic category: RCB grades residual disease, while the structural axis adds resolution within pCR. We present three convergent lines of evidence. First, structural entropy defines four clinically distinct response-quality archetypes spanning a 4.1-fold recurrence range across the external cohort and a 7.7-fold range at response extremes, enriching for the complete responders who carry residual recurrence risk within an otherwise reassuring category. Second, favourable structure corresponds to a PAM50-independent immune-engagement programme, including favourable-side EMT-pathway enrichment, directionally reproduced in a patient-non-overlapping HER2-positive pCR subset within the same I-SPY2 trial (n = 68), while the adverse tier contains a broadly immune-depleted substate, with exploratory evidence that the axis is selectively reconfigured by drug mechanism. Third, the within-pCR association is supported by a 27-variable confounder screen and manifests predominantly as distant recurrence. Separately, in the full Duke cohort (n = 908), the structural phenotype remained prognostic after eight-variable adjustment, supporting cross-institutional transportability. Together, these findings establish that pretreatment structural MRI adds a response-quality dimension that pCR and RCB do not resolve — refining rather than replacing the canonical neoadjuvant response endpoints.

## 2. Methods

### 2.1 Study design and cohorts

This study characterises the clinical consequences of a structural entropy axis previously identified in a companion study[1]. Four publicly available neoadjuvant breast cancer cohorts were used, each serving a distinct analytical role (Table 1). The I-SPY2 cohort (n = 701 patients with baseline structural entropy and pathologic response data) served as the discovery platform for framework construction, structural-axis pCR-orthogonality characterisation, drug-quality analysis, persistence evaluation, and deep learning convergence. I-SPY2 tier counts are reported throughout the main text using the locked master discovery substrate (n = 698, with T1 = 164, T2 = 55, T3 = 360, T4 = 119); the four-archetype mechanism substrate (n = 283), the original full-cohort discovery substrate (n = 701), and the entropy × sphericity crosswalk (n = 701) are referenced only where explicitly tagged. The I-SPY1 cohort (ACRIN 6657; n = 138 after bilateral exclusions, 42 recurrence-free survival events per Esserman et al.[11]) served as the primary external validation cohort for the four-tier framework's prognostic architecture, the formal structure × response interaction test, and the within-pCR recurrence analysis. The UCSF cohort (n = 49 projection-eligible, 15 events) provided frozen manifold transport validation without retraining or threshold re-optimisation. The Duke Breast Cancer MRI cohort[15] (n = 908 distant recurrence-free survival eligible, 76 events) provided large-scale entropy transportability validation, fourth-cohort pCR-orthogonality replication, and treatment-era confirmation. Of the 908 DRFS-eligible Duke patients, 312 received neoadjuvant therapy and are included in the primary neoadjuvant count (I-SPY2 701 + I-SPY1 138 + UCSF 49 + Duke NAT 312 = 1,200); the remaining 596 Duke patients contribute to prognostic transportability validation.

Two additional cohorts extended the validation architecture. The BreastDCEDL-ISPY2 dataset[35] (265 subjects from the same I-SPY2 trial, processed through an independent standardised imaging pipeline) provided cross-pipeline validation of the structural signal under a different preprocessing regime. The QIN-BREAST cohort (Vanderbilt University, Philips 3T, 39 baseline DCE-eligible subjects) provided a fifth-institution validation on a third scanner vendor, using protocol-informed enhancement localisation rather than segmentation-based extraction.

All analyses used publicly available de-identified datasets from The Cancer Imaging Archive (TCIA)[12,14] and associated public clinical data resources. No institutional review board approval was required for secondary analysis of these de-identified public datasets. The primary pathologic endpoint was pathologic complete response (pCR), defined as no residual invasive disease in breast and axillary lymph nodes (ypT0/is ypN0). Residual cancer burden (RCB) classification[10,27] was available for I-SPY1 (201 of 221 patients). Recurrence endpoints were recurrence-free survival (RFS) for I-SPY1 and distant recurrence-free survival (DRFS) for Duke. The I-SPY2 internal event indicator (event_work) was found to be perfectly aligned with hormone receptor status and cannot serve as a validated recurrence endpoint; this limitation is detailed in §2.6.

Residual cancer burden (RCB) is a continuous quantitative pathology score for residual invasive cancer after neoadjuvant chemotherapy[10], computed from four measurements on the surgical specimen — primary tumour bed size, residual invasive cellularity, number of involved axillary nodes, and largest nodal metastasis diameter — and reported either as the continuous RCB index or in four ordinal categories: RCB-0 (no residual invasive disease, equivalent to pCR), RCB-I (minimal residual), RCB-II (moderate residual), and RCB-III (extensive residual). Clinically, RCB is used for post-surgical prognostication[2], adjuvant intensification decisions (e.g., T-DM1 in HER2-positive residual disease[46]), and as a treatment-arm-stable surrogate endpoint in neoadjuvant trials, with hazard ratios stable in the 1.79–2.00 range across I-SPY2 treatment arms[27]. In this study, RCB enters analyses as a per-patient ordinal categorical variable from clinical metadata, parallel to the locked-nine architecture composite (§2.12) which provides the corresponding per-patient continuous biological score from bulk RNA-seq.

**Table 1 | Study cohorts and analytical roles.**

| Cohort | N | Events | Imaging | Role | Source |
|---|---|---|---|---|---|
| I-SPY2 | 701 | — | 1.5T/3T DCE | Discovery: framework, pCR-orthogonality, drug quality, DL, persistence, molecular characterisation (GSE194040) | TCIA (ISPY2) |
| I-SPY1 | 138 | 42 | 1.5T DCE | Validation: interaction, bilateral, DL transfer, within-pCR, GSE22226 transcriptomic bridge | TCIA (ACRIN 6657) |
| UCSF | 49 | 15 | 1.5T/3T DCE | Validation: frozen manifold transport | TCIA (Breast-MRI-NACT-Pilot) |
| Duke | 908† | 76 | 1.5T/3T DCE | Validation: entropy transport, pCR-orthogonality replication, treatment-era | TCIA (Duke-Breast-Cancer-MRI) |
| BreastDCEDL-ISPY2 | 265 | — | 1.5T/3T DCE | Cross-pipeline: independent entropy, DL carrier transport | TCIA (BreastDCEDL-ISPY2) |
| QIN-BREAST | 39 | — | Philips 3T | fifth-cohort pCR-orthogonality, DL-first validation, within-pCR heterogeneity | TCIA (QIN-BREAST) |

*All cohorts from TCIA. Events = validated recurrence endpoints (RFS or DRFS). I-SPY2 validated recurrence events not available (see §2.6). †908 DRFS-eligible for prognostic validation; 312 NAT-eligible included in the primary 1,200 neoadjuvant count.*

## 2.2 DCE-MRI feature extraction

Full IBSI-compliant[17] radiomics extraction was performed using PyRadiomics[16] (v3.0.1) on signal enhancement ratio (SER) derived DCE-MRI maps, as described in the companion study[1]. Voxel-level SER was computed from pretreatment and post-contrast dynamic series, and radiomic features were extracted from the resulting parametric maps within tumour regions of interest defined by functional tumour volume (FTV) segmentation. The extraction configuration used a fixed bin width of 0.1 for all discretisation-dependent features. A total of 107 features were computed spanning seven IBSI-defined families: first-order statistics (18 features), grey-level co-occurrence matrix (GLCM, 24), grey-level run-length matrix (GLRLM, 16), grey-level size-zone matrix (GLSZM, 16), grey-level dependence matrix (GLDM, 14), neighbouring grey-tone difference matrix (NGTDM, 5), and shape (14). The primary structural feature used throughout the study was baseline first-order Shannon entropy of the SER distribution (Entropy_T0). Because it uses the voxel-value histogram rather than voxel adjacency, Entropy_T0 is distinct from GLCM and other spatial-texture entropy features. Throughout this study, structural entropy refers specifically to the first-order Shannon entropy of the tumour signal-enhancement-ratio distribution; it is distinct from spatial texture-entropy features derived from voxel-neighbourhood relationships. It was previously identified in the companion study as the dominant scalar correlate of the first principal component of the longitudinal treatment-response manifold ($\rho = -0.933$)[1].

Four treatment timepoints were defined following the I-SPY2 imaging protocol: T0 (pretreatment baseline), T1 (early treatment, 3–4 weeks), T2 (inter-regimen, approximately 12 weeks), and T3 (presurgical, after completion of neoadjuvant therapy). Longitudinal persistence analyses used the visit-level radiomics table containing full 107-feature extractions at all four timepoints for n = 565 patients with matched T0–T3 measurements.

**2.3 The four-tier structural response-quality framework: motive, construction, interpretation, and significance**

Motive. Pathologic complete response (pCR) and the residual cancer burden (RCB) categorical scale remain the canonical neoadjuvant endpoints. Each is clinically valuable, but each compresses response-quality information that varies meaningfully within its own categories. Within pCR, some patients still recur; within RCB-0 — the most reassuring pathology category — recurrence rates are heterogeneous in ways that pathology alone does not resolve. The four-tier structural response-quality framework introduced here was developed to recover that within-category response-quality information from pretreatment imaging. The framework refines pCR and RCB rather than replacing them: pCR confirms whether residual disease was detected; RCB measures how much; the structural axis adds whether the response was complete in the architectural sense (Supplementary Note D19).

Construction. The four-tier framework was constructed by combining two pre-specified binary axes. The first axis is RESPONSE STATE — pathologic complete response versus residual disease — the canonical neoadjuvant breast cancer endpoint, anchored in three decades of clinical use and assigned by pathologist review at definitive surgery. The second axis is STRUCTURAL STATE — favourable versus adverse — defined by baseline DCE-MRI structural entropy at a locked q75 threshold within the cohort (q75 = 3.284 in the I-SPY2 training partition; selected before any external outcome analysis). Crossing these two axes yields four biologically distinct response-quality archetypes: Tier 1 (favourable structure, pCR — favourable complete response), Tier 2 (adverse structure, pCR — structurally adverse pCR), Tier 3 (favourable structure, residual disease — organised residual disease), and Tier 4 (adverse structure, residual disease — aggressive or inflamed residual disease). Threshold robustness was assessed by systematic sweep across the entropy distribution and by bootstrap stability analysis of tier assignments.

Feature identity and structural interpretation. The structural state used in this framework is carried by baseline structural entropy, defined as first-order Shannon entropy of the SER distribution within the pretreatment tumour ROI (Entropy_T0). This feature measures how broadly enhancement intensities are distributed within the tumour: lower entropy reflects more organised, homogeneous enhancement, whereas higher entropy reflects greater enhancement heterogeneity. In DCE-MRI, such first-order enhancement-intensity heterogeneity is interpreted as a tumour-level imaging proxy for architectural and vascular disorganisation. We therefore use "structural entropy" for the measured scalar feature, "structural axis" for the imaging-derived response-quality dimension organised by this feature, and "structural state" for the favourable/adverse partition created by the locked threshold. The axis is structural rather than volumetric because it is orthogonal to baseline functional tumour volume and non-redundant with molecular subtype, yet persistent through treatment and dominant in the longitudinal response geometry.

Feature selection and threshold choice. Baseline structural entropy was selected from 107 candidate IBSI-standardised radiomic features as the single feature satisfying three pre-specified properties: persistence through treatment (88.2% state retention; Spearman $\rho = 0.736$), reproducibility across institutions without retraining, and burden-orthogonality. The 75th-percentile threshold was chosen for clinical interpretability, robustness to ±5% perturbation, and sample-size balance (q75 = 3.284 in the I-SPY2 training partition; ~25% of pCR patients assigned to Tier 2). The 2×2 architecture combines the two best-validated response axes — pathology and structure — producing four biologically distinct archetypes; a candidate third axis (immune-engaged vs immune-depleted) did not discriminate at the tier-mean level (multivariable bulk-RNA AUC = 0.546), with within-tier substate decomposition capturing immune information at the right granularity inside Tier 2. In the present study, the structural axis is crossed with pathologic response to form the four-tier response-quality framework and then biologically interpreted through immune-recognition, cytotoxic, EMT, stromal, and drug-mechanism analyses. Feature identity, threshold sensitivity, and continuous-risk comparisons are provided in Supplementary Tables S15, S31, and S14, respectively.

Three cohort systems. The same four-tier framework is applied to three distinct cohort substrates throughout this manuscript, each answering a different question and each with different inferential affordances. We refer to these as System A, System B, and System C, and we tag every Results paragraph and figure caption with the system it draws from. SYSTEM A is the I-SPY2 discovery cohort (n = 696–701 by analysis substrate), where the framework was constructed and where structural-axis characterisation, immune programme analysis, drug-mechanism analysis, persistence, and molecular replication take place. I-SPY2 has no validated recurrence endpoint and is not used for recurrence claims. SYSTEM B is the I-SPY1 cohort (n = 138 with 42 RFS events; n = 120 framework-eligible; n = 42 within-RCB-0 imaging-outcome subset), where the framework is externally validated, where the RCB extension lives (Cox HR = 5.50 within RCB-0; HER2-positive subset HR = 9.04), and where the GSE22226 transcriptomic bridge operates. SYSTEM C is the pooled HER2-positive external synthesis of the I-SPY1 pathology-confirmed within-pCR cohort (n = 21) and the UCSF best-response proxy cohort (n = 12) (n = 33 with 10 RFS events), where the pooled responder recurrence anchor (pooled HR = 2.87, 95% CI 1.38–5.96) is established. Supplementary Table S57 maps the full three-system architecture; Methods Table 2 defines the four tiers.

*The full denominator and support / does-not-support mapping for the three cohort systems is provided in Supplementary Table S57.*

**Table 2 | Four-tier framework definitions.** Each row is one tier; tiers are produced by crossing the pathologic response axis with the structural response-quality axis.

| Tier | Pathology | Structural state | Clinical archetype | Core interpretation |
|---|---|---|---|---|
| Tier 1 | pCR | Favourable (≤ q75) | Favourable complete response | Organised clearance; most favourable complete-response state |
| Tier 2 | pCR | Adverse (> q75) | Structurally adverse pCR | Within-pCR recurrence-enriched; carried by an immune-depleted substate, not the whole tier; persistent adverse architecture despite pCR |
| Tier 3 | Residual disease | Favourable (≤ q75) | Organised residual disease | Architectural normalisation without clearance; immune-quiet |
| Tier 4 | Residual disease | Adverse (> q75) | Aggressive / inflamed residual disease | Persistent disorder with residual burden; inflammatory/stromal state-defining |

*q75 = 75th percentile of baseline first-order Shannon entropy of the SER distribution in the I-SPY2 training partition (q75 = 3.284). Detailed tier definitions, the forbidden-claim guardrails, and the full per-system evidence matrices are provided in Supplementary Tables S54–S57.*

### 2.4 Deep learning pipeline

A frozen deep learning representation was used to test whether the structural adversity phenotype could be independently recovered by a computationally unrelated image representation. A ResNet-18[18] convolutional neural network, pretrained on ImageNet[19], was used as a frozen feature encoder on single axial slices from baseline DCE-MRI. The tumour region was defined using a dilated mask that included peritumoural context. Frozen 512-dimensional feature vectors were extracted without fine-tuning.

For the locked internal analysis, 61 pCR patients with both radiomic structural classification and successful DL feature extraction were used. A simple linear readout (logistic regression) was trained to predict radiomic-defined structural adversity from frozen CNN features, with bootstrap stability-based feature selection (1,000 iterations). A four-cell consensus taxonomy was defined by crossing the radiomic and DL classifications. For scope assessment, all 220 pCR patients were uniformly re-extracted through the same frozen pipeline to test whether convergence generalised beyond the locked overlap. For external transfer, frozen ResNet-18 features were extracted from 133 I-SPY1 subjects across a 20-year imaging era gap.

### 2.5 External validation methodology

External validation followed the Representation-Family Commensurability Assessment (RFCA) framework established in the companion study[1]. Under RFCA, external validity is conditional on the commensurability of the structural representation between cohorts. Each external cohort was assessed for distributional compatibility (Cohen's d between source and target entropy distributions) before outcome analysis.

For UCSF, frozen manifold projection from the I-SPY2-derived trajectory space was applied without retraining. Restricted mean survival time (RMST)[24] was computed at $\tau = 60$ months with bootstrap confidence intervals and permutation-based significance (10,000 permutations each). For Duke, Entropy_Tumor was classified as regime-compatible under RFCA (Cohen's d = −0.020). The frozen I-SPY2 training-partition threshold was applied directly. Cox proportional hazards models[22] were fitted unadjusted, subtype-adjusted, and with full clinicopathologic adjustment. A known geometry-coupling boundary in the Duke cohort was documented as a predicted RFCA condition[1].

### 2.6 Responder recurrence methodology

Responder recurrence analyses for the pooled external HER2+ HR estimate were restricted to the HER2-positive subtype because that subgroup provided the necessary combination of pCR prevalence and recurrence event density for stable within-responder survival evaluation. The I-SPY2 discovery cohort contributed phenotype characterisation only; no recurrence claims were derived from I-SPY2. The I-SPY2 internal event indicator (event_work) was found to be perfectly aligned with hormone receptor status (36 of 36 HR-positive patients = event_work 1; 27 of 27 HR-negative patients = event_work 0; Supplementary Table S12) and cannot serve as a validated recurrence endpoint.

To test structural refinement within residual cancer burden classes, we analysed the locked I-SPY1 RCB-known denominator (n = 138; 42 RFS events; RCB-0 n = 42, RCB-I n = 11, RCB-II n = 57, RCB-III n = 28). The locked MAD-z structural threshold (1.133452), derived in the HER2-positive RCB-0 primary analysis, was applied unchanged. Because the I-SPY2 entropy q75 threshold was not numerically commensurable on the I-SPY1 native MAD-z feature scale, the locked MAD threshold was used as the canonical RCB-extension structural cut-point. UCSF response subgroup analyses used native path-size/proxy response fields and are therefore treated as supportive response-proxy analyses rather than pathology-locked pCR replication; full denominator reconciliation is provided in Supplementary Table S51 and Supplementary Notes D12 and D13.

Recurrence associations were established exclusively by external survival cohorts. The I-SPY1 HER2-positive pCR denominator was audited through a complete funnel: 30 raw HER2+ pCR patients → 22 with baseline MRI-linked projection data → 21 in the native MAD Cox model. Eight censored patients were excluded at the MRI-linkage step and one further censored patient at the native-MAD completeness step; zero recurrence events were lost. The UCSF within-responder cohort comprised 12 HER2+ best-response patients with 4 events. The pooled responder recurrence estimate was computed using DerSimonian–Laird random-effects meta-analysis[20] combining the I-SPY1 and UCSF results. Between-study heterogeneity was assessed using Cochran's Q, $I^2$, and $\tau^2$.

### 2.7 Statistical analysis

Group comparisons for continuous variables used the Mann–Whitney U test (two groups) or Kruskal–Wallis test (three or more groups). Correlations were assessed by Spearman's rank correlation. Categorical associations used Fisher's exact test or Pearson's $\chi^2$ test. Effect sizes included rank-biserial correlation, Cramér's V, epsilon-squared, and Cohen's d. Survival analyses used Cox proportional hazards regression and Kaplan–Meier estimation[21] with log-rank tests. For the four-tier framework, ordered survival trends were additionally evaluated using a 1-df log-rank test for trend[47], computed as the Cox score test with tier entered as an ordinal variable (1, 2, 3, 4) reflecting the pre-specified Tier 1 < Tier 2 < Tier 3 < Tier 4 hierarchy locked in §2.3. Interaction between structural quality and pathologic response was tested by likelihood-ratio comparison of additive and interaction Cox models. Model comparison used Akaike information criterion (ΔAIC) and Harrell's concordance index (ΔC). Sparse within-response survival and recurrence analyses were supplemented with small-sample robust estimators. For the HER2-positive RCB-0 survival analysis, standard Cox regression was paired with Firth penalized Cox regression to reduce small-sample bias under sparse-event conditions. Crude odds ratios from 2×2 recurrence tables were reported descriptively and supplemented with Firth penalized-likelihood logistic regression and exact conditional odds-ratio estimation as a distribution-free sensitivity estimate. The fragility index was calculated as the minimum number of event-status reassignments required to move Fisher's exact p-value above 0.05. Pooled responder recurrence estimates used DerSimonian–Laird random-effects meta-analysis; because between-cohort heterogeneity was absent ($I^2 = 0\%$), the fixed-effect inverse-variance estimate was numerically identical. Firth penalized Cox regression was implemented in Python from the Jeffreys-prior penalized Cox partial likelihood, with 95% confidence intervals obtained by profile penalized likelihood; the remaining small-sample and meta-analytic computations used R 4.5 (logistf, survival, and metafor).

Quality-adjusted pCR rates were computed for each treatment mechanism class as the product of the raw pCR rate and the structural retention rate. Three mechanism classes were defined: HER2-targeted monoclonal antibody therapy (pertuzumab ± trastuzumab), neratinib-containing regimens[28,29], and non-HER2-directed regimens. Pairwise differences in quality-adjusted rates were evaluated using bootstrap resampling (10,000 iterations). Structure-selectivity was quantified as the AUROC for entropy-based pCR discrimination within each mechanism class.

Multiple testing correction used the Benjamini–Hochberg false discovery rate (FDR) procedure[23] at $\alpha = 0.05$. Unless otherwise specified, bootstrap confidence intervals were computed from 10,000 resamples and permutation tests used 10,000 random permutations. All statistical analyses were performed in Python 3.10 using NumPy, SciPy, pandas, scikit-learn[26], lifelines, statsmodels, and PyTorch. Significance was set at $\alpha = 0.05$ unless otherwise specified. Directional concordance tests (sign and exact binomial tests) were one-sided; all other reported p-values are two-sided unless otherwise specified.

**2.8 Molecular cohort and pathway methodology**

Pre-treatment gene expression profiles were obtained from the publicly available I-SPY2 expression dataset (GEO accession GSE194040)[40] for 699 of 701 I-SPY2 patients (full-cohort discovery substrate) with structural classification (220 pCR, 479 non-pCR). The original discovery analysis used 63 pCR patients with matched Tier 1/Tier 2 assignments (49 Tier 1, 14 Tier 2). Expanded replication and extension analyses used the full 220-patient pCR cohort (Supplementary Note D11) (113 entropy-low, 107 entropy-high; including 68 HER2-positive) and 479 non-pCR patients (244 entropy-low, 235 entropy-high; including 95 HER2-positive). The replication strategy followed a five-stage design: discovery in the original tier-labelled cohort, expanded within-cohort replication, extension into non-pCR residual disease, cross-context synthesis, and arm-stratified mechanism analysis. This design applied conventional replication principles across sequentially wider biological contexts: pre-specified targets, an identical analytical framework, patient-level separation between the discovery and expanded within-I-SPY2 replication subsets, and directional concordance assessment. I-SPY1/GSE22226 supplied the external-platform consistency layer. Drug-mechanism convergence analyses combined pathway-level interaction testing in the discovery cohort with patient-level composition testing (Fisher's exact), pathway-profile correlation (Spearman), and structure-selectivity interaction in the expanded arm-stratified cohort. PAM50[25]-adjusted analyses used subtype indicators available for 219 of 220 pCR and 474 of 479 non-pCR patients.

Pathway enrichment was assessed using the 50 Hallmark gene sets from the Molecular Signatures Database (MSigDB)[38]. Single-sample gene set enrichment analysis (ssGSEA)[39] was performed to compute per-patient pathway scores. Full-rank pathway shift testing[42] compared Tier 1 versus Tier 2 distributions across the ranked transcriptome using a two-stage Mann–Whitney framework: gene-level signed ranking by differential-expression significance, followed by pathway-level in-set versus out-of-set comparison of the ranked gene statistics, with Benjamini–Hochberg false discovery rate correction across the 50 Hallmark pathways (Supplementary Note D10). The pathway-shift statistic tests whether genes belonging to a pathway are preferentially concentrated toward one side of the gene-level differential-expression ranking; it is not a patient-level pathway-activity score and should not be interpreted as one. PAM50 adjustment was performed by residualising expression matrices against PAM50 subtype indicators prior to pathway scoring.

Over-representation analysis (ORA) used the top 500 differentially ranked genes between Tier 1 and Tier 2, tested against Hallmark pathways using Fisher's exact test with Benjamini-Hochberg FDR correction. Gene-level differential expression used Mann-Whitney U tests with FDR correction. Drug-arm definitions for the mechanism-dependence analysis were: ADC arm (T-DM1 plus pertuzumab), neratinib-containing arm, and other HER2-directed arms. The delta-of-deltas permutation test compared the Tier 1 versus Tier 2 pathway separation magnitude between the ADC and neratinib arms using 10,000 permutations with Benjamini-Hochberg FDR correction.

Two complementary estimators were applied to the pooled 220-patient pCR cohort (Supplementary Note D9). The grouped entropy-split analysis tested whether high- and low-entropy groups separated on pathway-score distributions (§3.2.1), while the full-rank pathway-shift framework (defined above) tested pathway-direction enrichment along the ranked transcriptome for cross-context comparability (§3.2.3 and Figure 8). Both results are reported explicitly; neither is substituted for the other. The grouped analysis tests cohort-level two-group biological coherence; the full-rank analysis tests pathway-direction structure. Distinct null behaviour between estimators is expected when a cohort shows pathway-direction signal without group-level coherence.

### 2.8.1 Full-cohort structural-axis biological attribution

A separate exploratory analysis asked whether Entropy_T0 itself carries measurable molecular context across the full baseline cohort, independently of pathologic response. The MRI–RNA identifier bridge matched 704 subjects; 699 had a valid Entropy_T0 value together with expression data, and 698 had complete PAM50 and baseline functional tumour volume, giving 524 structurally favourable and 174 structurally adverse patients under the locked operator (adverse, Entropy_T0 > 3.284; favourable, Entropy_T0 ≤ 3.284). One patient had a genuine source-level missing PAM50 value. This full-cohort attribution substrate is distinct from the response-conditioned PAM50-adjusted analyses above, which used 219 of 220 pCR and 474 of 479 non-pCR patients (693 subtype-labelled patients in total); both denominators are correct for their respective estimands and are not interchangeable. Patient-level attribution used the nine exact precomputed pathway scores already present in the formal score table, together with a seven-pathway immune/inflammatory composite, modelled against continuous Entropy_T0 and against the locked binary structural state with adjustment for PAM50 subtype and log-transformed baseline functional tumour volume. Effects are reported as standardised coefficients with HC3 robust inference and incremental $R^2$, with Benjamini–Hochberg correction within the nine-pathway panel and leave-one-pathway-out robustness for the composite. The nine pathways had been defined previously for the response-quality molecular programme; their application to full-cohort Entropy_T0 attribution was exploratory, and a patient-level 50-Hallmark analysis was therefore retained as a broader sensitivity family, alongside a genome-wide gene-level scan and a direct patient-level EMT assessment. Recoverability of the structural phenotype from expression was assessed in the 698-patient MRI–RNA–PAM50–FTV substrate using five fixed stratified outer folds and four-fold inner cross-validation. Gene eligibility, rare-value imputation, log-transformed functional-tumour-volume standardisation, PAM50 and volume gene-wise nuisance regression and residual-variance filtering were estimated within training folds only and applied unchanged to held-out patients. The primary classifier selected the number of genes within inner cross-validation, followed by training-only scaling and L2-regularised logistic regression. Significance was assessed with 500 target-label permutations under the fixed nested partitions, repeating all target-dependent selection and fitting steps. No pathologic-response or recurrence information entered the model. Continuous Entropy_T0 recoverability and complementary raw-RNA, all-gene and principal-component representations were evaluated as sensitivity analyses. Full attribution, recoverability, estimand and direct-surrogate audits are provided in Extended Data Figure 6, Supplementary Tables S63–S75 and Supplementary Note D25.

### 2.9 Duke adjuvant and Duke 27 methodology

Adjuvant therapy records for the Duke cohort were obtained from linked clinical workbooks providing per-patient records for four modalities: adjuvant radiation, adjuvant chemotherapy, adjuvant endocrine therapy, and adjuvant anti-HER2 therapy. Sequential Cox proportional hazards modelling was performed with entropy as the variable of interest, adjusting for Nottingham histologic grade, nodal stage, molecular subtype, and all four adjuvant modalities (eight-variable model).

The Duke 27 subgroup comprised HER2-positive patients who received neoadjuvant chemotherapy and achieved strict pathologic complete response (institutional pCR code = 1, excluding patients with residual DCIS or LCIS). Sensitivity analyses included patients with DCIS/LCIS codes. Treatment-stack membership was modelled using logistic regression with neoadjuvant anti-HER2 therapy, neoadjuvant chemotherapy, adjuvant anti-HER2 therapy, adjuvant radiation, and baseline entropy as predictors.

### 2.10 I-SPY1 framework methodology

The I-SPY1 external validation cohort (ACRIN 6657) comprised 153 patients with available structural and outcome data. Framework analysis used 120 patients after applying locked eligibility criteria. The full-cohort framework crossed binary pCR status against binary structural classification (MAD above/below the locked I-SPY2-derived threshold). The bilateral framework restricted analysis to response extremes (RCB-0 and RCB-3 only, n = 70).

For pooled responder recurrence analysis, the pathology-confirmed I-SPY1 pCR denominator was locked at 21 patients and the UCSF best-response proxy denominator at 12 patients through a complete funnel audit documented in Supplementary Table S27. Pooled hazard ratios were estimated using DerSimonian-Laird random-effects meta-analysis. UCSF follow-up was harmonised from weeks to days. Follow-up distributions and descriptive cross-group follow-up comparisons are reported in Supplementary Table S30; no follow-up-based prefilter was applied to any survival analysis. Two further analyses characterised the external structural coordinates. First, within each cohort the relationship between the native continuous coordinate and the categorical crosswalk assignment was evaluated using categorical-assignment AUC, pairwise concordance, exact-threshold recoverability, median-split agreement and Cohen's κ, and the stability of the native recurrence ordering was assessed across all admissible descriptive cutpoints and by leave-one-event-out and leave-one-patient-out Cox refitting (Supplementary Tables S77 and S78). Cutpoint analyses assessed directional stability only and were not used to select or validate a clinical threshold. Second, because the two cohorts encode architecture on incompatible numerical scales, native MAD in I-SPY1 and the frozen two-dimensional structural distance in UCSF were converted independently to outcome-blind within-cohort percentile ranks, oriented so that higher ranks denote greater structural risk, and the common percentile coordinate was evaluated in a cohort-stratified Cox model expressed per 25-percentile increase, with 100,000 within-cohort association permutations, fixed administrative-censoring horizons and reverse-Cox assessment of the continuous score against censoring (Supplementary Tables S76 and S77). The rank analysis uses no recurrence information in the transformation, assumes no common raw-unit scale, selects no threshold and does not use the categorical crosswalk.

### 2.11 I-SPY1 BPE-matched pCR cohort and within-pCR biological consistency

The I-SPY1 BPE-matched pCR cohort comprises 83 I-SPY1 patients with both baseline structural-axis assignment and matched gene-expression data, of whom 23 carry a full Tier 1 / Tier 2 assignment (Tier 1 n = 18; Tier 2 n = 5; Supplementary Note D22). This cohort is distinct from the GSE22226 same-patient recurrence bridge (n = 19; the n = 16 GPL1708-only subset is a platform-sensitivity analysis, see §2.13/§2.14): the I-SPY1 cohort is a BPE-imaging-anchored expression subset; the GSE22226 bridge is a recurrence-event-anchored expression subset. The I-SPY1 cohort supports two complementary analyses: (a) a within-pCR biological consistency test (Tier 1 vs Tier 2 immune composite contrasts), and (b) a BPE-immune negative control (whether background parenchymal enhancement carries the immune programme that the structural axis indexes). Within-pCR contrasts used Mann-Whitney U with Cliff's δ effect size; pathway-level concordance was assessed by binomial sign test. PAM50 rank-residualisation tested whether subtype balance accounts for tier separation. BPE-immune correlation used Spearman's rank correlation, both unadjusted and PAM50-adjusted (partial correlation conditioning on PAM50 indicators); three sensitivity strategies were applied (unadjusted, PAM50-with-missing-category, PAM50-complete-case). Leave-one-out (LOO) stability was computed across all Tier 2 iterations.

### 2.12 Locked-nine immune–architecture panel construction and per-tier scoring

The locked-nine immune–architecture panel comprises seven immune/inflammatory Hallmark pathways from MSigDB[38] — IFNγ response, IFNα response, IL6/JAK/STAT3 signalling, inflammatory response, TNFα signalling via NF-κB, complement, and allograft rejection — together with two cell-state pathways, apoptosis and EMT, forming a structured 7+2 architecture. EMT was the strongest cell-state pathway-shift signal in the original discovery analysis. To keep the composite definitions explicit, we use three terms consistently: 'locked-nine architecture composite' for the unweighted mean of all nine pathway scores; 'non-EMT eight-pathway programme composite' for the seven immune/inflammatory pathways plus apoptosis; and 'seven-pathway immune/inflammatory module' for the immune/inflammatory pathways alone. The panel was pre-specified from the original pathway-shift discovery analysis in the I-SPY2 mechanism cohort and applied without modification across subsequent cohorts. Pathway scores were computed using single-sample gene set enrichment analysis (ssGSEA)[39] on log-transformed expression matrices, with tau = 0.25 and cohort rank-normalisation. Cohort-percentile positioning ranks each patient's score against the full I-SPY2 distribution. Within-tier substate decomposition (§3.2.4) used pre-locked k = 2 K-means clustering on the locked-nine architecture composite within I-SPY2 Tier 2; cluster assignments were locked from the original cohort-percentile analysis and were not re-derived in this study.

### 2.13 GSE22226 transcriptomic bridge methodology

Public expression data from the I-SPY1 / ACRIN 6657 GSE22226 dataset (149–150 baseline expression samples on Agilent G4502A arrays; platforms GPL1708 and GPL4133) were downloaded from GEO. The cross-walk between the locked-21 I-SPY1 HER2-positive pCR imaging-outcome cohort and GSE22226 captured 16 of 21 patients (events 1090 and 1098 absent from public expression). Probe-to-gene mapping used the most-variable-probe rule per gene; cross-platform harmonisation was applied where both GPL1708 and GPL4133 arrays were present. Pathway scoring used ssGSEA on the 50 Hallmark gene sets plus the locked-9 immune panel. The MAD-z structural axis reconciliation (the threshold-reconciliation procedure) projected the I-SPY2 entropy threshold onto the I-SPY1 expression cohort using a locked threshold (0.056217) that reproduces all 21 reference tier labels in the imaging-outcome cohort. Within-tier transcriptomic placement (Step 6 / Step 6.5) tested HER2-positive pCR Tier 1 vs Tier 2 (8 vs 8) and all-pCR Tier 1 vs Tier 2 (15 vs 14) using Mann-Whitney U with bootstrap stability (1,000 iterations) and leave-one-out audit. Step 6.6 tested Tier 4 inflammatory reactivation in the LumB non-pCR subgroup (n = 18). Step 7.1 tested within-RCB-0 transcriptomic shadow (n = 29: 15 favourable, 14 adverse). Step 7.2 tested across-RCB non-monotonic gradient (RCB-0 / I / II / III) with Spearman trend tests and per-class composite contrasts (Supplementary Note D24).

**2.14 Same-patient recurrence-immune bridge methodology**

The same-patient bridge provides the transcriptomic-recurrence link in the I-SPY1 GSE22226 substrate. Two cohort substrates were used: a primary HER2-positive pCR cohort (n = 19, 4 RFS events) and an all-pCR sensitivity cohort (n = 36, 6 events). Per-pathway recurrer-vs-non-recurrer contrasts used Mann-Whitney U with rank-biserial correlation; FDR was applied within the locked-9 panel. Absolute percentile positioning compared recurrer and non-recurrer composite scores against the full GSE22226 background distribution. A 50-Hallmark specificity check tested whether the locked-9 enrichment was specific or part of a broader transcriptomic shift. A GPL1708-only platform sensitivity (n = 16, 4 events) tested platform-independence. Joint tier × immune logistic models (n = 19, 4 events) tested whether structural-axis tier assignment and immune composite carry independent or redundant information about recurrence. The dedicated AUROC analysis applied contrasts (immune-only, structural-only, joint) with paired permutation testing for joint additivity. PAM50 Her2-enriched subset (n = 10, 3 events) tested subtype-conditional behaviour. All cohort denominators are reported with explicit substrate naming: every recurrence claim names its substrate.

**2.15 Bulk-transcriptomic hypothesis sub-analysis methodology**

The bulk-transcriptomic hypothesis sub-analysis tests pre-specified explanations for the broad immune depletion observed in Cluster 1 within recurrence-enriched Tier 2. Five hypotheses were declared in advance: H1 stromal-hypoxic exclusion (immune cells physically excluded by desmoplasia, hypoxia, or stromal remodelling); H2 immune-recognition/cytotoxic deficit (reduced antigen presentation and effector cytotoxicity); H3 pure cytotoxic deficit; H4 mixed/TGFβ-mediated mechanism; and NULL_RESULT. The pre-locked decision rule required cross-cohort directional concordance for hypothesis lock — a signature could not anchor a hypothesis on the strength of one cohort alone. Seven gene-level z-mean composites were computed: cytotoxic/TIL (CD8A, CD8B, GZMB, PRF1, GZMA, NKG7, IFNG); antigen presentation (HLA-A, HLA-B, HLA-C, B2M, TAP1, TAP2, HLA-DRA, HLA-DPB1); checkpoint/exhaustion (CD274, PDCD1, CTLA4, HAVCR2, LAG3, TIGIT); stromal/desmoplasia (COL1A1, COL1A2, COL3A1, FAP, ACTA2, S100A4, POSTN); TGFβ axis (TGFB1, TGFB2, SMAD2, SMAD3, SERPINE1); hypoxia/glycolysis (HIF1A, VEGFA, CA9, SLC2A1, LDHA, PGK1, ENO1, ALDOA); and angiogenesis (VEGFA, KDR, FLT1, ANGPT2, PECAM1, VWF). Each composite is the mean of within-cohort z-scored member genes; scores represent within-cohort standardised signature activity rather than absolute cross-platform expression. The analysis layer used was the combined evidence layers — continuous entropy gradient and gene-level mechanistic signatures; an exploratory RCB-augmented layer was excluded after a data-integrity check showed that its post-merge RCB distribution was incompatible with the verified source distribution. No result from this layer is interpreted here. The exclusion is confined to that exploratory layer and does not affect the primary within-RCB structural analyses (the within-RCB-0 structural ceiling, Cox HR = 5.50, and the across-RCB recurrence gradient), which derive from the independently verified I-SPY1 imaging-outcome RCB denominator (n = 138; RCB-0 = 42, RCB-I = 11, RCB-II = 57, RCB-III = 28) and stand on their own. Gene-level contrasts used Mann-Whitney U with FDR within the seven-signature panel and rank-biserial correlation as the effect size.

**3. Results**

Pretreatment structural entropy was independent of pCR status but informative about response quality. In I-SPY2, entropy distributions were indistinguishable between pCR and non-pCR patients, yet 55 of 219 complete responders (25.1%) were structurally adverse before therapy (Figures 2, 3). External outcome cohorts showed that this was not merely a descriptive imaging phenotype: pathology-confirmed I-SPY1 pCR carried elevated recurrence risk, the UCSF best-response proxy was concordant, and the pooled responder HR was 2.87, with stronger enrichment inside RCB-0. This adverse complete-response state was invisible to 27 conventional clinical variables and contained a broadly immune-depleted bulk-RNA substate. This within-pCR recurrence concentration is the central result of the paper.

The framework is built and tested across three cohort systems with distinct roles. System A (I-SPY2 discovery, n = 696–701) constructs the framework and its biology but makes no recurrence claim: its internal event indicator is hormone-receptor-aligned rather than a validated recurrence endpoint, so all recurrence inference is made in Systems B and C; pathology-confirmed within-pCR inference derives from System B. System B (I-SPY1, n = 138, 42 RFS events) provides external recurrence and RCB validation; System C (pooled HER2-positive I-SPY1 pathology-confirmed pCR plus UCSF best-response proxy, n = 33, 10 events) provides the primary pooled responder recurrence estimate. The full per-system evidence matrices are in Supplementary Tables S54–S56.

### 3.1 Pretreatment DCE-MRI defines a four-tier response-quality framework that adds a structural axis orthogonal to pathologic response

#### 3.1.1 The structural entropy axis reproduces across representations and institutions

Baseline structural entropy is orthogonal to pathologic response. In I-SPY2 (n = 701), entropy distributions were statistically indistinguishable between pCR and non-pCR patients (Mann–Whitney p = 0.966; r_b = 0.002) (Figure 1), yet the within-pCR structural range remained wide (0 to 4.26) — pCR collapses a continuous structural axis to a single point. Orthogonality held across three computational families (Table 3: first-order entropy, handcrafted texture, and learned carrier features) and was not a pipeline artefact: two positive controls (background parenchymal enhancement p = 0.010, baseline longest diameter) confirmed the pipeline detects associations where they exist. Adding all ten structural features to a model already containing functional tumour volume did not improve pCR discrimination (likelihood-ratio p = 0.994), and orthogonality held within every molecular subtype, after full covariate adjustment, and across age, race, and menopausal status (Supplementary Table S1). The null replicated in four further institutions — Duke (n = 908, p = 0.709), I-SPY1 (n = 138, p = 0.080), QIN-BREAST (n = 39), and UCSF — five institutions across three scanner vendors. The structural axis is therefore reproducible and carries information pathologic response does not.

**Table 3 | Structural-axis pCR-orthogonality battery across three computational families.**

| Feature | Family | p-value | Effect size | Verdict |
|---|---|---|---|---|
| Shannon entropy | First-order | 0.966 | r_b = 0.002 | pCR-orthogonal |
| Median absolute deviation | First-order | 0.080 | — | Borderline pCR-orthogonal (p = 0.08) |
| GLCM DifferenceAverage | Texture (GLCM) | 0.747 | r_b = 0.015 | Near-orthogonal |
| GLDM DependenceEntropy | Texture (GLDM) | 0.010 | r_b = −0.120 | Weak exception (disclosed) |
| QIN frozen structural score | Learned (cross-vendor) | 0.175 | AUROC 0.636 | No significant pCR alignment |
| Background parenchymal enhancement | Enhancement | 0.010 | — | Positive control |
| Baseline longest diameter | Burden | 0.002 | — | Positive control |

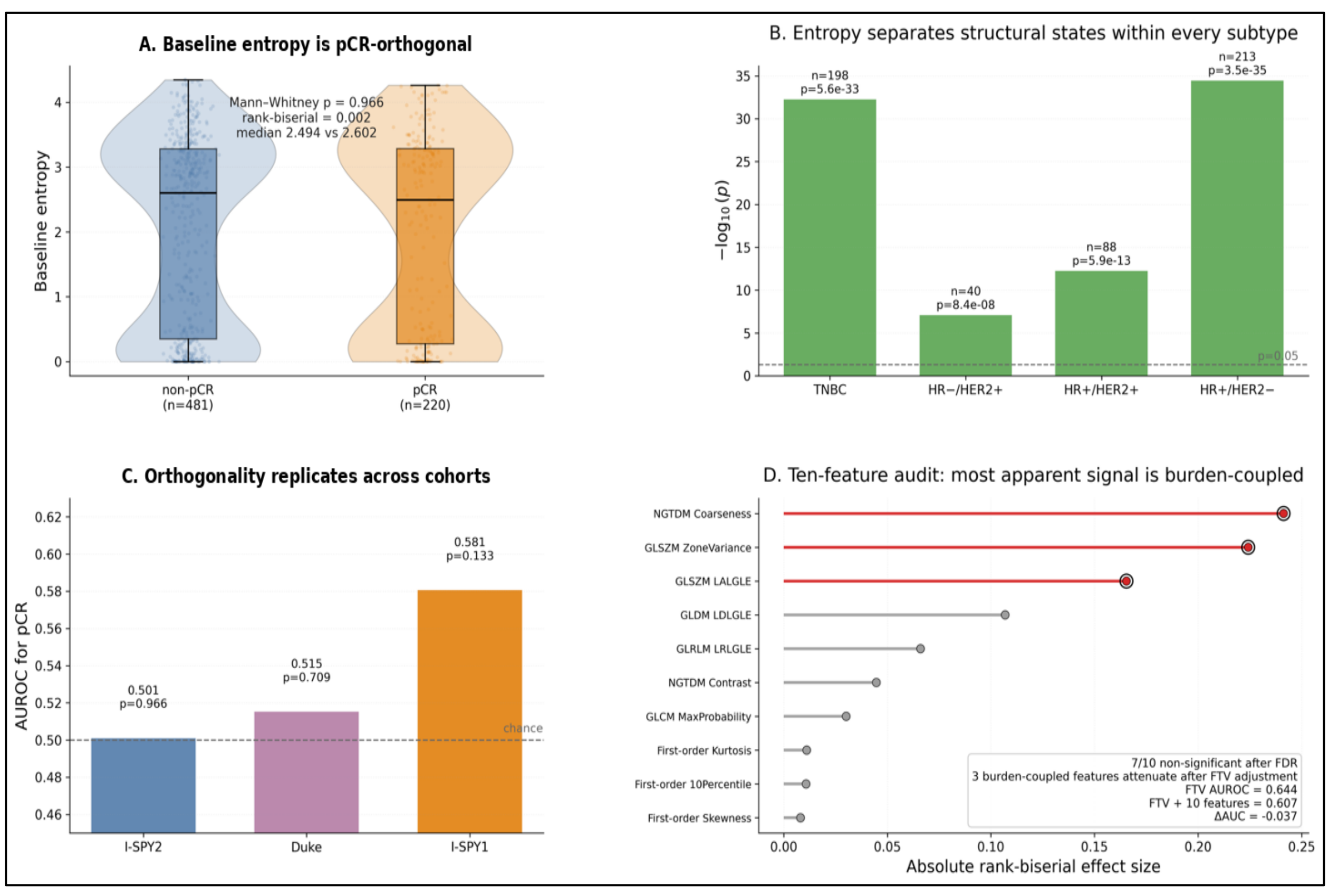

**Figure 1 |** ***The structural axis is orthogonal to pathologic complete response. a,*** *Baseline entropy in pCR (n = 220) and non-pCR (n = 481), I-SPY2. Mann–Whitney p = 0.966; rank-biserial rb = 0.002.* ***b,*** *Molecular non-redundancy: within-subtype entropy separation all p < 10−7 (companion structural-manifold analysis).* ***c,*** *Cross-cohort pCR-orthogonality with Duke NAT sensitivity: I-SPY2 p = 0.966, Duke p = 0.709, Duke NAT AUROC = 0.515, I-SPY1 p = 0.080.* ***d,*** *Ten-feature audit: 7/10 NS after FDR; 3 burden proxies lose significance after FTV adjustment. Adding all 10 to FTV reduced AUC by 0.037.*

### 3.1.2 Four-tier framework definition and within-axis prognostic structure

Crossing the structural axis with pathologic response defines four tiers (Table 2); within pCR, one in four complete responders is structurally adverse (Tier 2, 55 of 219 = 25.1%) (Figure 2). The framework's principal clinical claim is that structural quality cuts across the pCR boundary. In the I-SPY1 framework cohort (n = 120, 35 events), recurrence rises monotonically across tiers — 11.5%, 28.6%, 29.7%, 47.8% (4.1-fold, Tier 1 3/26 to Tier 4 11/23) — with a significant ordered trend (1-df log-rank $\chi^2 = 6.63$, p = 0.010) and a 5.09-fold separation between structural extremes (Tier 1 vs Tier 4 HR = 5.09, 95% CI 1.40–18.53, p = 0.006). Risk was not confined to one side of the pathology boundary. No prognostic separation was detected between structurally adverse complete response (Tier 2) and organised residual disease (Tier 3) (HR = 0.97, 95% CI 0.23–4.17, p = 0.968); however, the wide interval precludes a conclusion of equivalence. The comparison nevertheless shows that the binary endpoint alone does not resolve the response-quality boundary. At response extremes the gradient sharpens to 9.5% → 73.3% (bilateral framework, n = 70; log-rank $p = 8 \times 10^{-7}$; 7.7-fold range), and a structure × response interaction model outperforms the additive specification (likelihood-ratio p = 0.0018, ΔC-index = +0.071), confirming that the prognostic meaning of structure is response-context-dependent. Tier-level clinical interpretation is consolidated in Table 7.

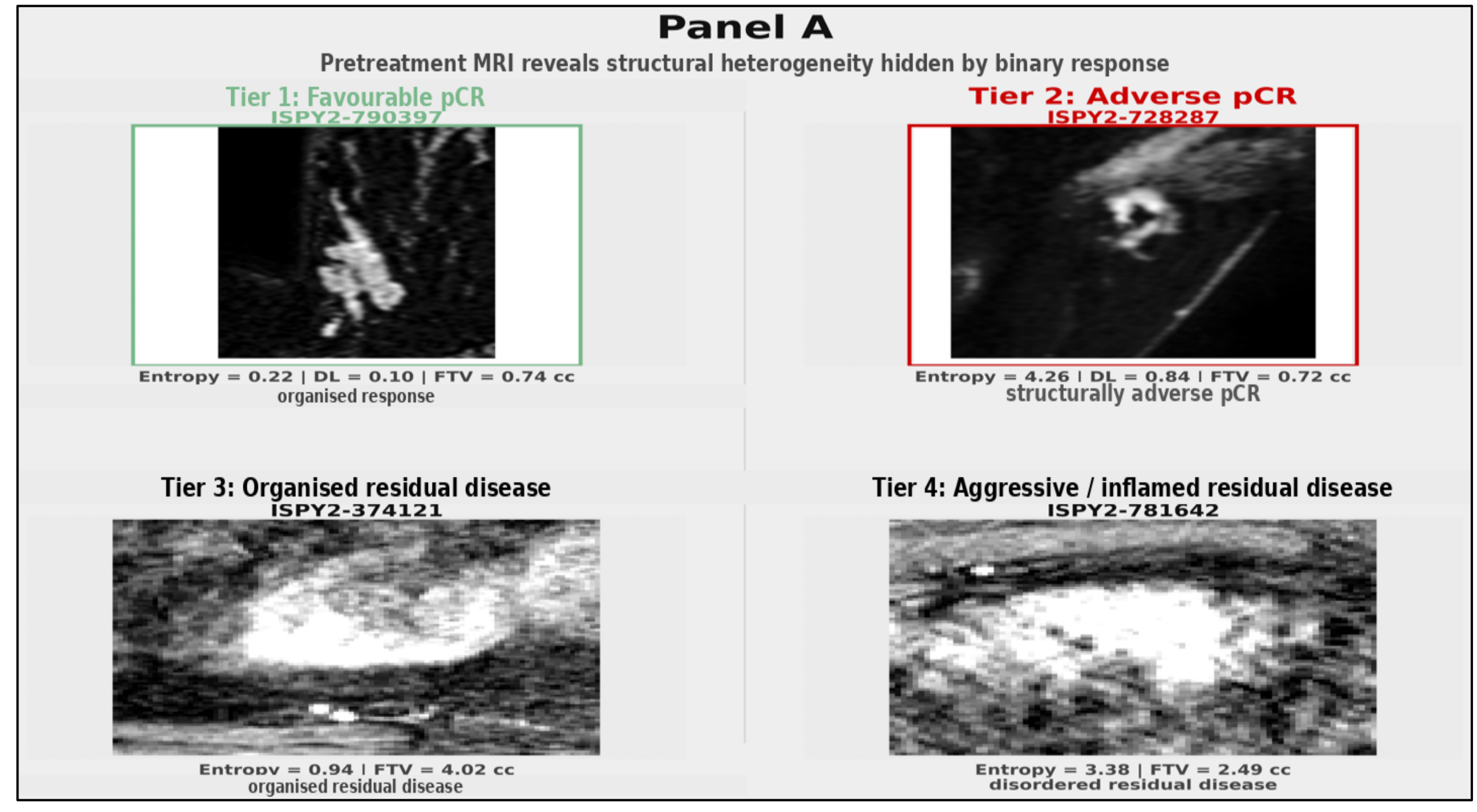

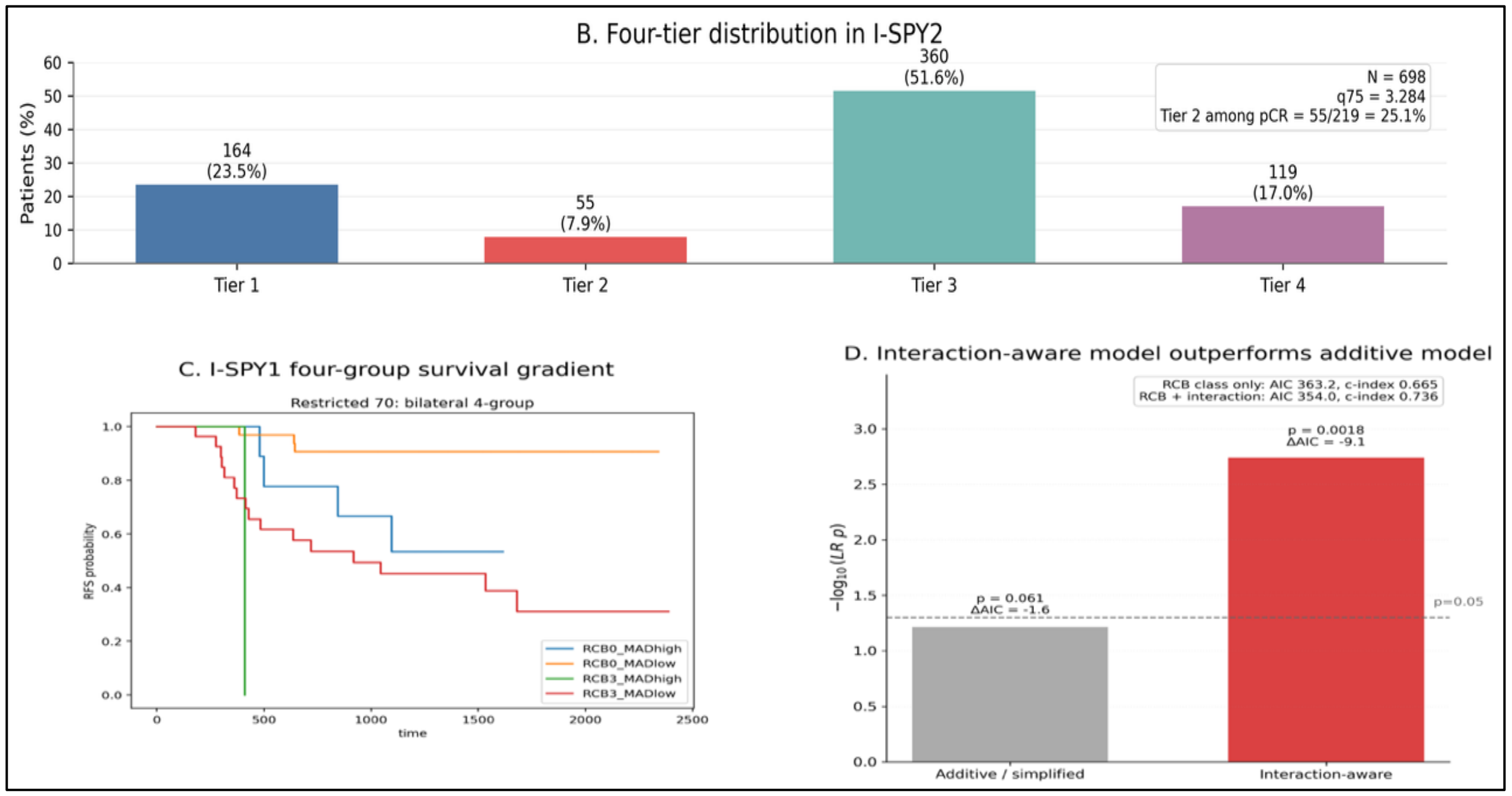

**Figure 2 |** ***The four-tier structural framework partitions four response-quality states within and beyond pathologic complete response. a,*** *Four-tier structural response-quality framework illustrated with representative pretreatment DCE-MRI from each tier. Top row: both patients achieved pCR with near-identical tumour burden (FTV = 0.74 vs 0.72 cc), yet baseline structural entropy differs nineteen-fold (0.22 vs 4.26). The binary endpoint labels them identically; the structural framework reveals organised response (Tier 1) versus structurally adverse response (Tier 2). Bottom row: within residual disease, structural quality separates organised residual disease (Tier 3, entropy = 0.94) from aggressive residual disease (Tier 4, entropy = 3.38). Framework defined in I-SPY2 at the training-partition 75th percentile (q75 = 3.284).* ***b,*** *Tier distribution in I-SPY2 (N = 698, master discovery substrate); Tier 2 comprises 25.1% of pCR patients (55 of 219).* ***c,*** *External validation in the I-SPY1 bilateral response-extreme framework (RCB-0/RCB-3, n = 70): four-group recurrence gradient 9.5% → 23.8% → 46.2% → 73.3% (log-rank $p = 8 \times 10^{-7}$). The binary endpoint marks the response/non-response boundary between groups two and three; the four-tier framework adds a within-category response-quality dimension at each end of the response axis.* ***d,*** *The four-tier framework is more than the sum of its parts: the interaction model (structure × response) significantly outperforms both the additive specification (LR p = 0.0018; ΔAIC = −9.1; ΔC-index = +0.071) and the simplified additive analogue (LR p = 0.061, non-significant), confirming that the prognostic meaning of structural quality is response-context-dependent.*

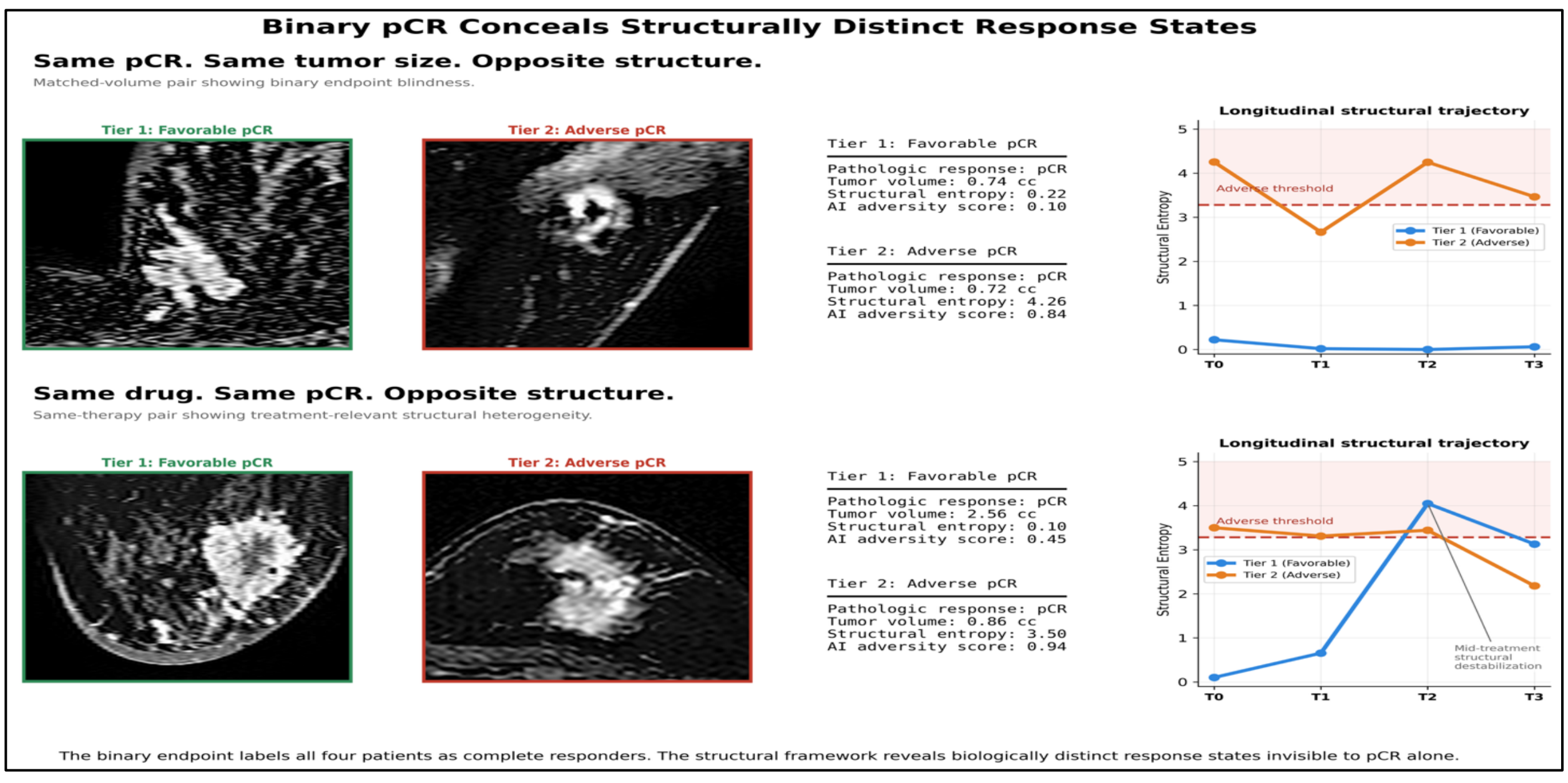

**Figure 3 |** ***Representative imaging examples of the structural phenotype.*** *Two matched pCR patient pairs illustrating structural heterogeneity despite identical pathologic outcome. Top row: burden-matched pair (baseline FTV difference, 0.028 cc). Left: Tier 1, entropy = 0.22, DL score = 0.10. Right: Tier 2, entropy = 4.26, DL score = 0.84. Both achieved pCR with nearly identical baseline tumour burden, yet their structural architecture differs markedly. Bottom row: treatment-matched pair (both treated with paclitaxel + neratinib). Left: Tier 1, entropy = 0.10. Right: Tier 2, entropy = 3.50. Despite the same treatment arm and the same pathologic outcome, the two patients exhibit opposite structural profiles. Pair 1 shows that pCR status and tumour burden alone do not distinguish these patients, whereas structural entropy does. Pair 2 shows that the same regimen can produce qualitatively different structural outcomes in different patients.*

### 3.1.3 Persistence, deep learning convergence, and cross-pipeline transport

The structural state is durable rather than transient. Across patients with complete longitudinal trajectories, baseline entropy persisted through treatment (within-pCR retention 81%, Spearman $\rho = 0.671$, $p = 4 \times 10^{-23}$; global structural-state persistence 88.2%) (Figure 4), and Tier 2 patients were categorically excluded from the persistently favourable low-to-low trajectory class (0/28; $\chi^2$ $p = 3.4 \times 10^{-7}$) — structurally adverse complete responders remained adverse at every timepoint despite identical pathologic clearance (Figure 4.1). A distinct deep-learning representation converged with the phenotype in the locked same-patient subset (conservative stability-aware AUROC = 0.738, permutation p = 0.007) (Extended Data Figure 1), with 87.5% of its signal non-redundant with entropy, and unsupervised clustering recovered the phenotype without outcome information. This is cross-representation support, not independent-patient validation. Full persistence, deep-learning convergence, and cross-pipeline transport analyses are in the Supplementary Information.

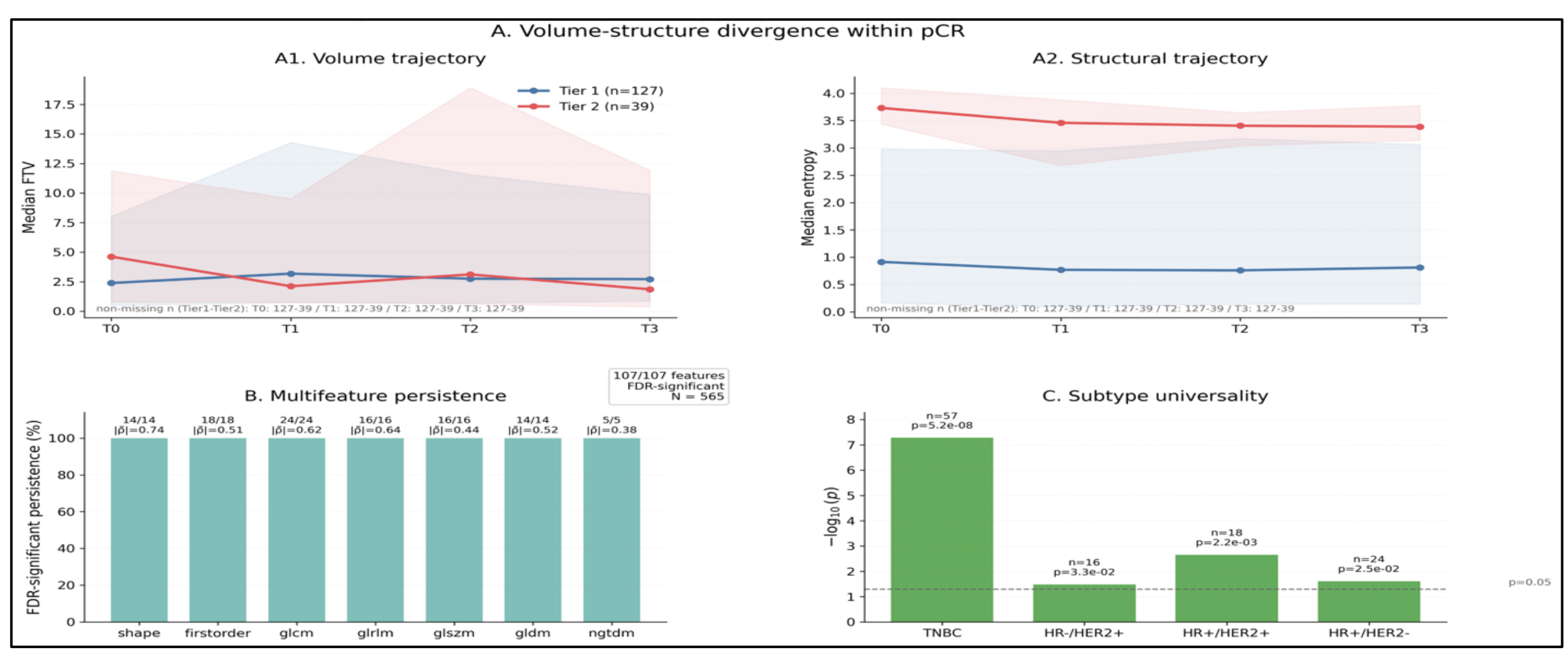

**Figure 4 | *Persistence and trajectory dissociation. a,*** *Within pCR, volumetric trajectories do not significantly separate structurally favourable from structurally adverse patients (all Mann–Whitney $p \geq 0.219$), whereas entropy remains strongly separated throughout treatment (T0 $p = 3.96 \times 10^{-21}$; T1 $p = 4.45 \times 10^{-9}$; T2 $p = 2.00 \times 10^{-8}$; T3 $p = 5.44 \times 10^{-8}$).* ***b,*** *Multifeature persistence across matched baseline and post-treatment scans (n = 565). All 107/107 features are FDR-significant. Shape is most persistent (median $|\rho| = 0.740$); NGTDM is least persistent (0.383).* ***c,*** *Subtype universality: all 4/4 molecular subtypes remain significant ($\rho = 0.457$–$0.673$).*

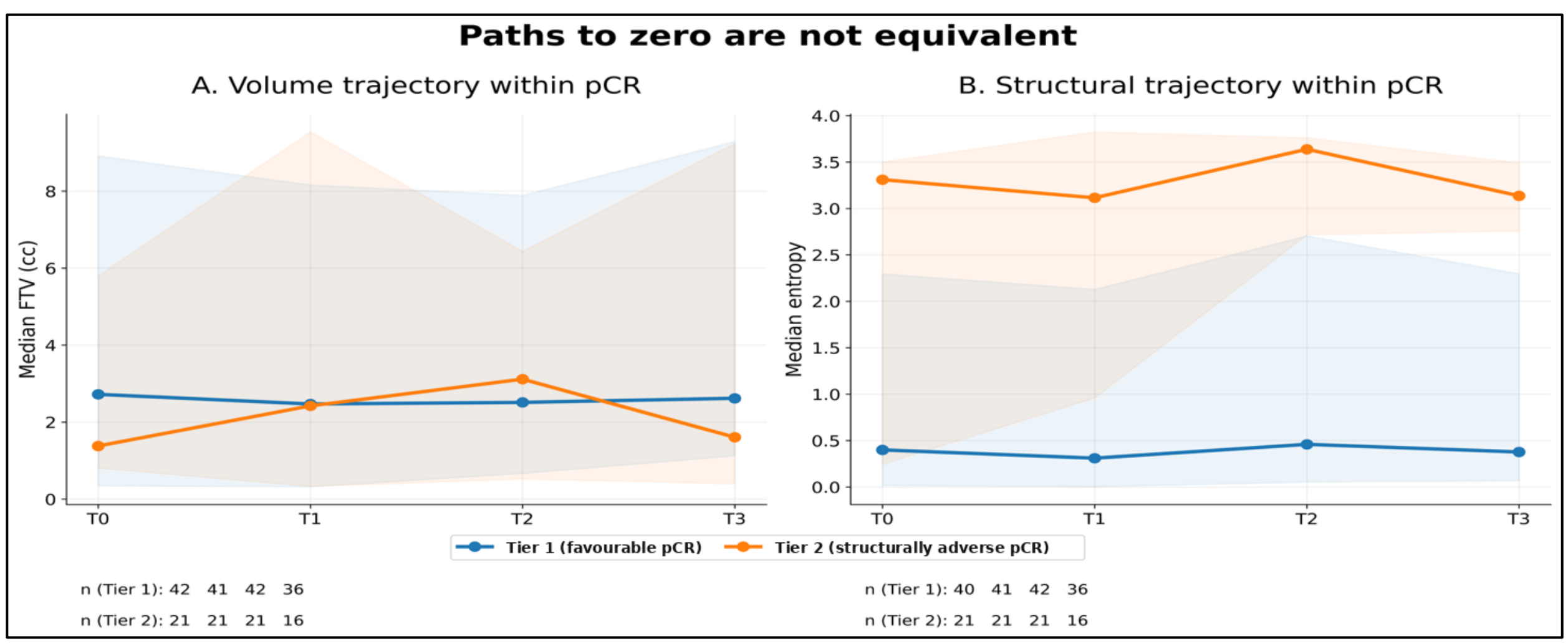

**Figure 4.1 | *Paths to zero are not equivalent: volumetric response converges while structural quality persists. a,*** *Median functional tumour volume (FTV) trajectories from baseline (T0) through pre-surgical assessment (T3) within the I-SPY2 within-pCR longitudinal substrate, stratified by structural tier (Tier 1, blue; Tier 2, orange). Both tiers converge to comparable volumetric end-states.* ***b,*** *Median structural entropy trajectories in the same patients. Tier 1 maintains low entropy throughout, whereas Tier 2 remains above the locked adverse threshold at every timepoint despite identical pathologic clearance (all between-tier Mann–Whitney $p < 10^{-7}$). Together the two panels show that patients reaching the same pathologic endpoint arrive along structurally different trajectories: the paths to zero are not equivalent. Solid lines, medians; shaded bands, IQR. Timepoint-level statistics are provided in Supplementary Figure S10.*

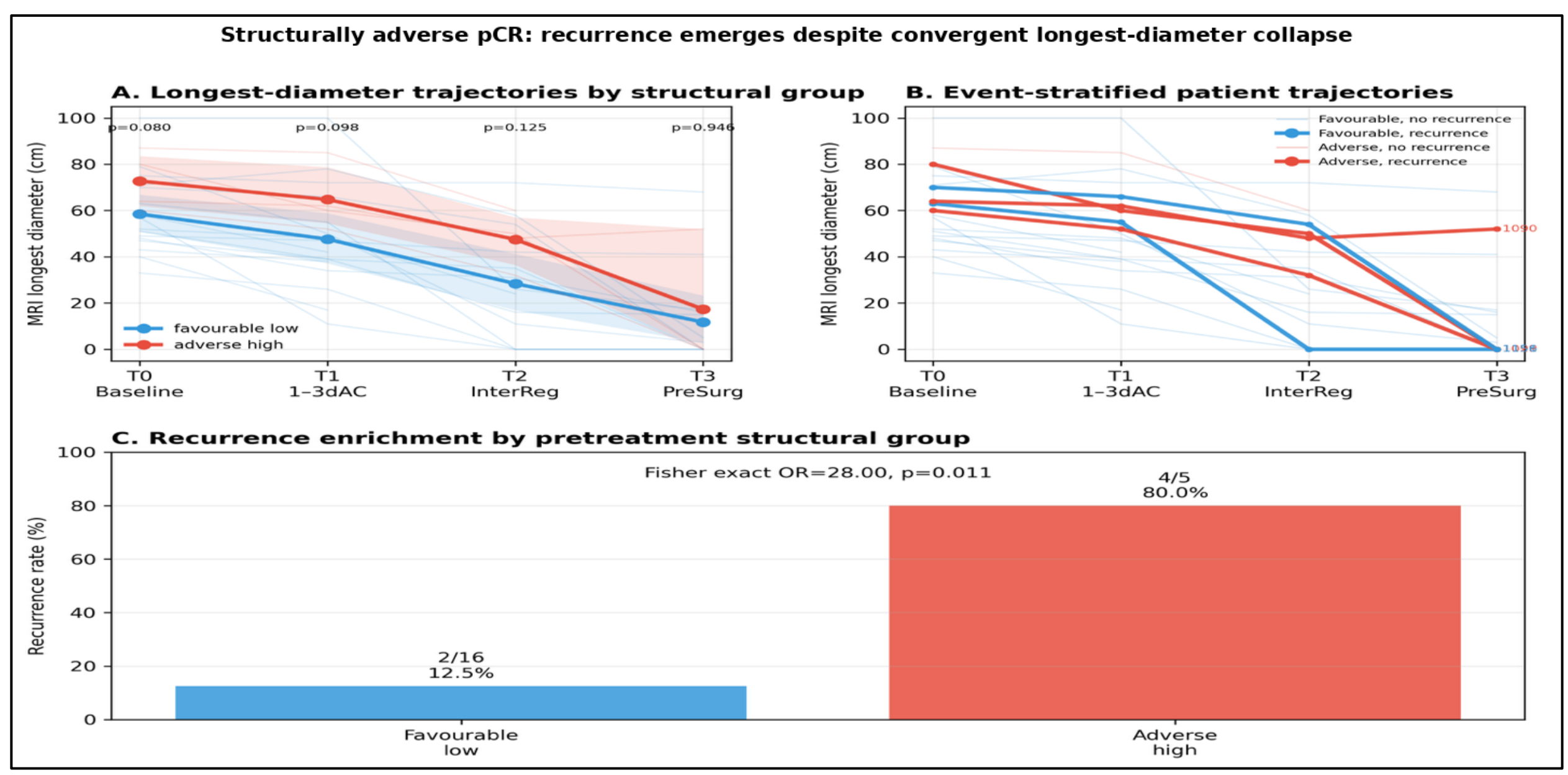

**Figure 5 | *Structurally adverse pCR: recurrence emerges despite convergent longest-diameter collapse.*** *Cross-cohort confirmation in the I-SPY1 HER2-positive RCB-0 concentration subset using the prespecified upper-tail MAD rule (n = 21; structurally favourable n = 16, blue; structurally adverse n = 5, red).* ***A,*** *Median MRI longest-diameter trajectories — the standard clinical metric — converge to ~0 at pre-surgical assessment (between-group $p \geq 0.08$ at every timepoint).* ***B,*** *The same patients plotted individually and stratified jointly by structural group and recurrence status; bold lines mark the recurrers within each group and faint lines the non-recurrers. Recurrence occurs despite marked or complete volumetric collapse, so the endpoint metric does not separate recurrers from non-recurrers.* ***C,*** *Recurrence enrichment within the same cohort: 12.5% (2 of 16) in structurally favourable versus 80.0% (4 of 5) in structurally adverse (Fisher exact OR = 28, p = 0.011). Despite the longest-diameter convergence in A and B, recurrence rate differs sharply between structural groups. Solid lines, medians; shaded bands, IQR; faint lines, individual patients.*

### 3.1.4 External cohort transport

A frozen I-SPY2-derived projection separated survival in UCSF (n = 49; restricted-mean-survival advantage 22.6 months, log-rank p = 0.0042, permutation p = 0.002), and in Duke (n = 908, 76 DRFS events) the favourable structural state remained associated with lower distant-recurrence risk under the frozen I-SPY2 threshold (HR = 0.466, 95% CI 0.232–0.935, p = 0.022; entropy regime-compatible under the representation-family framework, Cohen's d ≈ 0.02). A learned carrier feature reproduced the signal in QIN-BREAST (n = 39, a fifth institution on a third scanner vendor; AUROC = 0.790) (Figure 6). Full transport statistics appear in §3.4 and the Supplementary Information.

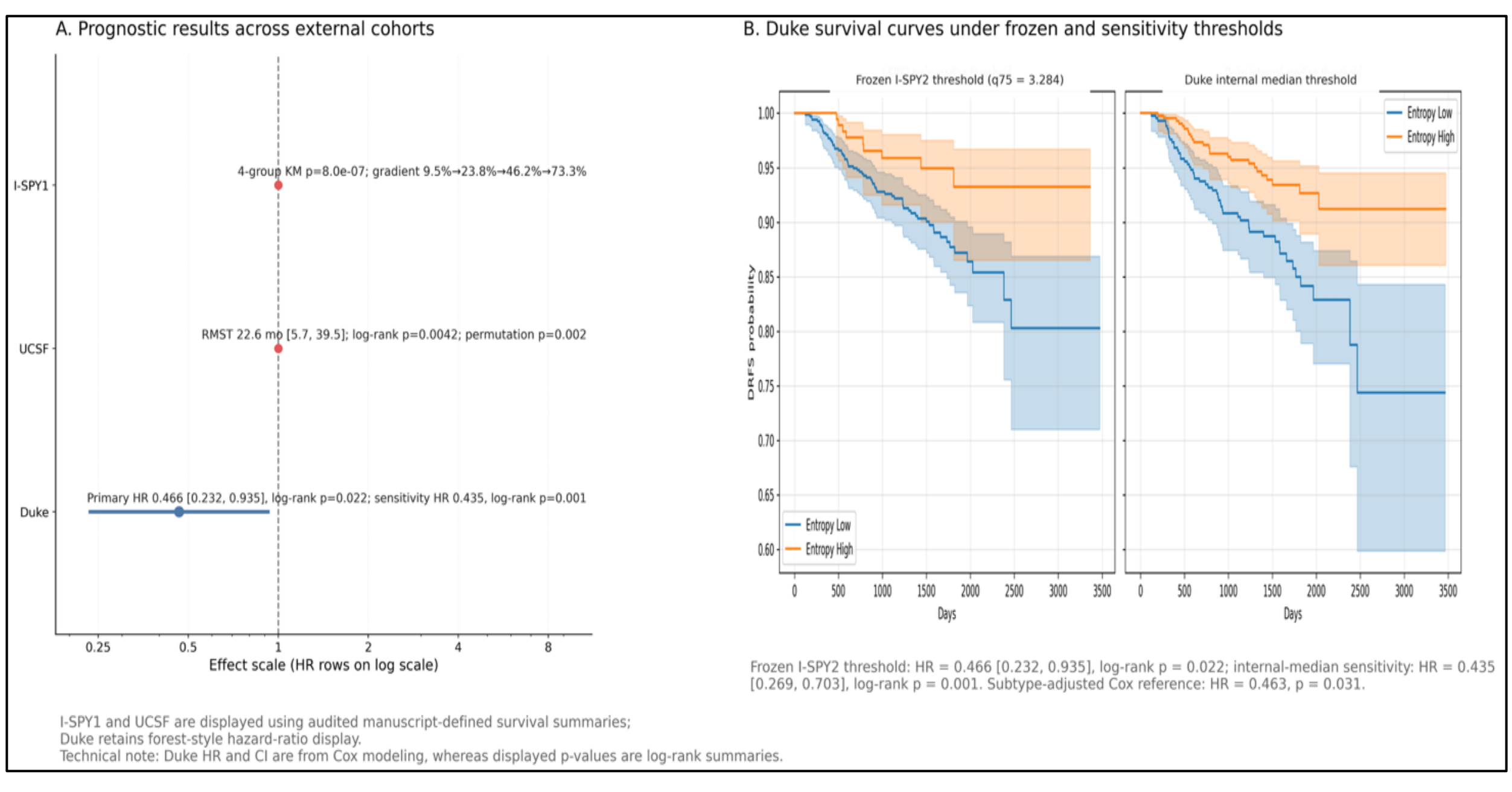

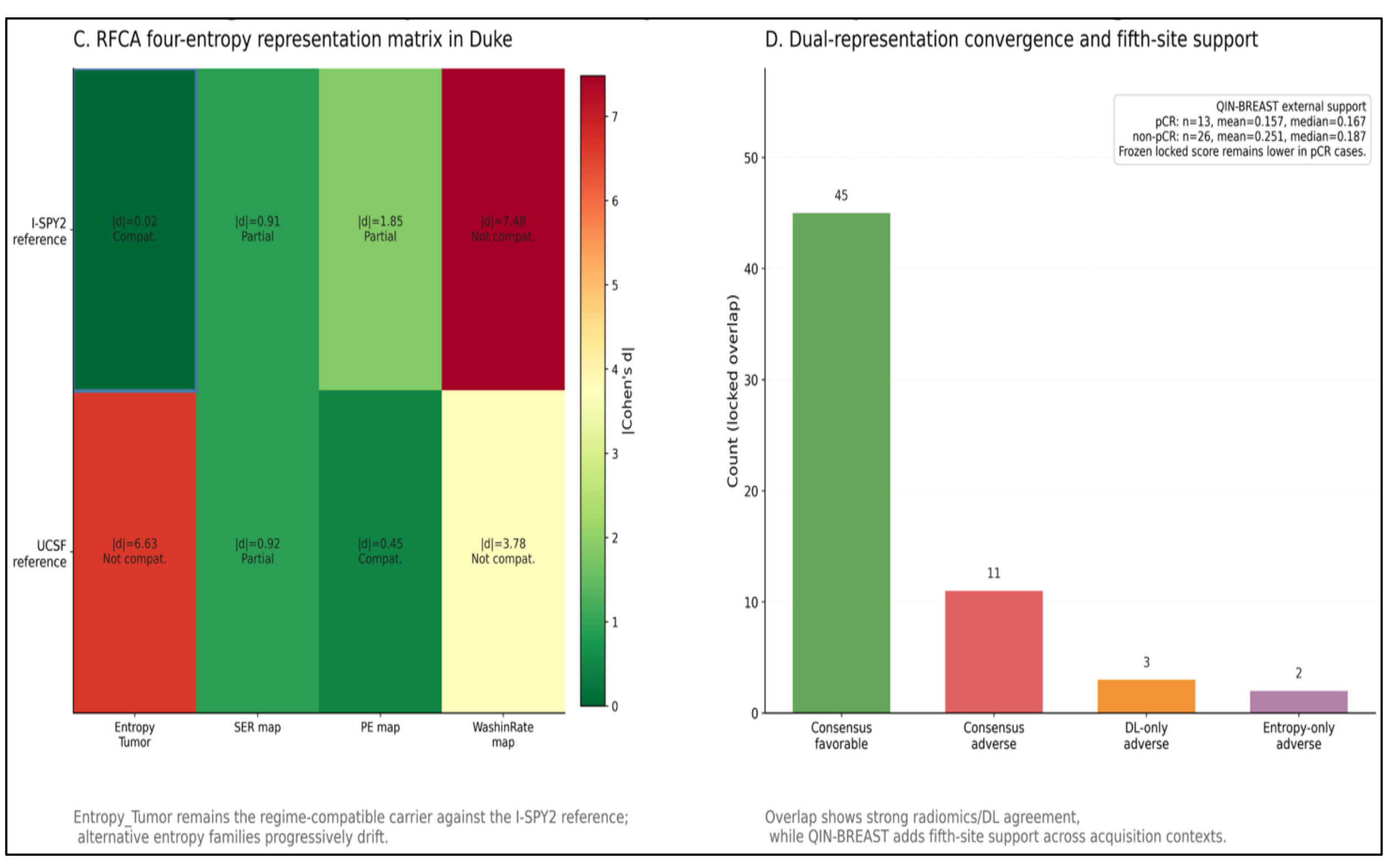

**Figure 6 | *External validation across independent cohorts.* *a,*** *Prognostic results across external cohorts. I-SPY1 is shown as a four-group bilateral survival gradient (log-rank p = 8.0 × $10^{-7}$; event-rate gradient 9.5% → 23.8% → 46.2% → 73.3%). UCSF shows convergent external survival association under frozen projection (RMST advantage 22.6 months [5.7, 39.5], log-rank p = 0.0042, permutation p = 0.002). Duke shows pre-specified external validation using the frozen I-SPY2 threshold (primary HR 0.466 [0.232, 0.935], log-rank p = 0.022), together with sensitivity-threshold results (HR 0.435 [0.269, 0.703], log-rank p = 0.001).* ***b,*** *Duke Kaplan–Meier survival curves using the frozen I-SPY2 threshold (primary) and Duke-internal median threshold (sensitivity).* ***c,*** *RFCA four-entropy representation matrix in Duke. Entropy_Tumor is regime-compatible against the I-SPY2 reference (|d| ≈ 0.02), whereas alternative entropy families show progressive representation shift.* ***d,*** *Dual-representation convergence and fifth-site support. Locked radiomics/deep learning overlap shows strong agreement within the structural framework, while QIN-BREAST provides external support that the frozen locked score remains lower in pCR than non-pCR cases across acquisition contexts.*

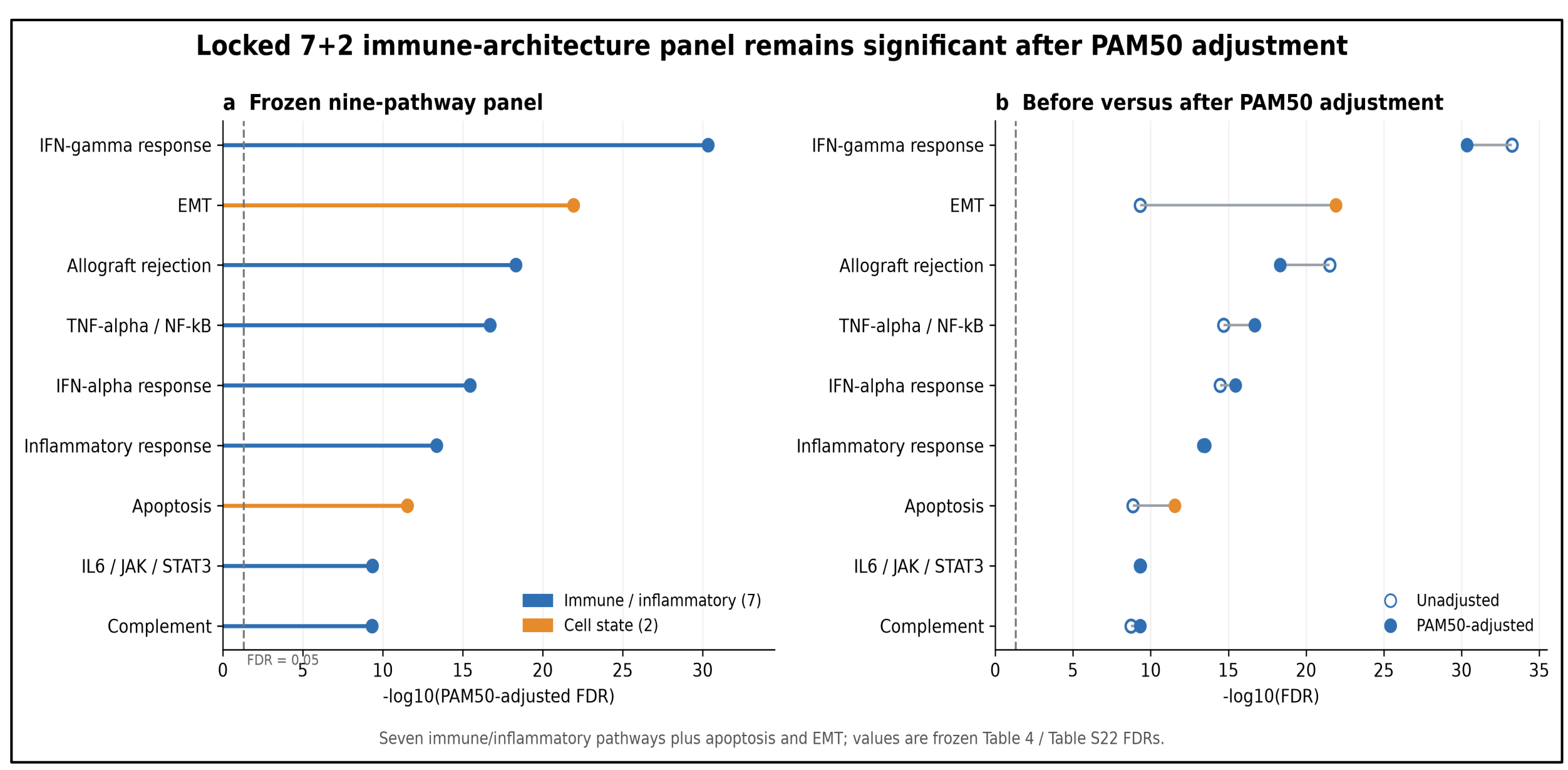

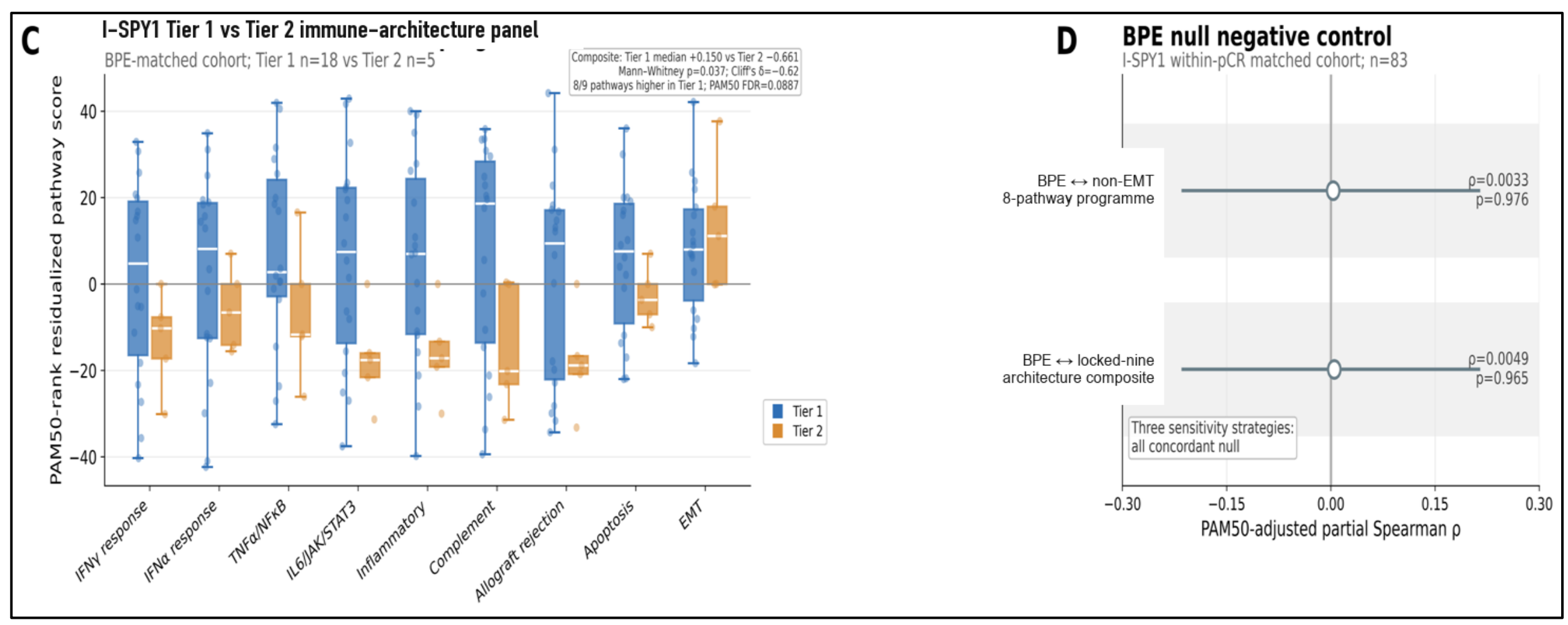

**Figure 7 | Within-pCR immune–architecture panel is PAM50-independent and BPE-independent.** ***a,*** *Locked Hallmark pathway-shift panel (System A I-SPY2 within-pCR; structurally favourable vs structurally adverse). The dominant signal is immune: IFNγ response, allograft rejection, TNFα/NF-κB, IFNα response, inflammatory response, IL6–JAK–STAT3, and complement remain enriched after PAM50 residualisation. The panel comprises seven immune/inflammatory pathways together with two cell-state pathways, apoptosis and EMT; all nine show favourable-side enrichment after PAM50 adjustment. Points, −log10(PAM50-adjusted FDR); dashed line, FDR = 0.05.* ***b,*** *Pathway-level significance before and after PAM50 adjustment (open circles, unadjusted; filled circles, PAM50-adjusted). The immune programme remains highly significant; EMT strengthens with adjustment.* ***c,*** *Cross-cohort consistency in the I-SPY1 BPE-matched cohort (System B; Tier 1 n = 18, Tier 2 n = 5). PAM50-residualised patient-level pathway scores; Tier 1 (blue) > Tier 2 (orange) in 8 of 9 pathways, with EMT the sole directional exception. Locked-nine architecture composite: Tier 1 median +0.150 vs Tier 2 −0.661 (Mann–Whitney p = 0.037; Cliff's δ = −0.62, large effect); pathway-level FDR = 0.0887.* ***d,*** *BPE-confounder negative control. PAM50-adjusted partial Spearman correlations between pretreatment BPE and the non-EMT eight-pathway programme composite (top) or locked-nine architecture composite (bottom) in the I-SPY1 BPE-matched cohort (n = 83): both essentially zero (ρ = 0.0033, p = 0.976; ρ = 0.0049, p = 0.965). Three sensitivity strategies yield concordant null results. Boxes (panel c), IQR; whiskers, 1.5× IQR; points, individual patients.*

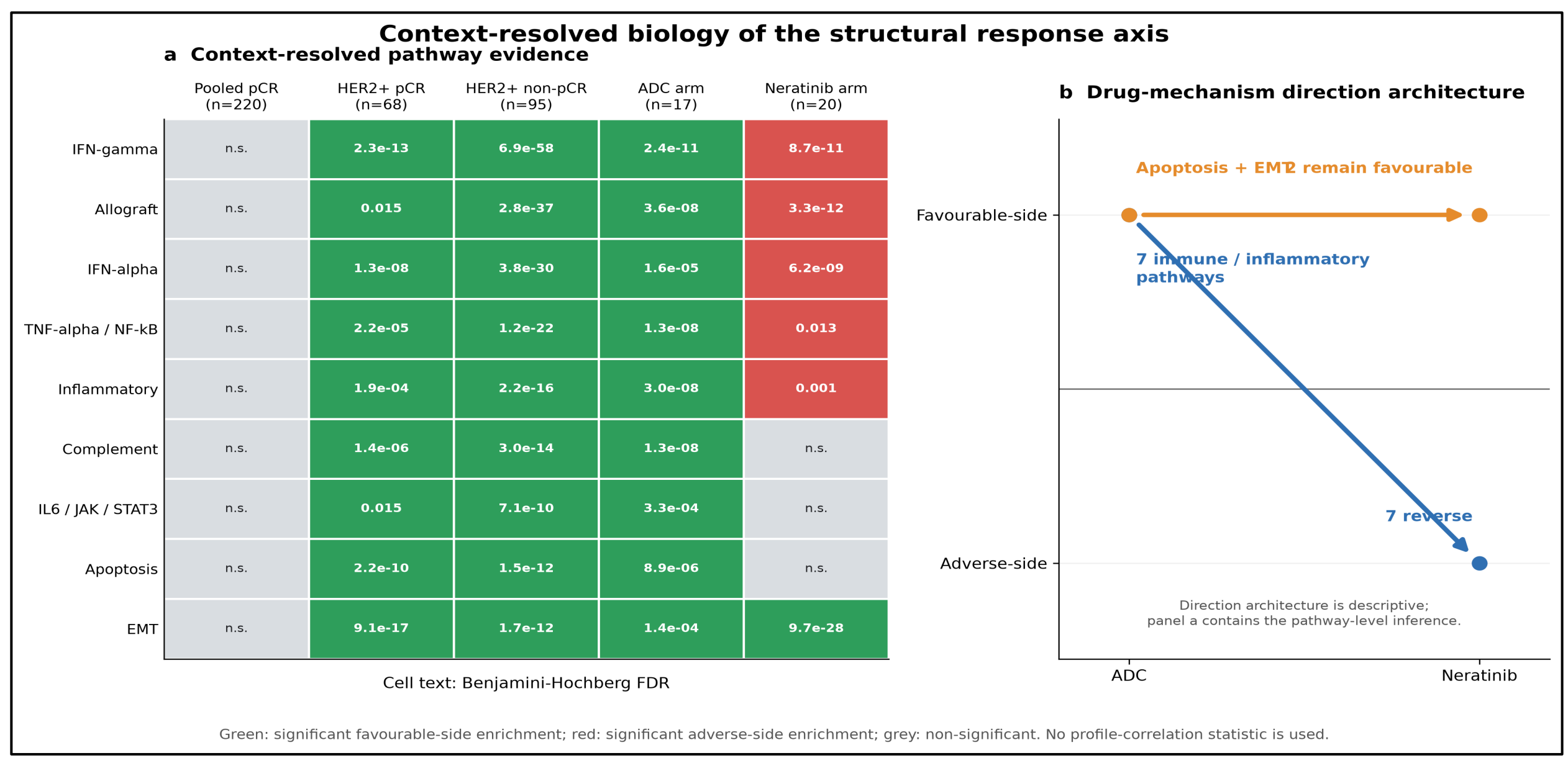

**Figure 8 |** ***Context-resolved biology of the structural response axis. a,*** *Nine-pathway × five-context heatmap showing pathway-level significance and direction across pooled pCR (n = 220, shown as full-rank pathway shifts for cross-context comparability; pooled grouped-analysis null reported in §3.2.1, all FDR = 0.892), HER2-positive pCR (n = 68), HER2-positive non-pCR (n = 95), ADC arm (n = 17), and neratinib arm (n = 20). All five contexts use the identical full-rank pathway-shift estimator; the first three are PAM50-adjusted, while the ADC and neratinib within-arm analyses are unadjusted owing to subgroup sample size and because PAM50 subtype composition is held constant within the HER2-directed arm subset. Columns are ordered left-to-right to display progressive biological resolution: pooled cohort, subtype-resolved response contexts, and mechanism-stratified arms within pCR. HER2-positive non-pCR retained favourable-side enrichment across all nine pathways and showed clearly larger effects for eight of nine relative to HER2-positive pCR. Across treatment arms, all seven immune/inflammatory pathways changed direction under neratinib relative to ADC, whereas apoptosis and EMT retained favourable-side orientation, producing a structured 7+2 architecture. Green denotes significant enrichment on the structurally favourable (entropy-low) side; red denotes significant enrichment on the structurally adverse (entropy-high) side; grey denotes non-significance (FDR ≥ 0.05).* ***b,*** *Drug-mechanism architecture: all seven immune/inflammatory pathways change direction between ADC and neratinib, while apoptosis and EMT retain favourable-side orientation, producing a coherent 7+2 architecture that indicates selective immune-module reconfiguration rather than wholesale programme disruption.*

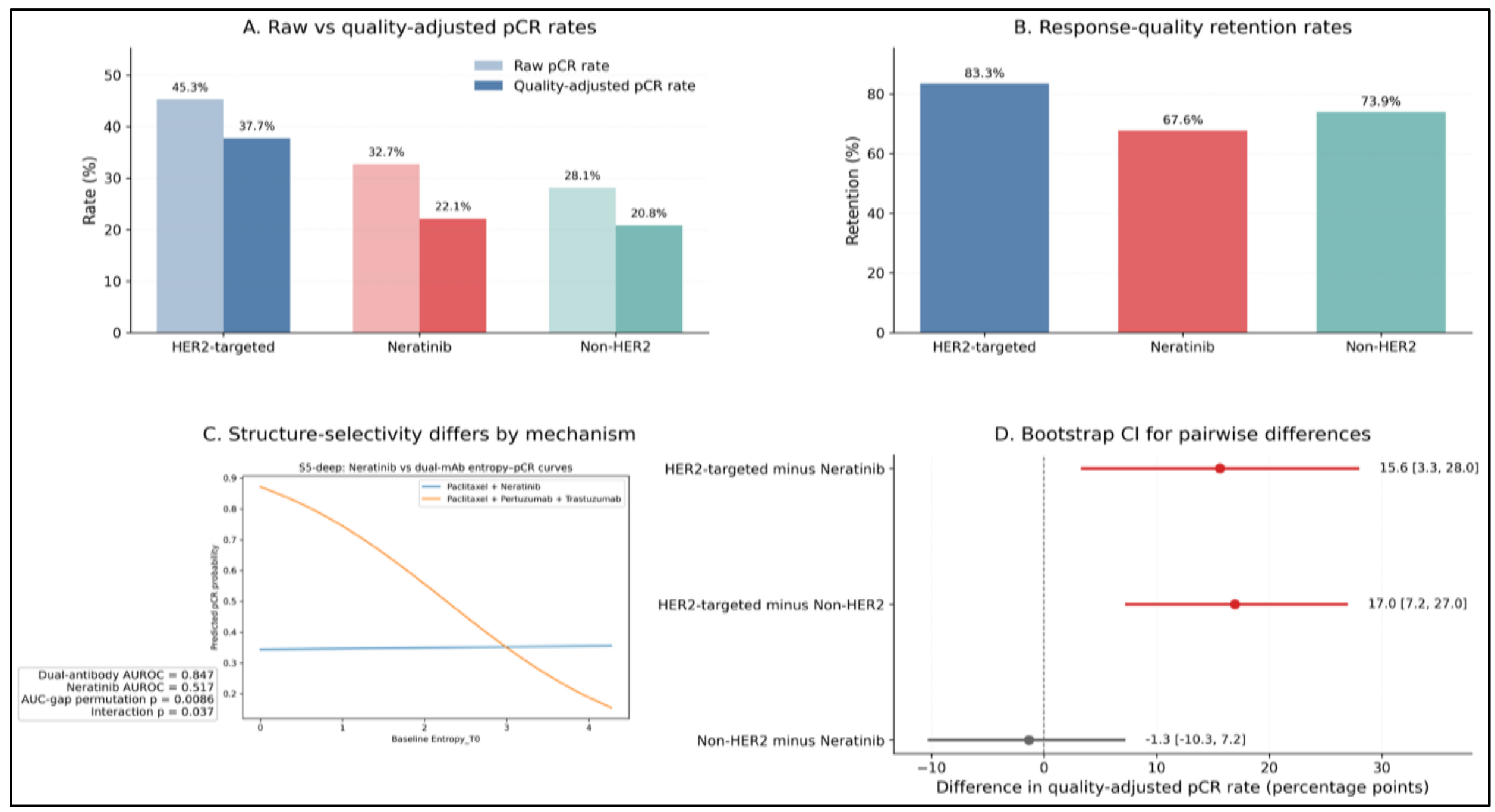

**Figure 9 |** ***Treatment mechanisms produce different-quality pathologic complete response*.** ***a,*** *Raw versus quality-adjusted pCR rates by mechanism class in I-SPY2. HER2-targeted therapy shows the highest raw pCR rate (45.3%) and the highest quality-adjusted pCR rate (37.7%),*

*followed by neratinib (32.7% raw; 22.1% adjusted) and non-HER2 therapy (28.1% raw; 20.8% adjusted).* ***b,*** *Response-quality retention rates by mechanism class. HER2-targeted therapy retains 83.3% of pCR patients as structurally favourable, compared with 67.6% for neratinib and 73.9% for non-HER2 therapy.* ***c,*** *Structure-selectivity differs sharply by mechanism. Baseline entropy predicts pCR strongly in dual-antibody HER2 therapy (AUROC = 0.847) but is essentially uninformative for neratinib (AUROC = 0.517); the entropy-by-mechanism interaction is significant (p = 0.037), and the AUROC gap is unlikely under permutation (p = 0.0086).* ***d,*** *Bootstrap confidence intervals for pairwise differences in quality-adjusted pCR rate. HER2-targeted therapy exceeds neratinib by 15.6 percentage points (95% CI 3.3 to 28.0) and non-HER2 therapy by 17.0 percentage points (95% CI 7.2 to 27.0), whereas non-HER2 does not differ clearly from neratinib (-1.3 percentage points, 95% CI -10.3 to 7.2).*

### 3.2 The structural axis corresponds to distinct immunological and clearance-mechanism states

The tiers occupy reproducible baseline immune-architectural contexts, not merely texture patterns. The biological analysis was deliberately staged: I-SPY2 defined the locked immune–architecture programme, while I-SPY1/GSE22226 tested whether the same programme appeared in recurrence-evaluable pCR and RCB-0 settings, linking structural state, immune biology, and outcome without deriving recurrence claims from I-SPY2. The biology resolves into several linked questions: what programme marks favourable complete response, why adverse complete response is not uniformly immune-cold, what substate resolves its immune heterogeneity, and how the residual-disease tiers and treatment mechanism fit.

Before conditioning on pathologic response, we asked whether the structural axis carried measurable molecular context or could be recovered from pretreatment bulk RNA. In 698 same-patient MRI–RNA cases, the previously defined immune/inflammatory programme showed modest positive associations with Entropy_T0 after adjustment for PAM50 subtype and baseline functional tumour volume: the seven-pathway immune composite was 0.27 SD higher in the locked structurally adverse group (HC3 $p = 0.0033$; incremental $R^2 = 0.013$) and rose with continuous entropy ($\beta = +0.098$, $p = 0.010$), with the direction preserved when each pathway was removed in turn. Six of nine associations survived correction within the previously defined panel for continuous entropy, whereas for the corresponding continuous-entropy analysis none survived correction across the exploratory 50-Hallmark sensitivity family; the coefficients are the same and only the inference family differs. EMT did not follow the immune pathways (continuous $\beta = +0.021$, FDR = 0.575). No individual gene reached FDR significance for either the structural state or continuous entropy (Extended Data Figure 6a,b,e; Supplementary Tables S63–S68).

Under fully fold-local nuisance adjustment and nested feature selection, discrimination of the locked structural state remained weak (out-of-fold AUROC = 0.537; 500-permutation $p = 0.112$). More decisively, continuous Entropy_T0 was not reconstructed from bulk RNA: sparse supervised modelling yielded out-of-fold $R^2 = -0.070$ (Spearman $\rho = -0.005$), and a training-local principal-component sensitivity likewise yielded $R^2 = -0.036$ ($\rho = 0.002$). Raw, all-gene and other low-dimensional transcriptomic sensitivities did not materially alter this interpretation. The structural axis therefore carries measurable bulk molecular context without being materially reconstructed under the tested frameworks (Extended Data Figure 6c,d; Supplementary Tables S69–S74). This full-cohort association does not overturn the responder-specific null: within the 220-patient pCR cohort, continuous entropy–pathway correlations remained negligible (maximum $|\rho| = 0.031$; all nine pathways at FDR > 0.9; Supplementary Table S35). The two analyses address different populations and different estimands, and the response-conditioned Tier 1 versus Tier 2 analyses below concern the latter. The positive full-cohort association should therefore not be read as a uniform within-response gradient: it was absent within pCR, ran in the opposite direction for continuous interferon-γ within the HER2-positive non-pCR subset (§3.2.2), and coexists with residual-tier contrasts estimated in different populations and under different estimands.

#### 3.2.1 Structurally favourable pCR is immune-engaged

A pre-specified nine-pathway immune–architecture panel — seven immune / inflammatory Hallmark pathways (led by IFN-γ response) plus two cell-state pathways, apoptosis and epithelial–mesenchymal transition (EMT) — separates structurally favourable (Tier 1) from structurally adverse (Tier 2) complete response (Figure 7). Full-rank gene-set analysis of this panel identified coordinated Tier 1/favourable-side transcriptomic organisation within pCR; all nine pathways were significant after FDR correction across the 50 Hallmark pathways tested, with IFN-γ dominant (FDR $\approx 10^{-31}$) and EMT second (Table 4). This estimator measures the relative placement of pathway-member genes within the transcriptome-wide differential-expression ranking and is distinct from patient-level pathway activity. The signal is programme-level, not single-gene — no individual gene reached FDR significance — identifying a systems-level state that conventional single-marker assays cannot see. It is PAM50-independent (all nine retained after PAM50 residualisation).

**Table 4 | Immune pathway enrichment in structurally favourable versus structurally adverse pCR.**

| Pathway | Full-rank pathway-shift direction | Unadjusted p | Unadjusted FDR | PAM50-adjusted p | Expanded within-I-SPY2 replication FDR (patient-non-overlapping within I-SPY2; n=68 HER2+, PAM50-adj) |
|---|---|---|---|---|---|
| IFN-gamma response | Tier 1 ↑ | $1.14 \times 10^{-35}$ | $5.69 \times 10^{-34}$ | $9.11 \times 10^{-33}$ | $2.32 \times 10^{-13}$ |
| Allograft rejection | Tier 1 ↑ | $1.19 \times 10^{-23}$ | $2.98 \times 10^{-22}$ | $2.86 \times 10^{-20}$ | $1.54 \times 10^{-2}$ |
| TNFα/NF-κB signalling | Tier 1 ↑ | $1.23 \times 10^{-16}$ | $2.04 \times 10^{-15}$ | $1.61 \times 10^{-18}$ | $2.18 \times 10^{-5}$ |
| IFN-alpha response | Tier 1 ↑ | $2.67 \times 10^{-16}$ | $3.34 \times 10^{-15}$ | $4.26 \times 10^{-17}$ | $1.33 \times 10^{-8}$ |
| Inflammatory response | Tier 1 ↑ | $3.25 \times 10^{-15}$ | $3.25 \times 10^{-14}$ | $6.02 \times 10^{-15}$ | $1.88 \times 10^{-4}$ |
| IL-6/JAK/STAT3 signalling | Tier 1 ↑ | $8.80 \times 10^{-11}$ | $4.89 \times 10^{-10}$ | $1.07 \times 10^{-10}$ | $1.54 \times 10^{-2}$ |
| Complement | Tier 1 ↑ | $4.52 \times 10^{-10}$ | $1.88 \times 10^{-9}$ | $1.22 \times 10^{-10}$ | $1.41 \times 10^{-6}$ |
| Apoptosis | Tier 1 ↑ | $3.07 \times 10^{-10}$ | $1.39 \times 10^{-9}$ | $4.60 \times 10^{-13}$ | $2.22 \times 10^{-10}$ |
| Epithelial-mesenchymal transition | Tier 1 ↑ | $7.47 \times 10^{-11}$ | $4.67 \times 10^{-10}$ | $4.75 \times 10^{-24}$ | $9.14 \times 10^{-17}$ |

Full-rank pathway-shift test across 50 Hallmark gene sets (MSigDB)[38]. Unadjusted: n = 63 (49 Tier 1, 14 Tier 2). PAM50-adjusted: n = 62 (49 Tier 1, 13 Tier 2) after residualisation. All seven immune/inflammatory pathways and apoptosis survived PAM50 adjustment; all nine pathways had PAM50-adjusted FDR $< 3 \times 10^{-9}$, and EMT was the second-strongest adjusted pathway. PAM50-adjusted FDR values are reported in Supplementary Table S22. The expanded within-I-SPY2 replication column uses a non-overlapping HER2-positive pCR subset (n = 68) analysed with the identical full-rank pathway-shift framework and PAM50 adjustment. The locked nine are IFN-γ response, IFN-α response, TNFα/NF-κB, IL6/JAK/STAT3, inflammatory response, complement, allograft rejection, apoptosis, and EMT. Apoptosis discovery PAM50-adjusted FDR = $2.88 \times 10^{-12}$ and replication FDR = $2.22 \times 10^{-10}$. IL-2/STAT5 reached FDR = $6.04 \times 10^{-14}$ in discovery but was not part of the pre-specified replication panel; it is reported as a discovery-stage extra consistent with immune engagement, not as a locked or replicated panel member. Full replication comparison is in Supplementary Table S32. Tier 1 ↑ denotes relative concentration of pathway-member genes toward the Tier 1/favourable side of the signed transcriptome-wide differential-expression ranking; it does not denote higher absolute patient-level pathway activity.

### 3.2.2 The structural axis is a state within pCR but a gradient within residual disease

Baseline entropy indexes different biology by response state: within HER2-positive pCR it is state-like (no meaningful continuous entropy–pathway gradient (0/9 pathways FDR-significant; maximum $|\rho| = 0.072$)), whereas within HER2-positive non-pCR it tracks a coherent immune-depletion gradient (higher entropy, lower pathway activity) (IFN-γ $\rho = -0.289$, FDR = 0.016). Across residual disease, Tier 4 is inflammatory-higher than Tier 3 in 18 of 18 concordant cohort–pathway observations (exact binomial $p = 3.81 \times 10^{-6}$) — yet neither residual tier's signature discriminates recurrers from non-recurrers within-state (Tier 4 p = 0.75; Tier 3 p = 0.17). The residual tiers are state-defining, not within-state prognostic — a deliberate boundary that keeps the recurrence claim on the within-pCR axis.

### 3.2.3 The immune programme replicates across discovery, replication, and cross-context

The programme replicated with full directional concordance (9/9) in an expanded, non-overlapping HER2-positive pCR patient subset within I-SPY2 (n = 68; IFN-γ FDR = $2.3 \times 10^{-13}$). Full per-pathway statistics and the discovery-versus-replication attenuation are in the Supplementary Information.

### 3.2.4 Adverse complete response is estrogen-programmed and harbours a broadly immune-depleted substate

Structurally adverse pCR is a distinct luminal-programmed biology rather than an absence of the favourable programme: estrogen-response pathways are enriched in Tier 2 and survive PAM50 adjustment (early FDR = 0.002; late FDR = $10^{-5}$). It does not track immune-checkpoint expression — CD274 (PD-L1), PDCD1, CTLA4, LAG3, and TIGIT are not differentially expressed between tiers — so it is not an exhausted-immune state. EMT ranks second among the nine PAM50-adjusted pathway-shift signals, with enrichment on the Tier 1/favourable side rather than the adverse side. The programme is most reproducibly recovered in HR-positive disease; one PAM50 subtype (PAM50-Her2, n = 21, not equivalent to clinical HER2-positive) shows a directional inversion that we report transparently and that does not affect the clinical-cohort claim (n = 68). Crucially, the tier mean conceals structure — Tier 2 is bimodal. Pre-locked unsupervised clustering (k = 2) resolves Tier 2 into an immune-depleted Cluster 1 (n = 22, ≈ 40%) and an immune-engaged Cluster 2 (n = 33, ≈ 60%); "immune-cold" is therefore a bulk-RNA description of Cluster 1, not of the whole tier; I-SPY2 cannot directly establish Cluster 1-specific recurrence because it lacks validated recurrence follow-up.

#### *3.2.4.1 Gene-level bulk-transcriptomic phenotype*

Cluster 1 is broadly immune-depleted at bulk-RNA resolution: cytotoxic/TIL, antigen-presentation, and checkpoint-associated signatures are all lower (FDR = $7.8 \times 10^{-7}$, $1.3 \times 10^{-6}$, and $2.8 \times 10^{-7}$). Their coordinated reduction is more consistent with an immune-cold state than classical exhaustion, but bulk RNA cannot distinguish fewer immune cells from lower per-cell function. The direction is concordant in the small I-SPY1 within-Tier-2 recurrence anchor (recurrers n = 3 vs non-recurrers n = 11), where checkpoint-associated expression reaches significance (rank-biserial = −1.000, p = 0.007). The tested bulk stromal/desmoplasia signature did not reproduce across cohorts; spatial stromal exclusion remains outside the resolution of these data. Under the pre-locked cross-cohort concordance rule, H2—the immune-recognition/cytotoxic-deficit hypothesis—was supported by concordant reductions in cytotoxic/TIL, antigen-presentation and checkpoint-associated signatures, whereas H1 and H4 were not supported (Supplementary Table S58); the licensed interpretation remains broad immune depletion at bulk-RNA resolution rather than a proved cellular mechanism. The gene-level signature dissection (Supplementary Table S58) and full hypothesis battery are in the Supplementary Information.

#### *3.2.4.2 Same-patient bridge*

The same-patient bridge is shown in Figure 11c, where the matched GSE22226↔I-SPY1 overlap cohort (n = 19; 4 recurrers vs 15 non-recurrers) preserves the 9/9 directional immune pattern.

### **3.2.5 Drug mechanism reconfigures structural response quality**

Tier generation is drug-mechanism-dependent: dual-antibody HER2 therapy produces structurally favourable complete response (structure-selectivity AUROC = 0.847) whereas neratinib does not (0.517; entropy-by-mechanism interaction p = 0.037), all seven immune/inflammatory pathways change direction between the two arms while apoptosis and EMT retain favourable-side orientation (Figure 8). The convergent four-method drug-mechanism analysis is detailed in Supplementary Table S59.

### **3.2.6 Quality-adjusted pCR exposes the pCR–survival disconnect**

HER2-targeted therapy retains 83.3% of complete responses as structurally favourable, widening the quality-adjusted pCR gap between mechanisms (Figure 9).

### **3.2.7 I-SPY1 BPE-matched transcriptomics provide directional support for the within-pCR immune contrast**

The tier-level contrast receives directional support on an external platform: in the I-SPY1 BPE-matched cohort (n = 23 tier-assigned) the locked-nine architecture composite is −0.661 in Tier 2 versus +0.150 in Tier 1 (Mann–Whitney p = 0.037; Cliff's $\delta = -0.62$; 8/9 pathways concordant, one-sided sign-test p = 0.0195), whereas background parenchymal enhancement substituted for the structural axis shows no separation (§3.4.4) — the signal is structural, not parenchymal.

## **3.3 Structurally adverse pCR concentrates within-pCR recurrence risk that is biologically resolved at the substate level**

Two independent external cohorts with validated recurrence endpoints — I-SPY1 (ACRIN 6657; n = 21 HER2-positive pCR patients, 6 recurrence events) and UCSF (n = 12, 4 events) — provided the prognostic evidence. No recurrence claims in this study derive from the I-SPY2 discovery cohort, whose internal event indicator was found to be perfectly aligned with hormone receptor status rather than validated recurrence (Supplementary Table S12; Supplementary Note D1; Methods §2.6). All recurrence analyses below are external; pathology-confirmed within-pCR claims derive from I-SPY1. The analysis focuses on HER2-positive pCR because this subtype provides the largest pCR population with validated long-term follow-up across multiple independent cohorts (Methods §2.6); extension to other subtypes is discussed in §4.6.

### 3.3.1 Pooled responder recurrence and deep characterisation

The decisive clinical test of the framework is whether pretreatment structure still carries prognostic information after the best conventional outcome — pathologic complete response — has already been achieved. In a prespecified external synthesis combining the I-SPY1 pathology-confirmed HER2-positive pCR cohort ($n = 21$) with a UCSF HER2-positive best-response proxy cohort ($n = 12$) (System C; 33 patients, 10 recurrence events), structurally adverse responders recurred at substantially higher rates than structurally favourable responders (pooled hazard ratio = 2.87, 95% CI [1.38, 5.96], $p = 0.005$; $I^2 = 0\%$; Figure 10). Independent institutional confirmation underlies the pooled result: I-SPY1 (ACRIN 6657) established the within-pCR structural signal (HR 2.58, 95% CI 1.02–6.52, $p = 0.045$), and UCSF — a second, independent institution (the Breast-MRI-NACT-Pilot collection) with different scanners and acquisition era — showed a concordant, individually significant association in a HER2-positive best-response proxy cohort (HR 3.41, 95% CI 1.03–11.27, $p = 0.044$), the two pooling without detectable heterogeneity ($I^2 = 0\%$). Within I-SPY1, the native MAD coordinate reproduced the categorical structural assignment exactly (categorical-assignment AUC = 1.00; $\kappa = 1.00$; a single exact threshold recovers every label), and the adverse recurrence direction was retained across all 17 admissible descriptive cutpoints, all 6 leave-one-event-out refits and all 21 leave-one-patient-out refits (Supplementary Tables S77 and S78). The association therefore follows the graded MAD ordering rather than a particular dichotomisation or an influential observation. In UCSF the categorical crosswalk assignment was not recoverable by thresholding the frozen continuous 2D coordinate (categorical-assignment AUC = 0.07; $\kappa = -0.50$), although that continuous coordinate was itself directionally stable across all 9 admissible cutpoints, all 4 event deletions and all 12 patient deletions. Within this pooled cohort, structurally favourable responders ($n = 14$; 11 I-SPY1, 3 UCSF) contributed 3 recurrence events, whereas structurally adverse responders ($n = 19$; 10 I-SPY1, 9 UCSF) contributed 7 events; descriptive follow-up comparisons are reported in Supplementary Table S30. This pooled estimate reflects direction and approximate magnitude of responder-associated recurrence risk; precision remains limited by sample size and should be interpreted alongside the larger-cohort convergent analyses reported below. The structural reading concentrates risk rather than predicting it: within System C it flagged the 19 structurally adverse responders and captured 7 of the 10 recurrences in the pooled responder cohort (sensitivity 70.0%, 95% CI 39.7–89.2; number-needed-to-screen 2.71), at a quantified cost — specificity 47.8% (11 of 23 non-recurrers correctly classified favourable), positive predictive value 36.8% (7 of 19 flagged patients recurred), and negative predictive value 78.6% (11 of 14); threshold operating characteristics across all three cohort scopes are tabulated in Supplementary Table S60.

A structurally adverse reading therefore identifies a subgroup enriched for, not destined for, recurrence; a formal net-reclassification index was not computed because at ten events it would be unstable. The responder-associated recurrence signal is directionally supported across three convergent scales: pooled responder clinical signal (System C: pooled HR = 2.87, 33 patients, 10 events), prognostic information gain beyond pCR status (System B full-cohort nested model $\Delta C = 0.031$, HR = 1.92, 120 patients, 35 events; this section), and adjuvant-confounder closure (Duke adjuvant cohort 8-variable Cox HR = 0.610, $p = 0.015$, 908 patients, 76 events; §3.4.2). Effect magnitude attenuates with increasing cohort size, consistent with a real but more modest signal than the smallest-cohort estimate alone would suggest. Because the two cohorts encode architecture using different native structural measures, we next asked whether transport survived removal of representation-specific numerical scale. With each cohort's native coordinate converted independently to outcome-blind within-cohort structural-risk ranks, each 25-percentile increase in rank was associated with a 2.42-fold increase in recurrence hazard across the 33 responders and 10 events (95% CI 1.24–4.76; $p = 0.010$), and censoring-aware permutation supported the pooled ordering (concordance = 0.763; 100,000 permutations; $p = 0.0045$). The adverse direction was preserved across every admissible cutpoint, every event and patient deletion and every estimable administrative-censoring horizon in both cohorts, and the continuous score showed no detectable association with censoring (Supplementary Tables S76 and S77). Because this analysis uses the same patients, events and native coordinates as the pooled estimate above, it is a scale and representation robustness analysis of that evidence rather than an additional independent validation. A cohort-stratified binary sensitivity using the canonical categorical crosswalk was directionally consistent but not significant (Mantel–Haenszel common OR = 0.55, 95% CI [0.13, 2.36], $\chi^2 = 0.78$, $p = 0.376$; Supplementary Table S46), reflecting non-uniform categorical separation across the two small external cohorts. The categorical crosswalk is therefore retained descriptively rather than as a common cross-cohort threshold; the primary pooled inference remains the random-effects synthesis of the cohort-specific continuous structural measures, which this sensitivity does not modify. The denominator was locked through a complete funnel audit: I-SPY1 narrowed from 30 to 21 (zero recurrence events lost); UCSF narrowed from 13 to 12 (one event-bearing patient lost, disclosed). Excluded patients had a higher event rate (41.2%) than included patients (30.3%). To test whether structural quality adds prognostic information beyond pCR, we fit nested Cox models in the locked I-SPY1 framework-eligible cohort ($n = 120$, 35 events; Methods §2.10).

pCR alone showed modest prognostic discrimination (Harrell's C-index = 0.575); adding pretreatment structural status improved discrimination to 0.606 (ΔC = 0.031, 95% CI −0.009 to 0.117; likelihood-ratio $\chi^2 = 3.18$, p = 0.075; ΔAIC = 1.18). In the joint model, structural adversity carried an approximately two-fold hazard beyond pCR (HR = 1.92, 95% CI 0.96–3.83, p = 0.065). A continuous encoding of the latent structural axis yielded stronger discrimination (C-index = 0.635; optimism-corrected 0.613), consistent with a continuous prognostic axis rather than a threshold-dependent construct. Descriptive event rates confirmed the pattern in four numbers: within pCR, recurrence was 28.6% in structurally adverse versus 11.5% in structurally favourable patients; within non-pCR, 47.8% versus 29.7%. Within I-SPY1 pCR achievers alone, the concentration sharpened further — structurally adverse responders recurred at 66.7% (2/3) versus 16.7% (2/12) in favourable responders (log-rank p = 0.024). The information-gain analysis is consistent with the representation-independent DL convergence (§3.1.3, AUROC = 0.738, permutation p = 0.007), indicating the prognostic increment does not reflect a radiomic-specific artefact. Full model battery and term-level estimates are provided in Supplementary Tables S42–S43; the within-pCR meta-analysis (frequentist and Bayesian) is reported in Supplementary Table S8, and individual projected-analysis leave-one-out iterations in Supplementary Table S10; leave-one-out robustness, threshold stability, denominator funnel, and Bayesian posterior analyses are provided in Supplementary Figure S16; bridge-derived sensitivity analyses are provided in Supplementary Table S20.

Within the I-SPY1 subset (n = 21, 6 events), three of six recurrers died; zero of fifteen non-recurrers died. Volumetric shrinkage did not distinguish recurrers from non-recurrers at any treatment timepoint (all p > 0.35). Tumours that recurred shrank just as much as tumours that did not. The univariable continuous structural-score HR was 2.68 (p = 0.045), remaining directionally stable across four adjusted specifications (HR range 1.93–2.68; Supplementary Table S29). Figure 10a displays the I-SPY1 survival-evaluable within-pCR subset (n = 15, 4 events) — the subcohort with complete baseline structural-score coverage used for the Kaplan–Meier panel (12 structurally favourable, 3 structurally adverse; Supplementary Table S43 for the full cohort architecture). The pooled n = 33 cohort underlies the hazard ratio reported above. The two within-HER2+-pCR I-SPY1 substrates used in this section are operationally distinct and should not be summed: (a) the locked HER2+ pCR modelling cohort (n = 21, 6 RFS events; 11 structurally favourable, 10 structurally adverse) is the System C constituent and underlies the univariate Cox HR = 2.68 (p = 0.045), the OR = 28 within-RCB-0 finding, and the System C pooled HR = 2.87; (b) the HER2+ supporting four-tier framework subset (n = 15, 4 events; 12 structurally favourable, 3 structurally adverse) is the survival-evaluable subset displayed in Fig. 10a, restricted to patients with complete four-tier framework classification. The n = 15 Kaplan–Meier cohort and the n = 21 modelling cohort are separately locked, overlapping analytical sets; they are not strictly nested (patient 1182 appears in the n = 15 set but not the n = 21 pool) and must not be summed. The three within-cohort estimators are distinct: the native-MAD median dichotomisation yields the 11 favourable / 10 adverse split and the binary HR = 2.582 meta-analysis input; the continuous structural score yields HR = 2.68; and the locked upper-tail MAD-z threshold (1.133452) yields the 16 favourable / 5 adverse RCB-0 split underlying OR = 28 and Cox HR = 9.04. The pooled responder HR = 2.87 combines the n = 21 pathology-confirmed I-SPY1 modelling cohort with the UCSF best-response proxy; the pathology-confirmed within-pCR estimates (HR = 2.68, OR = 28, HR = 9.04) derive from I-SPY1 alone. Fig. 10a is the visual subset only.

The volumetric convergence finding replicated in UCSF: neither SER volume change (p = 0.416) nor longest-diameter change (p = 0.413) discriminated events from non-events. UCSF_BR_49 showed 98.9% SER volume reduction yet recurred with distant metastasis at 182 weeks; UCSF_BR_51 showed 99.0% reduction yet recurred locally at 25 weeks.

Recurrence was early (median 431 days; all ten within three years) and predominantly distant — compatible with, but not proof of, impaired immune surveillance in the immune-depleted Cluster 1 substate within Tier 2.

### 3.3.2 Clinical concentration and distant-metastasis specificity

Within the highest-reassurance conventional category — HER2-positive patients achieving RCB-0 in I-SPY1 (System B) — recurrence concentrated sharply in the structurally adverse subgroup. A prespecified upper-tail MAD rule identified five structurally adverse patients among 21 HER2-positive RCB-0 cases. Four of five adverse patients recurred, compared with two of sixteen favourable patients (crude OR = 28.0, Fisher p = 0.011, log-rank p = 0.003, Cox HR = 9.04, 95% CI 1.57–51.89, p = 0.0136; Figure 10e). Because this analysis contained only 21 patients and 6 events with one sparse contingency cell, both the survival estimate and the crude odds ratio are reported descriptively and supplemented with small-sample robust estimators that preserved the direction and magnitude of the effect. A Firth penalized Cox model preserved the survival signal with a bounded small-sample estimate (Firth HR = 8.13, 95% CI 1.71–49.21, p = 0.0089); the recurrence table gave Firth penalized logistic OR = 17.4 (95% CI 2.13–246.9, p =

0.0068), concordant with exact conditional estimation (OR = 21.6, 95% CI 1.38–1452, p = 0.011). The fragility index was 1 of 6 events (reassigning one event moved Fisher's exact p from 0.011 to 0.063); accordingly, the pooled System C estimate, rather than this single contingency table, is treated as the primary pooled responder recurrence anchor (Supplementary Table S62). Standard volumetric metrics converge to near-zero in both groups while recurrence diverges sharply between structural states (Figure 5a–c). In this 21-patient, six-event subset, the structurally adverse group contained four of six recurrences despite comprising five of 21 patients. Because the fragility index was 1, this finding is supportive and hypothesis-strengthening rather than a standalone recurrence estimate.

We next asked whether structural heterogeneity persisted when pathologic response was represented by residual cancer burden rather than binary pCR. In the locked I-SPY1 RCB-known denominator, recurrence increased across RCB classes as expected (RCB-0, 16.7%; RCB-I, 18.2%; RCB-II, 28.1%; RCB-III, 60.7%), consistent with RCB as a residual-burden axis. However, RCB-0 remained structurally heterogeneous. Among 42 RCB-0 patients, the locked MAD threshold identified 9 structurally adverse patients, of whom 4 recurred, compared with 3 recurrences among 33 structurally favourable patients (44.4% versus 9.1%; Cox HR 5.50, 95% CI 1.23–24.66; p = 0.026; log-rank p = 0.0123; Fisher p = 0.0281; Spearman ρ = 0.322, p = 0.0378). This association received concordant support across time-to-event, categorical, and rank-based tests (Supplementary Table S48). Thus, even complete pathologic response by RCB does not collapse the pretreatment structural risk state; the HER2-positive RCB-0 ceiling is summarised in Figure 10e, with the broader all-subtype RCB-0 analysis retained in Supplementary Table S48.

The broader all-subtype RCB-0 result contains the previously reported HER2-positive RCB-0 concentration as its sharpest expression, indicating that the within-RCB-0 response-quality stratification is not restricted to HER2-positive disease but is most clinically concentrated there. The RCB-0 analysis provides the strictest test of non-redundancy: even when pathology reports no residual invasive disease — the most reassuring category in current practice — pretreatment architecture separates recurrence risk.

RCB-III showed complementary continuous structural severity gradients (continuous Cox HR = 2.72, p = 0.000917; Supplementary Tables S47, S49, and S50; Supplementary Figure S17), whereas RCB-I and RCB-II were sparse or not robustly stratified, supporting a two-axis interpretation in which RCB quantifies residual burden and pretreatment MRI structure captures response quality.

The structural axis preferentially tracked systemic rather than local failure. In the Duke cohort, the structural phenotype was associated with distant recurrence-free survival (DRFS HR = 0.466, p = 0.022) but not with local recurrence-free survival (LRFS p = 0.361). In UCSF, ten of fifteen full-cohort recurrences were distant metastases; among HER2-positive patients specifically, three of four recurrences were distant. Structural adversity preferentially tracks systemic failure — consistent with the immune-surveillance-failure biology characterised in §3.2.

Among I-SPY1 pCR patients with greater than 50% presurgical longest-diameter shrinkage — the patients a clinician would judge as the most dramatic volumetric responders (Figure 5) — structural adversity remained common: 18 of 31 (58.1%) were structurally adverse, despite near-complete volumetric regression in both groups (mean LD decline 89.3% in structurally adverse versus 93.8% in structurally favourable; median 100% in both). Shrinkage-based refinement was not merely uninformative — it was actively harmful: selecting on conventional imaging response criteria preferentially retained the structurally adverse subgroup. Response magnitude does not determine response quality.

### 3.3.3 Within-pCR response-quality stratification and the faces of the framework

Across increasingly restrictive composite reassurance criteria — pCR, major volumetric response, favourable genomic risk, and favourable treatment context — the structurally adverse fraction within HER2-positive I-SPY2 pCR patients remained approximately constant at ~21% (trend p = 0.97; Supplementary Note D20). The structural phenotype is therefore orthogonal to the conventional reassurance dimension itself, not merely to any single filter: no combination of pathology, volumetrics, genomics, and therapy context excludes the adverse-pCR phenotype (Supplementary Figure S6 documents stability of the structurally adverse fraction across composite reassurance-score thresholds). Because I-SPY2 lacks a validated recurrence endpoint (§2.6), this is interpreted as phenotype persistence rather than recurrence stratification; the recurrence consequence of this phenotype is established by external I-SPY1, UCSF, and Duke analyses.

Patient ISPY2-268937 embodies this within-pCR response-quality stratification. She achieved pCR with 95.1% volume decline, MammaPrint-low, T-DM1 plus pertuzumab treatment — yet baseline entropy of 3.50 and DL adversity probability of 0.734 placed her firmly in the structurally adverse state. Every conventional reassurance criterion was satisfied. The structural axis — and only the structural axis — identified her as belonging to the adverse-pCR phenotype externally associated with residual recurrence risk (Supplementary Table S5).

Patient 1090 illustrates the same principle from I-SPY1. RCB-0 confirmed, yet presurgical longest diameter was 52 mm — the highest of any HER2-positive pCR patient. Pathology said the tumour was eliminated. Imaging said the tissue never structurally normalised. She recurred at 843 days. Clearance without reorganisation, in one patient.

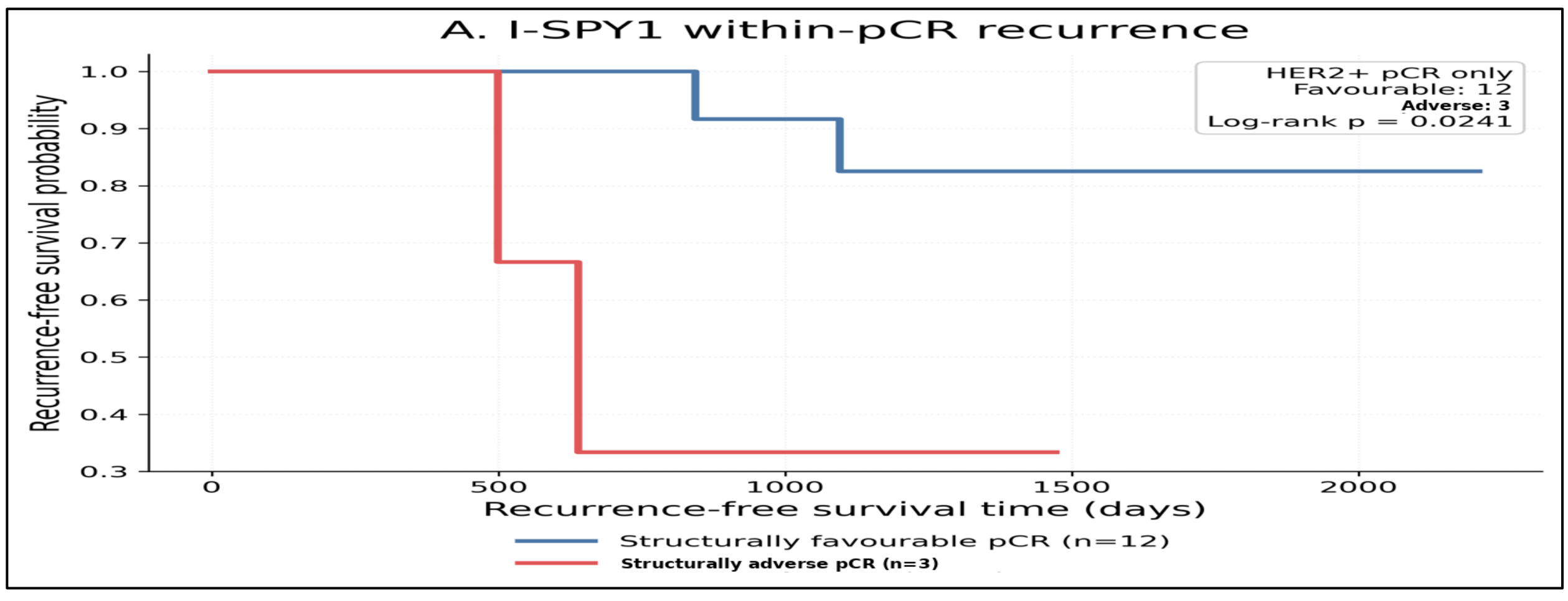

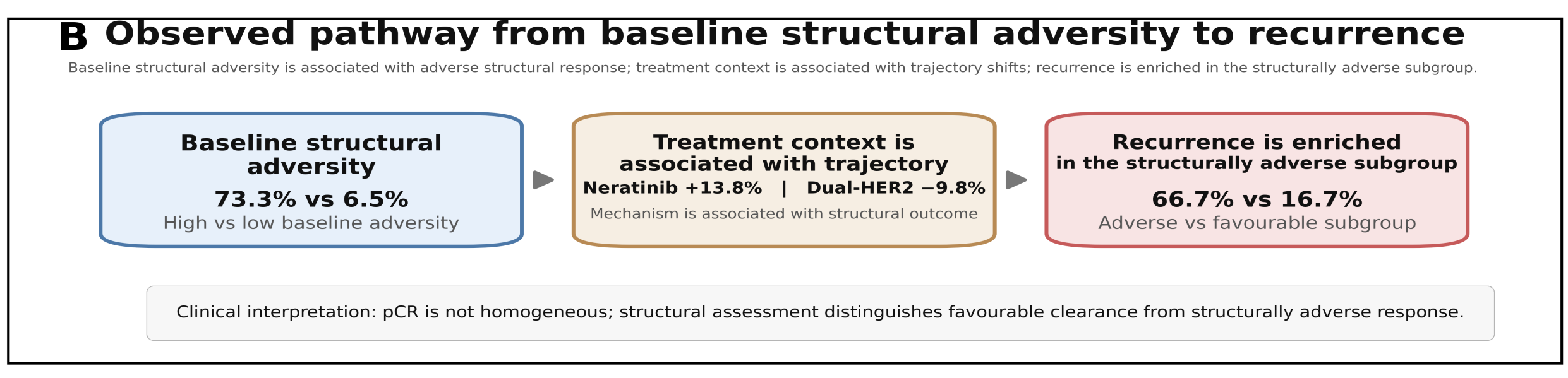

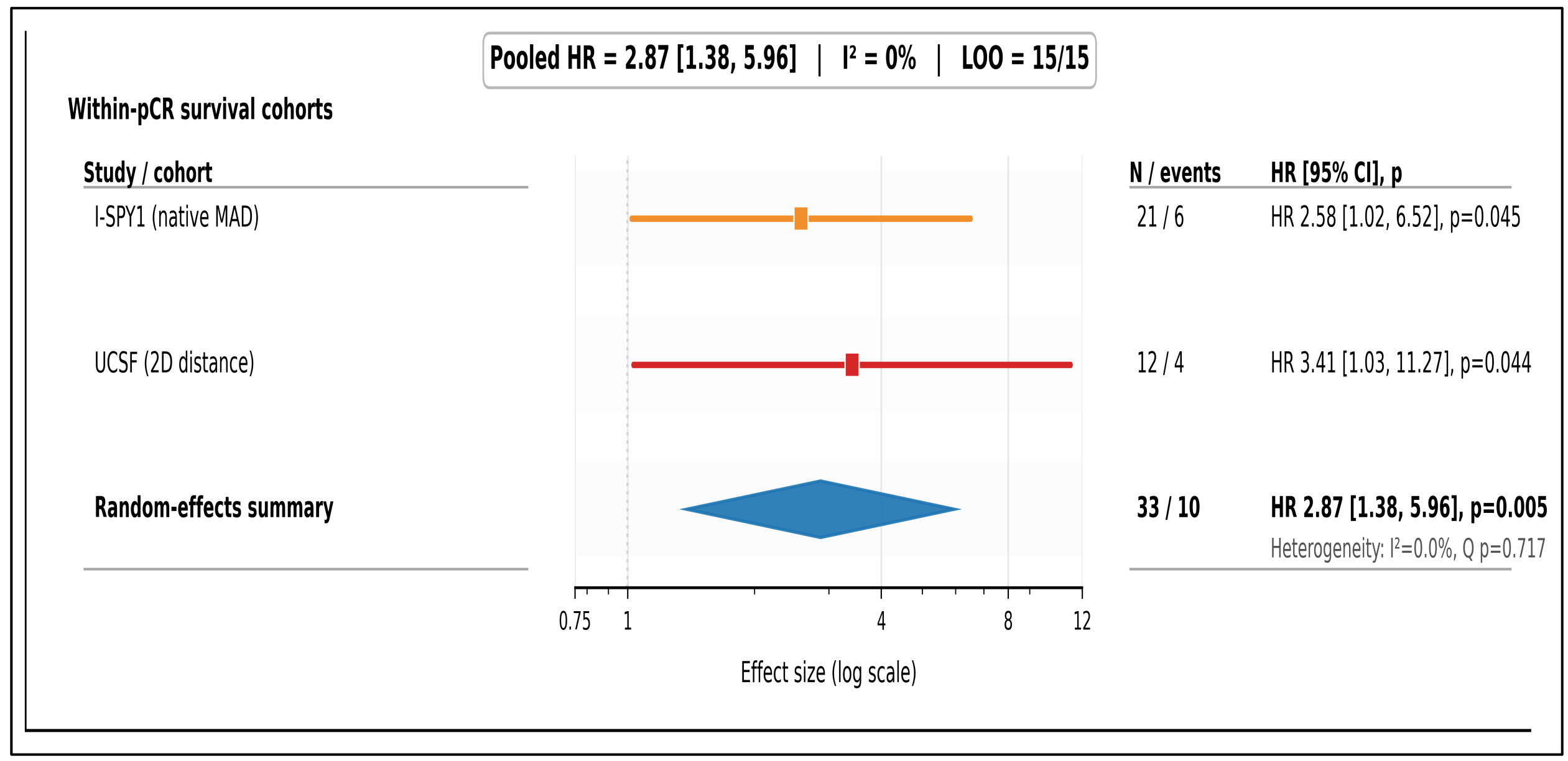

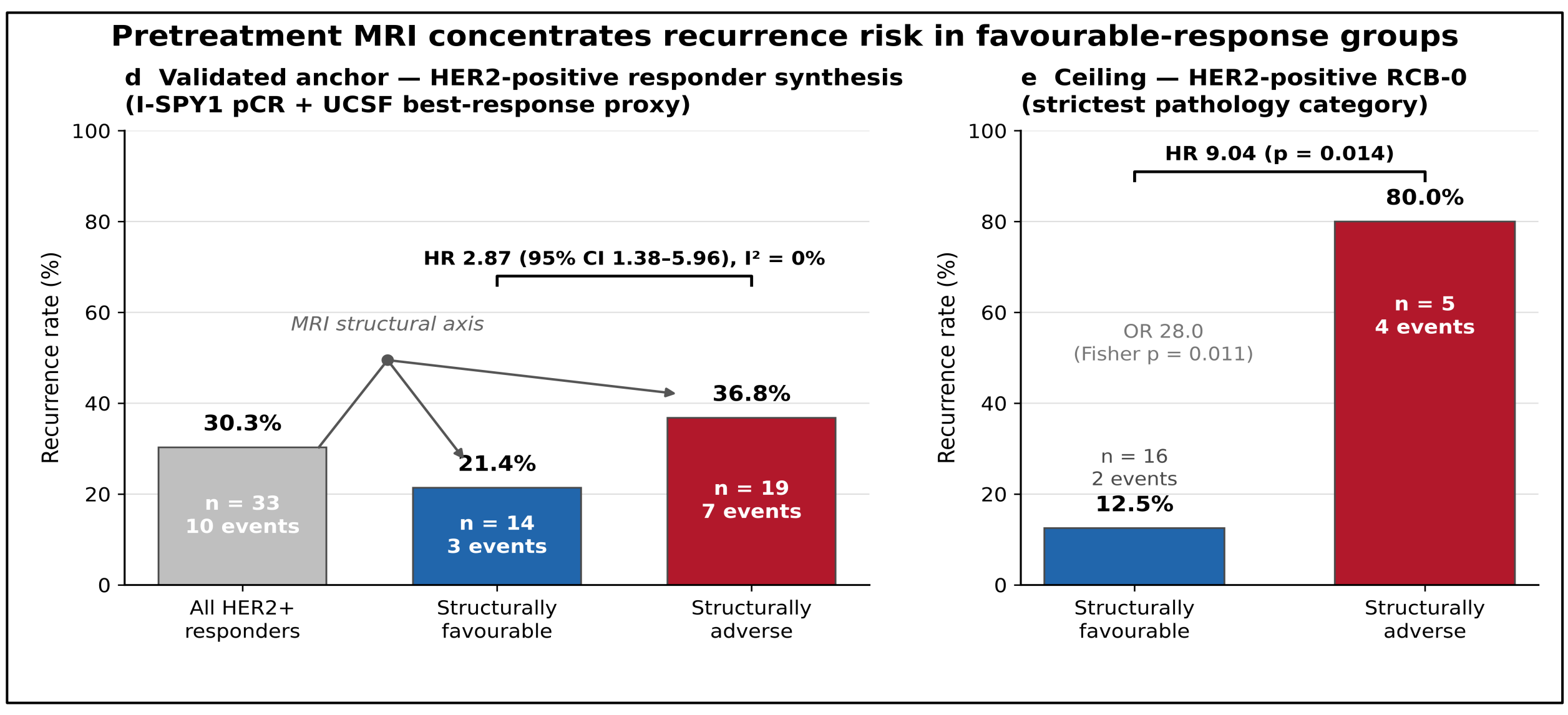

**Figure 10 |** ***Recurrence in favourable-response groups: external evidence, clinical mechanism, and pooled estimate. a****, In the I-SPY1 HER2-positive pathologic complete responder survival-evaluable subset (n = 15, 4 events), structurally favourable patients (n = 12) showed better recurrence-free survival than structurally adverse pCR patients (n = 3); log-rank p = 0.0241, indicating that pathologic complete response and pretreatment structural quality carry complementary information about residual recurrence risk. Fig. 10a uses the HER2+ supporting four-tier framework subset (n = 15, 4 events) with complete tier-classification data; numerical recurrence claims throughout §3.3 use the locked HER2+ pCR modelling cohort (n = 21, 6 events) — see §3.3.1 substrate disclosure.* ***b****, Condensed observational pathway summarising the proposed clinical mechanism: baseline structural adversity is associated with adverse structural response (73.3% vs 6.5%), persists through therapy, is associated with treatment class (neratinib +13.8%, dual-HER2 −9.8%), and is enriched in patients with subsequent recurrence in I-SPY1 (66.7% vs 16.7%).* ***c,*** *Convergent evidence across independent external cohorts shows that structural adversity within responders is associated with elevated recurrence risk. In I-SPY1, the hazard ratio is 2.58 (95% CI 1.02–6.52; p = 0.045), and in UCSF it is 3.41 (95% CI 1.03–11.27; p = 0.044). The pooled random-effects estimate is HR 2.87 (95% CI 1.38–5.96; p = 0.005), with no detectable between-cohort heterogeneity (I² = 0%). I-SPY2 is not used for recurrence claims (see §2.6); the pathology-confirmed within-pCR consequence is anchored in I-SPY1, with concordant support from the UCSF best-response proxy. I-SPY1 HR = 2.58 reflects the native-MAD meta-analysis input; the standalone I-SPY1 Cox model is reported separately in §3.3.1.* ***d****, Validated anchor — HER2-positive responders (System C: I-SPY1 pathology-confirmed pCR + UCSF best-response proxy; n = 33, 10 events): overall responder recurrence 30.3%, resolving into 21.4% (3 of 14) structurally favourable versus 36.8% (7 of 19) structurally adverse (pooled HR 2.87, 95% CI 1.38–5.96; I² = 0%); the pretreatment structural axis captures 7 of 10 recurrences in the pooled responder cohort (sensitivity 70%; number-needed-to-screen 2.71).* ***e****, Ceiling — HER2-positive RCB-0 (strictest pathologic category; System B; n = 21, 6 events): 12.5% (2 of 16) structurally favourable versus 80.0% (4 of 5) structurally adverse (standard Cox HR 9.04, p = 0.0136; Firth penalized Cox HR 8.13, p = 0.0089; Fisher exact OR = 28.0, p = 0.011); the longest-diameter convergence underlying this ceiling is shown in Figure 5a–c. Panels d–e present the reclassification as absolute responder recurrence rates; their shared header summarises that pretreatment MRI concentrates recurrence risk in favourable-response groups.*

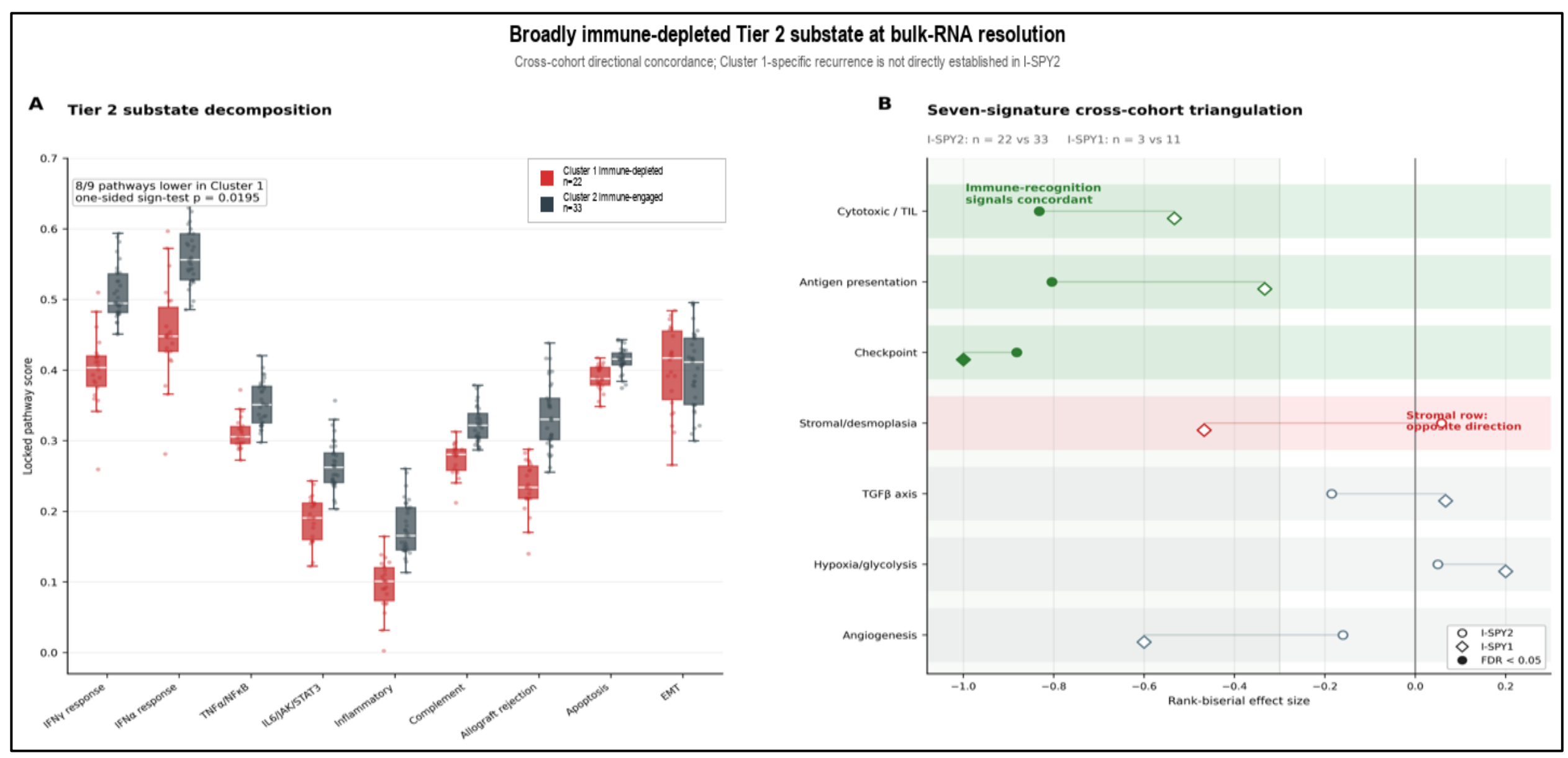

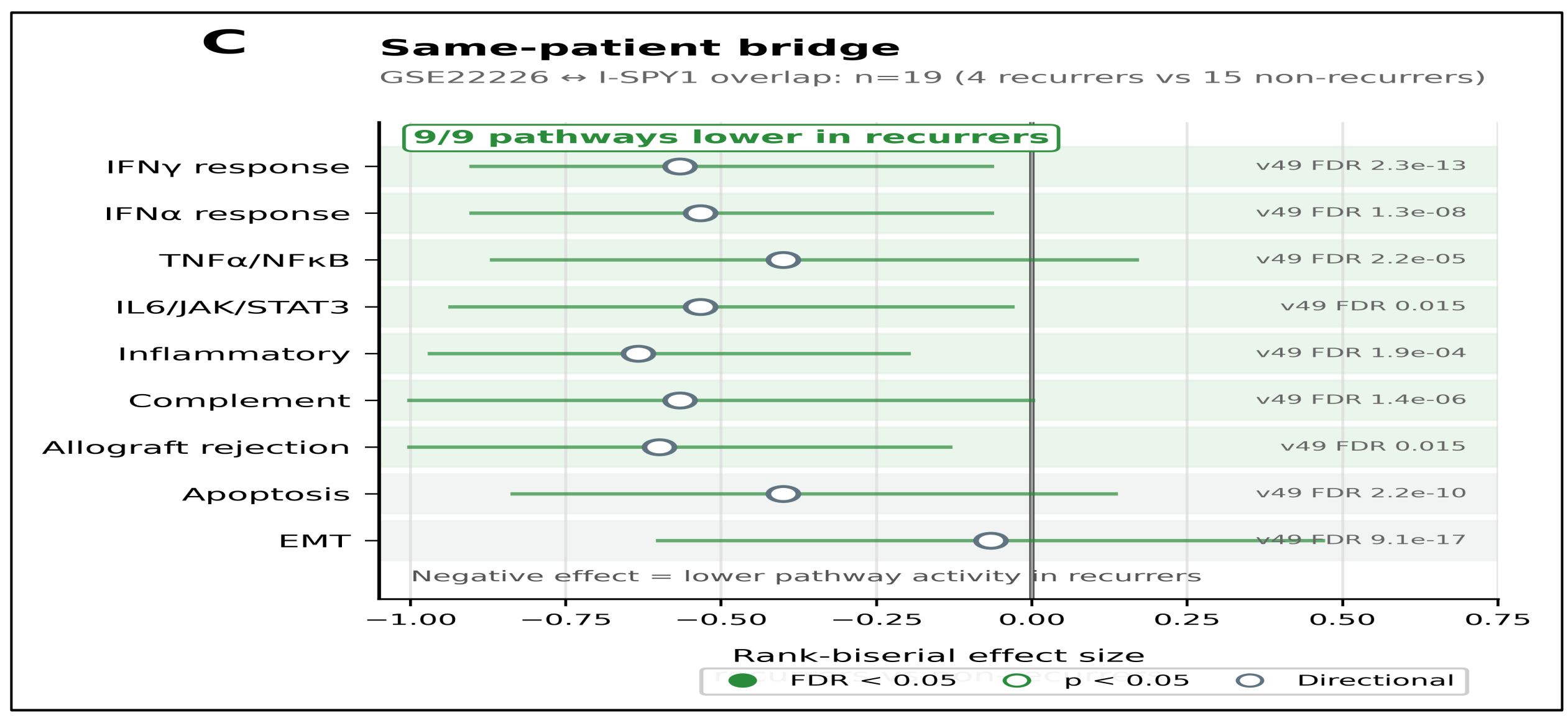

**Figure 11 | Bulk-transcriptomic phenotype of the immune-depleted Cluster 1 substate within Tier 2.** ***a***, *Within Tier 2 of the I-SPY2 master discovery substrate (n = 55), unsupervised k = 2 clustering on the locked-nine architecture composite partitions patients into Cluster 1 (n = 22, immune-depleted) and Cluster 2 (n = 33, immune-engaged). Across the locked-9 panel, Cluster 1 shows lower median scores in 8 of 9 pathways (only EMT non-concordant; one-sided sign-test p = 0.0195).* ***b***, *Seven-signature cross-cohort triangulation comparing I-SPY2 Cluster 1 vs Cluster 2 (circles) with I-SPY1 within-Tier-2 recurrers vs non-recurrers (diamonds). Cytotoxic/TIL, antigen-presentation, and checkpoint-associated signatures are concordantly lower; checkpoint reaches FDR < 0.05 in both. The bulk stromal/desmoplasia signal did not reproduce across cohorts (I-SPY2 +0.058, FDR = 0.764; I-SPY1 −0.467), and TGFβ, hypoxia/glycolysis, and angiogenesis remain null. These results do not exclude spatial exclusion.* ***c***, *Same-patient GSE22226↔I-SPY1 bridge (n = 19; 4 recurrers vs 15 non-recurrers): all 9 locked pathways shift lower in recurrers; in the sensitivity cohort (n = 36, 6 events), 6 of 9 reach FDR < 0.05. Filled markers, FDR < 0.05; open markers, p < 0.05; grey, directional only. Horizontal lines, 95% CI. Boxes (panel a), IQR; whiskers, 1.5× IQR. The licensed conclusion is broad immune depletion at bulk-RNA resolution, not a proved cellular mechanism or direct Cluster 1-specific survival effect.*

### 3.4 Confounder closure and treatment context

The pathology-confirmed I-SPY1 cohort establishes clinically consequential within-pCR recurrence risk, with concordant support from the UCSF best-response proxy. We next tested whether this risk could be attributed to the clinical variables that currently guide treatment decisions. The closure battery operates across all available cohorts. The three cohort systems (System A I-SPY2, System B I-SPY1, System C I-SPY1+UCSF pooled) carry the framework's discovery, external validation, and pooled responder recurrence anchoring claims; Duke (n = 908) operates outside the three-cohort-system architecture as the adjuvant-confounder closure cohort, addressing whether structural-axis prognostic information survives adjustment for the four adjuvant treatment modalities used in clinical practice (radiation, chemotherapy, endocrine therapy, anti-HER2 therapy).

#### 3.4.1 Baseline clinicogenomic closure

We evaluated a multi-cohort closure battery testing association between the structural phenotype and baseline clinicogenomic variables across I-SPY1 (System B), I-SPY2 (System A), UCSF (System C constituent), and Duke (adjuvant-confounder closure cohort) (Table 5). All fifteen resolved baseline variables showed no evidence of association at $\alpha = 0.05$, including age, receptor-defined subtype (HR × HER2 and PAM50 intrinsic subtype in I-SPY2, n = 704 in the companion-study GSE194040 cohort, which is larger than the present study's 701-patient imaging cohort), genomic risk (MammaPrint in I-SPY2; Oncotype DX in Duke, n = 261), race, menopausal status in I-SPY2 and Duke, histologic grade, molecular subtype, and UCSF lymph node status (Supplementary Table S44), with subtype non-redundancy and PAM50-not-a-proxy checks consolidated in Supplementary Table S61. Variables reflecting tumour burden or anatomic extent (baseline longest diameter, baseline functional tumour volume, AJCC T stage in Duke, AJCC N stage in Duke, pathologic residual size in UCSF), treatment assignment, and downstream care-pathway decisions are reported separately in Supplementary Table S45 as coherence rather than baseline-confounder analyses, because associations in these categories are mechanistically expected under the structural phenotype's biological interpretation and do not challenge the baseline non-redundancy interpretation. Formal adjusted closure spanning both non-redundancy and coherence variables is provided by the Duke 8-variable Cox model in §3.4.2, which retains an independent structural effect (HR = 0.610, 95% CI 0.41–0.91, p = 0.015) after explicit adjustment for nodal stage, histologic grade, molecular subtype, and the full adjuvant therapy stack.

**Table 5 | Baseline association battery: structural phenotype versus baseline clinicogenomic variables.**

| Variable | Cohort | Test | Statistic | p-value | Verdict |
|---|---|---|---|---|---|
| HR status | I-SPY1 | Fisher | OR = 1.87 | 0.137 | No evidence |
| HER2 status | I-SPY1 | Fisher | OR = 0.87 | 0.842 | No evidence |
| Age at diagnosis | I-SPY1 | Spearman | ρ = −0.067 | 0.466 | No evidence |
| Molecular subtype (HR × HER2) | I-SPY2 | Cramér's V | V = 0.070 | 0.329 | No evidence |
| PAM50 intrinsic subtype | I-SPY2 | Cramér's V | V = 0.083 | 0.301 | No evidence |
| MammaPrint genomic risk | I-SPY2 | Cramér's V | V = 0.061 | 0.104 | No evidence |
| Age | I-SPY2 | Spearman | ρ = −0.043 | 0.256 | No evidence |
| Race | I-SPY2 | Kruskal–Wallis | H = 1.97 | 0.579 | No evidence |
| Menopausal status | I-SPY2 | Kruskal–Wallis | H = 7.21 | 0.301 | No evidence |
| Lymph node status | UCSF | Cramér's V | V = 0.000 | 1.000 | No evidence |
| Age at MRI1 | UCSF | Spearman | ρ = −0.157 | 0.235 | No evidence |
| Oncotype DX recurrence score | Duke | Spearman | ρ = 0.036 | 0.565 | No evidence |
| Nottingham histologic grade | Duke | Cramér's V | V = 0.048 | 0.483 | No evidence |
| Molecular subtype | Duke | Cramér's V | V = 0.077 | 0.143 | No evidence |
| Menopausal status | Duke | Fisher | OR = 0.96 | 0.789 | No evidence |

All fifteen baseline clinicogenomic variables tested across four cohorts returned no evidence of association with the structural phenotype at α = 0.05. Tumour burden, anatomic extent, treatment-assignment, and care-pathway variables are evaluated separately in Supplementary Table S45 as coherence rather than confounder analyses, because associations in these categories are mechanistically expected under the structural phenotype's biological interpretation. Formal adjusted closure spanning both non-redundancy and coherence variables is provided by the Duke 8-variable Cox model (§3.4.2; Extended Data Figure 2; Supplementary Table S26), which retains an independent structural effect (HR = 0.610, 95% CI 0.41–0.91, p = 0.015). Complete 27-variable battery documented in Supplementary Tables S44 and S45.

### 3.4.2 Duke adjuvant closure

Baseline entropy showed no significant correlation with any of the four adjuvant modalities in the Duke adjuvant cohort (outside the three-cohort-system architecture; full-cohort prognostic and confounder closure substrate, n = 908). In an eight-variable Cox model (n = 908, 76 DRFS events; entropy + grade + nodal stage + subtype + four adjuvant modalities), the structural signal remained independently prognostic (HR = 0.610, 95% CI 0.41–0.91, p = 0.015). The structural phenotype survives adjustment for every available adjuvant therapy modality. Additional GLCM texture and geometry-coupling sensitivity analyses for the Duke prognostic substrate are provided in Supplementary Table S11.

### 3.4.3 Duke 27 treatment-intensification context

Twenty-seven HER2-positive Duke patients (Duke adjuvant cohort, outside the three-cohort-system architecture) received neoadjuvant therapy and achieved strict pCR. All 27 received anti-HER2 therapy; 19/27 (70.4%) received adjuvant radiation. Over 3.85 years median follow-up, zero distant recurrences occurred. This subgroup was not enriched for structurally favourable entropy (median percentile 35.8%) and had substantially higher tumour burden (median volume 6,801 vs 2,834 mm³, $p = 0.0006$). Treatment-stack intensity predicted membership (OR = 10.38, $p = 0.004$) while entropy did not ($p = 0.677$; Extended Data Figure 3). The Duke 27 finding suggests that durable pCR can emerge in structurally heterogeneous patients when mechanism-appropriate multimodal therapy is delivered at sufficient intensity. This is not a fatalistic framework. It is a tractable one.

### 3.4.4 Background parenchymal enhancement is not a confounder of the structural-immune axis

A separate concern is whether the structural axis is a proxy for background parenchymal enhancement (BPE), an MRI-visible measure of fibroglandular tissue activity in the contralateral breast that has been independently associated with breast cancer biology in some cohorts. We tested this question directly in the I-SPY1 BPE-matched pCR cohort (System B; n = 83 with paired imaging-axis BPE measurements and matched gene expression). After PAM50 rank-residualisation, BPE showed no association with the non-EMT eight-pathway programme composite (Spearman $\rho = 0.0033$ (Figure 7d), $p = 0.976$; n = 83) or with the locked-nine architecture composite ($\rho = 0.0049$, $p = 0.965$), and all nine pathways were null at FDR > 0.99. Three sensitivity strategies (unadjusted, PAM50-with-missing-category, PAM50-complete-case) produced concordant null results. This is a planned negative control: the structural-axis result in the same n = 23 tier-assigned subset shows separation (Mann-Whitney $p = 0.037$, Cliff's $\delta = -0.62$), while BPE in the same cohort shows none. The structural axis is not background parenchymal enhancement.

### 3.4.5 Causal architecture

**Table 6 | Driver/Confounder Map: outcome × variable class matrix.**

| Variable class | pCR occurrence | pCR quality | Overall recurrence | Distant recurrence | Local recurrence |
|---|---|---|---|---|---|
| Structural phenotype | pCR-orthogonal (r_b = 0.002) | Indexes (Tier 1 vs 2) | HR = 2.87 within pCR | HR = 0.466 (Duke DRFS) | NS (Duke LRFS) |
| Immune programme | Not tested (non-pCR excluded) | Indexes (IFNγ FDR = $10^{-31}$) | Candidate mechanism | Distant-specific pattern | — |
| Drug mechanism | Known driver | Quality-dependent (83.3% vs 67.6%) | Candidate framework | — | — |
| Tumour burden | Known driver | Not confounded | Not confounded | — | — |
| Molecular subtype | Known driver | Not confounded (PAM50-independent) | Adjusted: HR stable | Adjusted: stable | — |
| Histologic grade | Known modifier | Not tested | Adjusted: independent | — | — |
| Genomic risk (MammaPrint) | Associated | pCR-orthogonal (V = 0.061) | Not discriminating | — | — |
| Adjuvant therapy (4 modalities) | — | — | 8-var Cox: independent | — | — |
| Demographics (age, race, menopause) | — | pCR-orthogonal (all $\varepsilon^2 \leq 0.003$) | — | — | — |

The target was zero untested cells for major clinical variables. The map confirms (Table 6; Extended Data Figure 4) that the structural-immune axis indexes pCR quality and within-pCR recurrence independently of all tested confounders.

### 3.5 Clinical synthesis

**Table 7 | Clinical synthesis: four-tier response-quality framework with integrated biology and clinical implication.**

| Tier | Label | Biology | Recurrence | Clinical implication |
|---|---|---|---|---|
| 1 | Structurally favourable pCR | Immune-engaged; IFNγ-dominant; favourable-side EMT pathway-shift enrichment | 9.5–11.5% | Most favourable complete-response phenotype; prospective management relevance |
| 2 | Structurally adverse pCR | Tier-mean immune-engaged with bimodal substates (Cluster 1 ~40% immune-depleted + Cluster 2 ~60% engaged); adverse architectural state; estrogen-programmed; drug-mechanism-dependent | 23.8–28.6% (HR ≈ 2.87) | Not identifiable as low-risk by pCR alone |
| 3 | Organised residual disease | Context-dependent gradient biology | 29.7–46.2% | Standard adjuvant; favourable architecture despite residual disease |
| 4 | Aggressive residual disease | Inflammatory/stromal-elevated state-defining programme | 47.8–73.3% | Potential escalation or intensification priority |

Together, Systems A–C define four clinically interpretable response-quality states that map structural measurement, biological identity, and clinical consequence onto a single decision architecture (Table 7).

Tier 1 (structurally favourable pCR; n = 164 in System A) is the structurally favourable complete-response archetype and is associated with an immune-engaged programme (IFNγ replication FDR = $2.32 \times 10^{-13}$; allograft rejection replication FDR = $1.54 \times 10^{-2}$). System B recurrence is 11.5% in the full-cohort framework and 9.5% in the bilateral framework; The System C responder synthesis records 21.4% (n = 14, 3 events). Clinically, Tier 1 represents the most favourable complete-response phenotype for prospective management studies (Supplementary Note D17).

Tier 2 (structurally adverse pCR; n = 55 in System A) is the within-pCR response-quality stratum. Recurrence is 23.8–28.6% in pathology-confirmed System B; the System C responder synthesis records 36.8%, with an external pooled responder HR of 2.87. The association is strongest in the highest-reassurance pathology category (RCB-0), and observed events are predominantly distant, often fatal (three of six I-SPY1 recurrers died), and early (median 431 days). The signal remains directionally supported across the prespecified clinicogenomic, adjuvant-treatment, tumour-burden, and treatment-context analyses (§3.4). Within Tier 2, unsupervised k = 2 decomposition identifies Cluster 1 (n = 22, broadly immune-depleted) and Cluster 2 (n = 33, immune-engaged). This establishes bulk-transcriptomic heterogeneity within the recurrence-enriched tier but does not validate Cluster 1-specific recurrence risk (§3.2.4.1).

Tier 3 (structurally favourable residual disease; n = 360 in System A) is the organised-residual archetype: residual invasive disease at surgery within a structurally favourable architecture, with context-dependent gradient biology rather than one dominant programme. System B recurrence is 29.7% in the full-cohort framework and 46.2% in the bilateral framework. No prognostic separation was detected between Tier 3 and Tier 2 (n = 71, 21 events; HR = 0.97, 95% CI 0.23–4.17, p = 0.968), but the wide interval prevents a conclusion of equivalence. The structural extremes remain strongly separated (Tier 1 vs Tier 4 HR = 5.09).

Tier 4 (structurally adverse residual disease; n = 119 in System A) is the aggressive-residual archetype: highest-risk structural state. System B recurrence reaches 47.8% in the full-cohort framework and 73.3% in the bilateral framework. The structural axis further stratifies within RCB-3 (categorical-tier Cox HR = 3.58, p = 0.015; continuous per-SD HR = 2.72, §3.3.2), and five structural features reach FDR-corrected significance — the most replicated structural survival signal in the study. Molecular profile shows continuous immune-inflammatory variation with LumB-driven inflammatory reactivation.

The structural axis functions as a gateway rather than a complete biological classifier. A single pretreatment DCE-MRI scan identifies Tier 2, while bulk RNA resolves two substates within it: Cluster 1 is broadly immune-depleted and Cluster 2 is immune-engaged. The resolver is not routinely available at clinical decision time, and the available data do not estimate a Cluster 1-specific recurrence rate. Cross-cohort substate alignment (8 of 9 pathway directions concordant) and the GSE22226 same-patient bridge (n = 19, 4 events) support this interpretation, but neither substitutes for direct within-cohort outcome validation. The gateway+resolver model is therefore a prospectively testable architecture, not current clinical guidance.

The four-tier framework does not replace pCR or RCB; it assigns response quality within and across those categories, and is proposed as a hypothesis-generating architecture for prospective evaluation rather than validated clinical decision guidance (Extended Data Figure 5).

## 4. Discussion

### 4.1 The structural axis adds a response-quality dimension that pCR and RCB do not resolve

Complete response is not a single biological state: the structural axis adds a response-quality dimension that cuts across the pCR boundary. In the full I-SPY1 framework (n = 120, 35 events), no prognostic separation was detected between structurally adverse pCR (Tier 2) and organised residual disease (Tier 3) (n = 71, 21 events; HR = 0.97, 95% CI 0.23–4.17, p = 0.968). The wide interval prevents a conclusion that the tiers are equivalent, but the comparison shows that pathology alone does not resolve this boundary. By contrast, the structural extremes were strongly separated (Tier 1 vs Tier 4: HR = 5.09, 95% CI 1.40–18.53, p = 0.006). The clinically important direction is within pCR, where no residual-disease-based escalation signal exists; the non-pCR direction principally shows that the axis reflects response biology rather than a pCR-specific artefact. This addresses a question framed, but not resolved, by CTNeoBC[8], pooled meta-analyses[9], and I-SPY2[43]: pCR is strongly favourable for individuals, yet trial-level gains in pCR do not always translate into proportional survival benefit. The framework offers one candidate explanation—structurally different forms of pCR within the same endpoint label—without challenging pCR's individual-level validity.

This reframing distinguishes the present work from three related but fundamentally different literature streams. Prior imaging work in neoadjuvant breast cancer has largely treated pCR as the target to be predicted[33,34], and a parallel pathology and biomarker literature has focused on identifying baseline histopathologic predictors of pCR — including receptor status, proliferation indices, and tumour-infiltrating lymphocytes associated with pathologic clearance particularly in HER2-positive and triple-negative disease[50]. The present study inverts both lines of inquiry, asking whether the binary endpoint itself is biologically complete once pCR or RCB-0 has been reached. A related literature refined the non-pCR side of the response spectrum through the residual cancer burden classification[10,27], but left the pCR side undifferentiated; the four-tier framework addresses that missing half. The structural quality framework thus completes a symmetry: residual cancer burden graduated the non-pCR side of the response spectrum; the present work graduates the pCR side. The largest systematic attempt to identify recurrence-risk markers within pCR[7] found no significant associations for conventional variables; the structural dimension is not a refinement of those markers but an orthogonal axis they were not designed to detect.

The framework's evidentiary claim rests on a single structural axis tested across three cohort systems with distinct inferential roles. The axis operates as a gateway: where structural quality is adverse, the population is enriched for an immune-depleted substate within the recurrence-enriched tier (Cluster 1, ~40% of Tier 2), but the coupling is at the substate level rather than the tier-mean level — the two-layer model (structural axis as clinical screening, immune programme as biological resolver) in §4.4. The structural axis therefore contributes information that pathologic response and tier-mean bulk pathway summaries did not recover, while molecular decomposition remains informative within structurally defined states. System A characterises the axis biologically; System B reproduces it prognostically and provides the within-RCB-0 stratification anchor; System C anchors the pooled responder recurrence estimate. The GSE22226 same-patient transcriptomic bridge connects System A's molecular depth to System B's recurrence endpoint, closing the inference gap that no single cohort spans.

Three convergent lines of evidence support this claim. First, a persistent structural axis organises four response-quality tiers, remains orthogonal to pCR, and transports across five institutions under frozen inference. Deep learning and unsupervised clustering provide cross-representation support on overlapping patients rather than independent-patient confirmation (§3.1). Second, the axis is associated with a PAM50-independent immune programme that varies by treatment mechanism (§3.2). Third, structurally adverse pCR is associated with elevated external recurrence risk, with predominantly distant events and prespecified confounder analyses providing bounded support (§3.3–3.4). The added information is therefore structural, biological, and clinical, while the causal relationships among these layers remain unresolved.

Pathologic complete response and residual cancer burden remain powerful clinical endpoints. Complete pathologic clearance is prognostically meaningful and will continue to guide treatment decisions. But pathologic clearance does not, on its own, distinguish a structurally complete response from a structurally adverse one. The structural framework does not replace pCR — it adds a response-quality axis that pCR alone does not resolve, separating Tier 1 (structurally favourable pCR) from Tier 2 (structurally adverse pCR with residual recurrence risk) and Tier 3 (organised residual disease) from Tier 4 (aggressive residual disease, RCB-3 HR = 3.58). The framework refines pCR and RCB rather than replacing them.

The structural axis is complementary to RCB rather than redundant. RCB measures residual cancer burden — quantity and pattern of invasive disease at surgery — and is validated as a prognostic tool independently of treatment context (hazard ratio range 1.79–2.00 across I-SPY2 treatment groups). What RCB does not measure is response quality — the architectural and immunological properties of the tissue state that produced the observed burden. The manuscript's architectural claim is that pCR, RCB, and the structural axis together resolve within-category recurrence heterogeneity that any single endpoint compresses.

The structural axis is supported by five complementary evidence strands. It is orthogonal to pCR across five institutions and three scanner vendors; persists across treatment timepoints and molecular subtypes; varies with drug mechanism; shows external prognostic association across three institutions; and converges with a distinct frozen deep-learning representation in a locked same-patient subset. These strands do not all provide independent-patient validation—the deep-learning analysis is explicitly cross-representational—but they differ in cohort, representation, or inferential role. No single strand carries the framework. Their convergence makes a single-feature or single-pipeline artefact less plausible while preserving the limitations of each component.

### 4.2 Drug efficacy and treatment context

The drug-mechanism dependence of pCR quality offers a candidate explanation for why drugs that increase pCR rates do not always increase survival[37]. Residual cancer burden, the complementary pathologic endpoint, is prognostic independently of the treatment that produced it[27] (hazard ratio range 1.79–2.00 across graduated, non-graduated, and control treatment groups in I-SPY2). Pathologic complete response, by contrast, is drug-mechanism-dependent in both its rate and its biological quality — a dissociation the present analysis makes explicit at the level of immune engagement. The four-tier framework maps onto a clinical shorthand: structurally favourable pCR (Tier 1, immune-engaged at the tier-mean level) and structurally adverse pCR (Tier 2, with a bimodal substate distribution containing the immune-depleted Cluster 1 substate within recurrence-enriched Tier 2 alongside the immune-engaged Cluster 2). This response-quality distinction, complementary to the binary endpoint, is drug-mechanism-dependent. HER2-targeted monoclonal antibodies can engage complement-dependent cytotoxicity and phagocytosis[41]. Within I-SPY2, this treatment class produced predominantly Tier 1 structurally favourable pCR (83.3% retention; original tier-based estimates), while neratinib produced Tier 2 structurally adverse pCR at substantially higher rates (67.6% retention; interaction FDR = 0.004). The quality gap between mechanism classes widened after structural adjustment, from 12.6 to 15.6 percentage points. This finding is restricted to the HER2-positive mechanisms represented in I-SPY2. Whether the principle that endpoint quality is mechanism-dependent extends to other drug classes and subtypes is a question for future investigation. Adaptive platform trials using binary pCR as the graduation criterion may therefore systematically favour arms producing higher rates of structurally adverse response, potentially advancing therapies whose apparent pCR-rate efficacy does not translate into durable disease control. These findings do not invalidate pCR as a surrogate, but they identify a structural axis that may partially explain the pCR-rate-versus-survival-gain disconnect documented in CTNeoBC[8] and subsequent meta-analyses[36]; whether structurally-stratified pCR improves trial-level surrogacy properties is an empirical question testable in existing pooled datasets. The expanded arm-stratified analysis revealed a structured pattern across four convergent methods: all seven immune/inflammatory pathways changed direction between ADC and neratinib while apoptosis and EMT retained favourable-side orientation, producing a structured 7+2 architecture (patient composition Fisher p = 0.049), consistent with selective immune-module reconfiguration rather than wholesale programme disruption. Drugs do not merely differ in pCR rate; they produce different biological versions of pCR. The tier-level decomposition makes the hypothesis testable: a drug that increases the Tier 1 fraction of pCR may enrich for more durable responses, whereas a drug that increases the Tier 2 fraction may raise the pCR rate without a proportional gain in durable disease control. Under this framework, the pCR-rate-versus-survival-gain disconnect documented in CTNeoBC and subsequent meta-analyses is a candidate tier-composition disconnect.

This offers a candidate resolution for the long-standing disconnect between trial-level pCR rate gains and survival improvements[8,37]: trials that graduate arms based on raw pCR rates may systematically favour mechanisms that produce a higher fraction of structurally adverse pCR (the immune-depleted Cluster 1 substate within recurrence-enriched Tier 2) without producing more durable disease control. Symmans et al. 2021 (the I-SPY2 RCB-EFS analysis) explicitly hypothesised, on the basis of the ARTEMIS bevacizumab observation, that a pCR achieved with the addition of bevacizumab might not carry the same good prognosis as a pCR achieved with other therapies[27] — bevacizumab significantly raised pCR rates without improving disease-free survival, and pCR was not prognostic within the bevacizumab-treated subgroup. The present analysis operationalises that hypothesis biologically, identifying the immune-engagement programme that distinguishes structurally favourable from structurally adverse pCR and showing that the distinction is drug-mechanism-dependent. Whereas Symmans et al. reported that RCB retained similar prognostic value across I-SPY2 treatment arms[27], indicating that residual burden captures a mechanism-agnostic dimension of response, the within-pCR structural axis identified here is drug-mechanism-dependent. These measures are therefore complementary: RCB quantifies residual disease magnitude, whereas structural quality distinguishes the biological route by which pathologic clearance was achieved. Phase 3 evidence from KATHERINE supports an analogous mechanism asymmetry in the residual-disease setting[46], although that trial did not address within-pCR biology directly.

The Duke 27 subgroup illustrates a complementary principle: treatment intensity is associated with reduced clinical expression of structural risk. Among HER2-positive patients receiving intensive multi-agent neoadjuvant and adjuvant therapy, zero distant recurrences occurred over 3.85 years — despite structurally heterogeneous baseline profiles (not enriched for favourable entropy) and elevated tumour burden. Treatment-stack intensity, not baseline entropy, predicted membership in this durable-response subgroup (OR = 10.38, $p = 0.004$). Confounding by indication cannot be excluded: intensification likely reflects clinical features that themselves modify outcome, and randomisation would be required to attribute the effect to therapy alone. Clinical evaluation of neoadjuvant treatment success should therefore consider not only whether pCR was achieved, but what kind of pCR was achieved, by what drug mechanism, and in what treatment-intensity context.

**4.3 Patient-level response-quality stratification**

The immune programme of §3.2 offers a candidate biological interpretation for the within-pCR persistence pattern. The immune-depleted Cluster 1 substate may mark weaker surveillance after pathologic clearance, but neither bulk RNA nor the small external cohorts establish that mechanism causally. The predominantly distant, early recurrence pattern (median 431 days) is compatible with impaired systemic immune surveillance and motivates direct spatial and longitudinal testing.

**4.4 Biological interpretation**

The structural phenotype, initially defined purely by imaging, is associated with a reproducible immune-architecture programme with prominent EMT-set differential-expression organisation rather than a single molecular identity. Within pCR, the structural distinction corresponds to coordinated changes across the nine pre-specified pathways of the locked panel, with IFNγ response dominant and EMT ranking second after PAM50 adjustment (IFNγ FDR = $4.55 \times 10^{-31}$; EMT FDR = $1.19 \times 10^{-22}$); in the gene-level pathway-shift ranking, EMT-set organisation concentrates on the Tier 1/favourable side, whereas patient-level EMT scores do not reproduce this favourable-side relationship. This is a pathway-shift contrast and does not classify Tier 2 as absolutely EMT-low. No individual gene reaches FDR significance; the signal is a systems-level state, not a marker. Entropy_T0 therefore indexes an architectural category associated with this transcriptomic programme; it does not directly measure EMT activity.

Three properties of this window are worth making explicit. First, it is a state, not a gradient: within responders, continuous entropy–pathway correlations collapse (0/9 pathways were FDR-significant, with a maximum $|\rho|$ of 0.072; Supplementary Tables S38 and S40), while categorical Tier 1 versus Tier 2 contrasts reach FDR = $2.32 \times 10^{-13}$. Within non-responders the same panel behaves as a continuous gradient — exemplified by interferon-γ ($\rho = -0.289$, FDR = 0.016; §3.2.2) — while the full-rank pathway-shift analysis retains favourable-side enrichment across all nine pathways, with larger rank-biserial effects than within pCR for eight of the nine ($n = 95$; Supplementary Table S34). The architecture of the immune programme depends on whether tumour has been cleared — discrete when clearance has succeeded, continuous when it has not. Second, it is drug-selective, not drug-erasable: antibody-drug conjugate therapy preserves the programme, while neratinib reverses all seven immune/inflammatory pathways, whereas apoptosis and EMT retain favourable-side orientation. A structural phenotype this persistent would be uninteresting if therapy could not touch it; what makes it clinically relevant is that mechanism-specific therapy engages it differently. Third, response-conditioned biology is not reducible to canonical immuno-oncology readouts. Within the 220-patient pCR cohort, checkpoint-gene expression, CIBERSORT cell-abundance proxies and continuous entropy–pathway correlations were null (immune-cell proxies and checkpoint genes, all FDR = 0.986; maximum $|\rho|$ = 0.031 with all nine pathways at FDR > 0.9; Supplementary Table S35). In the exploratory 698-patient full cohort, by contrast, Entropy_T0 carried modest immune/inflammatory patient-level context after adjustment for PAM50 subtype and baseline functional tumour volume, while corrected transcriptomic modelling did not materially reconstruct the underlying continuous imaging phenotype. The full-cohort association and the responder-specific null are therefore complementary rather than contradictory: within responders, response-quality biology is resolved more strongly by coordinated programme organisation and immune-substate architecture than by scalar pathway amount.

These properties support a layered association, not a one-way causal cascade. The full-cohort analysis further separates molecular association from patient-level reconstruction. Immune/inflammatory pathway activity carried modest same-patient context for Entropy_T0, but this context was insufficient to reproduce the underlying continuous imaging coordinate in held-out patients, and discrimination of the locked structural category remained weak under the corrected primary classifier. This distinction matters: biological covariance does not imply that the imaging phenotype is simply a molecular assay in another form. Bulk RNA instead provides partial biological decomposition of an imaging-defined tissue-level phenotype. Pretreatment MRI identifies a macroscopic response-quality category; bulk RNA resolves the associated pathway programme and the distinct immune substates within it. The clinical contribution is that complete response can arise within biologically different baseline contexts, some of which are enriched for recurrence despite pathologic clearance. This boundary is clarified by the companion tissue-biology analysis.[52] That analysis defines a recurrent ECM-myCAF-aligned matrix-construction programme across molecular and spatial measurement layers and shows that stromal identity, immune abundance, spatial allocation, epithelial contact and functional context are not interchangeable. Although the matrix-construction programme is associated with EMT-pathway signal in bulk expression, that relationship cannot be composed transitively with the MRI result: in 692 same-patient MRI–RNA cases from the same I-SPY2 discovery cohort, imaging entropy was essentially unrelated to matrix-construction, pan-CAF and related stromal quantities. The structural axis identified here is therefore an imaging-defined response-quality phenotype associated with EMT-pathway organisation and immune-substate heterogeneity, not a direct measurement of stromal quantity, ECM-myCAF or matrix activity (Extended Data Figure 6e; Supplementary Table S75).

At bulk-transcriptomic resolution, Cluster 1 is broadly immune-depleted: cytotoxic/TIL, antigen-presentation, and checkpoint-associated programmes are all lower in I-SPY2, with concordant directions in the small I-SPY1 recurrence anchor. This coordinated fall is more consistent with immune scarcity or broad immune disengagement than classical exhaustion, but bulk RNA cannot distinguish fewer immune cells from lower activity per cell. The structural axis should accordingly be interpreted as an imaging-derived index of response architecture, not as a direct measurement of spatial texture, any single histological compartment, or any individual molecular programme. The tested stromal/desmoplasia signature did not reproduce across cohorts, while hypoxia and TGFβ remain null; spatial stromal exclusion is not excluded.

Important caveats limit how far this interpretation extends. The molecular analyses use bulk RNA and cannot resolve cell number, cell state, or spatial allocation. The discovery subset contains 63 pCR patients; the expanded replication uses 68 non-overlapping HER2-positive pCR patients within the same I-SPY2 platform, while I-SPY1/GSE22226 supplies external-platform consistency. Tier 2 Cluster 1 vs Cluster 2 contains 22 vs 33 patients and has no validated I-SPY2 recurrence endpoint. What the present data establish is the existence and clinical relevance of the structural phenotype; what the molecular data establish is an associated immune-substate architecture with EMT-set organisation rather than a causal pathway.

This biological framework anchors the constrained interpretation of structural refinement across RCB classes. The data support a precise claim: structural information is most clinically consequential where conventional pathology is most reassuring (RCB-0/pCR), and within heavy residual disease (RCB-III) it behaves as a continuous severity modulator rather than a binary classifier. RCB and pretreatment MRI structure are therefore complementary axes — RCB quantifies residual burden, while pretreatment structure captures a complementary response-quality dimension.

The cross-cohort biological architecture is asymmetric.

The bimodal substate decomposition demonstrated in I-SPY2 (Cluster 1 vs Cluster 2; sign-test p = 0.0195) predicts the cohort-scale pattern of bulk-RNA visibility. In the pooled within-pCR cohort of I-SPY2 (n = 219, Tier 1 + Tier 2), the within-Tier-2 bimodal mixture (~40% Cluster 1 immune-depleted + ~60% Cluster 2 immune-engaged) dilutes the tier-mean separation between Tier 1 and Tier 2, which is why the within-pCR bulk-RNA test finds AUC = 0.546. The smaller I-SPY1 BPE-matched subset (Tier 2 n = 5) may sample disproportionately from the immune-depleted substate, and tier-mean separation becomes visible (Mann-Whitney p = 0.037, Cliff's δ = −0.62). Both observations are consequences of the same substate-resolved model, not contradictory findings. The GSE22226 same-patient bridge (n = 19, 4 events) provides the same-patient anchoring: it connects System A's molecular depth to System B's recurrence endpoint in matched patients — the architectural piece that licenses the within-pCR molecular-recurrence inference at scale.

The within-pCR immune architecture is supported by complementary, non-equivalent evidence layers: the Tier 1-versus-Tier 2 pathway-shift signal in I-SPY2; within-Tier-2 Cluster 1-versus-Cluster 2 decomposition (Cluster 1 immune-depleted, n = 22 of 55) as a specificity and resolver analysis; directional Tier 1-versus-Tier 2 support in the small I-SPY1 BPE-matched cohort; and the GSE22226 same-patient bridge (immune AUROC = 0.896 within the same-patient cross-walk subset, n = 16). In the residual-disease setting, the framework now resolves two convergently supported state signatures with architectural symmetry. The earlier Tier 3-up immune pattern did not replicate in its original form; what emerged instead is a cleaner asymmetric architecture. Tier 3 — organised residual disease — shows reproducibly low locked-9 immune/stromal activity relative to Tier 4 across both cohorts: 9/9 pathways Tier 3-lower in the I-SPY2 residual-disease analysis (n = 477; 7/9 BH-FDR-significant; IPW-adjusted preserves direction) and 9/9 Tier 3-lower in the I-SPY1 residual-disease analysis (n = 68; preserved after PAM50 residualisation; PAM50 $\chi^2$ subtype-balance p = 0.930), yielding 18 of 18 cohort-pathway observations concordantly Tier 3-lower (exact binomial $p = 3.81 \times 10^{-6}$; Supplementary Table S53). Tier 4 — aggressive residual disease — carries the inverse state signature: an inflammatory/stromal-elevated programme over the same locked-9 panel, with the strongest within-subtype FDR-significance localising to LumB residual disease in I-SPY1. Within both residual-disease tiers, the locked-nine architecture composite does not discriminate recurrence in the cohorts where this could be tested (Tier 3: I-SPY1 rank-biserial −0.262, p = 0.170; Tier 4: I-SPY1 rank-biserial −0.086, p = 0.747; I-SPY2 −0.171, p = 0.501): the immune/stromal programme is state-defining for residual-disease architecture but not within-state prognostic, in cross-cohort consistent fashion. In clinical-translation terms, the immune programme is therefore not a substantial within-state biomarker in Tier 3 or Tier 4 residual disease, where pathology and RCB anatomy already provide prognostic resolution; its candidate biomarker resolution concentrates within pCR, where pathology calls the cancer cleared but the immune programme distinguishes an immune-engaged favourable state from an immune-depleted substate within recurrence-enriched Tier 2. This is the framework's core biological claim: that the convergently supported immune programme adds biomarker resolution exactly where pathology is least informative. The within-pCR Tier 1 vs Tier 2 contrast remains the clinical claim and the candidate biomarker direction. The residual-disease cross-cohort signal characterises two biologically distinct non-pCR states — immune-quiet organised residual disease and inflammatory/stromal-elevated aggressive residual disease — establishing framework-validity by demonstrating that pCR/non-pCR is too coarse to capture the structural-immune state space the four-tier framework resolves, while preserving the asymmetric framework that places clinical actionability in the within-pCR axis.

### 4.5 Clinical translation and future directions

The structural quality assessment is implementable on existing clinical infrastructure. The primary measurement — baseline Shannon entropy of the SER-derived DCE-MRI — requires only the pretreatment dynamic contrast-enhanced scan already acquired as part of standard neoadjuvant breast cancer imaging protocols. No additional imaging visits, contrast agents, or hardware are needed. The longitudinal slope analysis confirmed that the structural phenotype does not require serial monitoring: a single baseline scan reflects what persists through treatment (slope p = 0.614). In the I-SPY1 HER2-positive within-pCR cohort (n = 21), native structural assessment flagged two patients for every recurrence identified (NNS = 2.0; Supplementary Table S28) with 83% event sensitivity (5 of 6 events captured). Pooled external validation (I-SPY1 + UCSF, n = 33) yielded NNS = 2.71 and 70% sensitivity. Prospective evaluation could take the form of structural quality stratification at randomisation in platform trials such as I-SPY2.2, where Tier 2 could be evaluated as a stratification factor for surveillance, treatment-escalation, or trial-enrichment strategies targeting residual structural risk.

This uncertainty reduction operates on both sides of the pathologic-response boundary. Within pCR, pathology alone treats all complete responders as belonging to the same favourable category. The pathology-confirmed I-SPY1 arm establishes that premise, while the UCSF best-response proxy provides concordant responder support; across the pooled System C cohort, 10 recurrences occurred among 33 responders. Adding pretreatment structural MRI concentrated 7 of these 10 recurrences in the structurally adverse subgroup, reducing the missed-recurrence fraction from all pooled responder events under response-only reassurance to 3 of 10 after structural stratification, at a quantified specificity cost. The same principle holds in the stricter HER2-positive RCB-0 subset: RCB-0 alone carried a 28.6% recurrence rate (6 of 21), whereas structurally favourable RCB-0 carried a 12.5% recurrence rate (2 of 16) and structurally adverse RCB-0 carried an 80.0% recurrence rate (4 of 5). Conversely, pCR was not necessary for recurrence-free follow-up: in the I-SPY1 four-tier framework cohort, 45 of 64 patients with favourable-structure residual disease remained recurrence-free. Thus, structural MRI does not make pCR wrong; it reduces the uncertainty left by a binary endpoint, identifying complete responders who still carry risk and residual-disease patients whose favourable architecture suggests a different risk profile than pathology alone implies.

The four-tier framework augments three clinically established neoadjuvant readouts, each on its own well-validated and independently used in clinical practice: clinician-assigned response assessment, pathologic complete response, and residual cancer burden. Structural quality adds non-redundant response-quality information to each. In UCSF (n = 49), baseline structural entropy correlates with clinician-assigned response assessment at $\rho = 0.527$ ($p = 6.8 \times 10^{-4}$), confirming that the imaging axis carries information consistent with what experienced clinicians already recognise qualitatively, then quantifies it. In the pooled external responder synthesis, the structural axis identifies the recurrence-enriched subgroup central to this paper: HR = 2.87, 95% CI 1.38–5.96 in System C, combining pathology-confirmed I-SPY1 pCR with the UCSF best-response proxy. Within RCB-0 — the most reassuring pathology category — the structural axis stratifies recurrence in I-SPY1, with Cox HR = 5.50 overall and HR = 9.04 in the HER2-positive subset, where conventional pathology offers no further resolution. Each of these three readouts is independently used in clinical practice; each is independently strengthened by the addition of pretreatment structural quality. The framework therefore augments the existing endpoint hierarchy at three independent levels, not at a single pCR-specific boundary. In the clinical decision pathway it would sit upstream of pathology, providing a baseline structural-quality reading available at the time of MRI acquisition; the surgical pathology determination remains the primary endpoint at definitive surgery; and the structural axis adds a within-category response-quality flag for patients whose pathology category alone does not resolve their residual recurrence risk. No single readout is sufficient. Together, they provide complementary views of which complete responders remain at elevated recurrence risk.

Beyond the primary I-SPY1 external validation, the structural framework received additional external transport support in UCSF, Duke and QIN-BREAST, spanning two decades of acquisition, three scanner vendors, and five institutions under frozen inference. External recurrence transport was strongest at the level of graded structural risk. In I-SPY1 the native MAD coordinate reproduced the categorical structural assignment exactly and preserved the adverse recurrence direction across every admissible cutpoint and every single-patient or single-event deletion, so the association reflects a graded ordering rather than one convenient dichotomisation. Once representation-specific numerical scale was removed, that ordering was preserved across both external cohorts, whereas the categorical crosswalk assignments were not interchangeable between representations. This supports transport of a graded structural-risk dimension rather than of a universal radiomic feature, an absolute numerical scale or a common binary boundary. The representation-aware validation framework (RFCA) developed in the companion study[1] (Supplementary Note D8 extends RFCA to learned representations) provided a principled basis for assessing when frozen transport is appropriate: Duke was classified as regime-compatible (Cohen's d = −0.020), while QIN-BREAST required carrier-feature transport (Supplementary Note D8) due to vendor-induced distributional shift. This five-level commensurability taxonomy offers a reusable infrastructure for structural biomarker validation beyond the present study.

Extension to non-HER2 subtypes represents the most immediate next step toward clinical translation. Spatially resolved tissue characterisation remains an important future test of which tissue compartments produce the MRI-visible architecture. Prospective validation of the structural phenotype as a clinical decision tool — particularly for post-pCR surveillance intensity and adjuvant treatment escalation — requires interventional studies with pre-specified structural assessment protocols — including dedicated cohorts for within-pCR recurrence validation and drug-mechanism interaction testing — that this retrospective analysis cannot provide. As mature recurrence endpoints become available, QLHC I-SPY2 and I-SPY2.2 provide natural next-stage validation cohorts. The existing pretreatment entropy definition and prespecified transport framework should be applied without outcome-driven feature selection, threshold optimisation, or model retraining. This prospective handoff will test transportability across a later treatment era and provide the event count needed to improve the precision of the within-pCR recurrence estimate; the current power analysis indicates that approximately 28 events would provide 80% power if the observed hazard ratio of 2.87 is maintained (Figure S9). Until such validation is completed, the present findings should be interpreted as recurrence-enriching and hypothesis-strengthening rather than as a validated clinical decision rule.

The pairwise Tier 3 vs Tier 2 finding raises a complementary clinical question on the residual-disease side. Tier 2 patients have achieved pCR but retain adverse structural response quality, whereas Tier 3 patients have residual disease but favourable structural response quality. This does not support treatment de-escalation or escalation by itself, but it suggests that future trials with pre-specified structural assessment would help clarify whether structural response quality refines post-neoadjuvant management in both directions: identifying complete responders who may need closer follow-up or additional therapy, and residual-disease patients whose organised structural phenotype may carry a different risk profile than pathology alone implies.

The endpoint-compression principle demonstrated here for pCR in breast cancer may apply to other binary clinical endpoints that claim biological completeness. Across epithelial solid tumours, pCR after neoadjuvant therapy is widely used but varies substantially in frequency, prognostic strength, immune correlates, and relationship to long-term outcome[51] — consistent with the broader principle that binary clearance endpoints can be clinically useful while remaining biologically compressed. Concrete extension settings include pathologic complete response in rectal cancer, RECIST-complete response in solid tumours, RANO-stable disease in glioblastoma, and MRD-negative status in haematologic malignancies. The two-axis response-space construct and the replication strategy may provide a portable template that can be tested in other endpoint-compression settings.

### 4.6 Limitations

Several limitations should be considered. I-SPY2 recurrence endpoints are not available through the public TCIA release (the public event_work field is aligned with hormone receptor status and cannot be interpreted as validated recurrence-free survival; §2.6), restricting pathology-confirmed within-pCR recurrence analysis to I-SPY1; UCSF contributes a supportive best-response proxy. A direct next validation step is nonetheless available: mature I-SPY2 event-free and distant recurrence-free survival follow-up has been reported for 950 patients across treatment arms[43], and de-identified subject-level endpoint data can in principle be requested through the I-SPY2 concept-proposal and data-use process; linking those endpoints to the locked structural-tier assignments would test whether the present response-quality framework generalises from the external I-SPY1/UCSF/Duke validation architecture into the larger adaptive-platform cohort. The pooled responder recurrence analysis rests on 33 patients with 10 events — sufficient for the pooled HR but underpowered for subgroup-specific adjusted analyses (Supplementary Figure S9). The within-RCB-0 structural concentration analysis is further limited by small event counts (HER2-positive RCB-0 n = 21, 6 events; pooled responder System C n = 33, 10 events), so the RCB-0 result should be interpreted as a concentrated external signal supported by small-sample robust estimators, requiring validation in a larger mature-follow-up cohort rather than serving as a standalone survival endpoint. Long-horizon context is provided by the 5,161-patient pooled meta-analysis[2], within which the present pooled estimate represents the early portion of risk accumulation. One event-bearing patient (UCSF_BR_13) was lost during denominator locking and is disclosed. Adjuvant endocrine therapy is not adjusted for in the pooled external analysis because I-SPY1 and UCSF pre-date standardised adjuvant records; the Duke eight-variable Cox (§3.4.2) provides independent adjuvant-confounder closure. Within-RCB-class structural refinement was not uniform across the RCB hierarchy: RCB-II showed no refinement and RCB-I was too sparse for inference (n = 11, 2 events), although continuous structural severity remained prognostic within RCB-III. Within-pCR external replication was not estimable in Duke neoadjuvant pCR patients because no DRFS events occurred in the Duke NAT pCR subset (n = 64); Duke therefore serves as a full-cohort prognostic and adjuvant-confounder closure cohort. UCSF response labels derive from proxy and native path-size fields rather than the locked I-SPY pCR annotation framework, and UCSF response subgroup results are treated as supportive sensitivity analyses.

Framework-level limitations are as follows. The full-cohort framework's unordered four-group log-rank test (3 df) was borderline (p = 0.057), reflecting designed Tier 2/Tier 3 overlap that the binary pCR endpoint does not separately resolve. The framework-aligned ordered alternative (1-df log-rank for trend, p = 0.010, HR = 1.66 per tier) and the pairwise structural-axis contrasts (Tier 1 vs Tier 4; the pooled responder HR) are the appropriate primary tests for the ordered four-tier hypothesis. Deep-learning convergence is support-bounded ($\kappa = 0.069$ at 220-patient scale). Cross-section multiplicity was not formally adjusted beyond within-family FDR. The Tier 2 estrogen biology is hypothesis-generating. Tumour-infiltrating lymphocyte (TIL) counts were not available; individual canonical immune markers (CD274/PD-L1, CD8A) did not separate Tier 1 from Tier 2, consistent with a programme-level state distinction. Arm-level subgroups remained small (ADC n = 17, neratinib n = 20), and the drug-mechanism finding should be regarded as a convergent, specific, testable claim that requires prospective replication. The pooled responder recurrence anchor (System C, n = 33 with 10 events) is the smallest of the framework's three primary recurrence substrates. Several convergent analyses support the direction of this estimate: cross-cohort meta-analytic concordance ($I^2 = 0\%$); within-RCB-0 stratification in System B; the GSE22226 same-patient transcriptomic bridge; cross-cohort substate alignment; and the framework-aligned ordered trend test. These analyses motivate, but do not replace, validation in a larger dedicated within-pCR recurrence cohort before any clinical decision use. Access to validated I-SPY2 survival endpoints and I-SPY2.2 outcome data would provide the natural next test of this framework: whether the structural response-quality states defined here retain prognostic value within the same adaptive-platform setting in which pCR is used for treatment evaluation. Such data would shift the recurrence analysis from convergent external anchoring toward larger same-platform outcome validation, while preserving the present study's separation between discovery, biology, and recurrence inference.

The molecular characterisation was discovered in 63 pCR patients with matched Tier 1/Tier 2 assignments, replicated in a non-overlapping 68-patient HER2-positive within-pCR subset of I-SPY2 (IFNγ FDR = $2.32 \times 10^{-13}$), expanded in the full 220-patient I-SPY2 within-pCR cohort, and extended to 479 non-pCR patients. Spatial and single-cell resolution remain unavailable. The drug-mechanism interaction includes ADC Tier 2 n = 2 as the smallest subgroup. Non-pCR patients used the rebuilt Entropy_group definition (Entropy_T0 > 3.284 from the locked I-SPY2 training-partition threshold) rather than formal tier assignments, since these were unavailable for the GSE194040 non-pCR subset — biologically corresponding to but not formally equivalent to the Tier 3/Tier 4 partition. Subtype-stratified analyses within pCR revealed context-dependent pathway structure across PAM50 subtypes, but per-subtype sample sizes (n = 21–114) preclude definitive within-subtype claims. Prospective validation with pre-specified structural assessment is required before clinical deployment.

The full-cohort structural-axis attribution was exploratory. Although it applied a previously defined nine-pathway response-quality panel, that panel was not originally established for full-cohort Entropy_T0 attribution: six of nine continuous associations survived panel-restricted correction, whereas none survived the corresponding exploratory 50-Hallmark continuous-entropy correction. Exact and independently reconstructed pathway scores use overlapping patients and expression data and therefore establish estimator consistency rather than independent replication. Transcriptomic recoverability was examined under several complementary linear and low-dimensional frameworks, but these analyses do not establish that no conceivable molecular model could reconstruct the imaging phenotype. The corrected primary binary classifier remained weak rather than definitively null (out-of-fold AUROC = 0.537; 500-permutation p = 0.112), whereas the cleaner continuous analyses yielded no out-of-fold reconstruction. The all-gene ridge model selected the strongest regularisation boundary tested and is therefore interpreted only as a sensitivity analysis. The predefined pathway-summary predictors used overlapping expression data and were not independently permutation-tested; they are descriptive representations rather than independent validation. Bulk RNA also lacks the spatial and cellular resolution required to identify the tissue substrate generating the imaging phenotype, and the q75 threshold is retained as a locked structural classifier rather than as a demonstrated biological transition.

### 4.7 Concluding statement

Pretreatment DCE-MRI entropy identifies a structural response-quality axis that is orthogonal to pCR, persists through treatment, varies by drug mechanism, and transports across institutions. External cohorts associate adverse structural response with elevated recurrence risk within pCR: in the pooled HER2-positive responder synthesis the hazard ratio was 2.87, with 7 of 10 recurrences concentrated in the adverse subgroup, and within the small HER2-positive RCB-0 subset (n = 21) recurrence was 12.5% with favourable versus 80.0% with adverse structure. Bulk RNA resolves a favourable-side immune/EMT pathway-shift programme and a broadly immune-depleted substate within the adverse tier. Bulk RNA supplies bounded biological context, while the underlying continuous imaging phenotype was not materially reconstructed under the tested transcriptomic frameworks. Deep learning supplies cross-representation convergence on overlapping patients, not independent-patient validation. Together, these findings support a reproducible response-quality dimension while preserving the observational and small-event limits of the clinical estimates.

pCR and RCB remain essential post-treatment endpoints; the framework refines rather than replaces them. Pretreatment MRI provides a baseline structural reading, pathology establishes clearance or residual burden at surgery, and bulk RNA supplies bounded biological context. The recurrence-enriched adverse-pCR tier contains a broadly immune-depleted substate, while the tested bulk stromal-hypoxic signatures do not explain the within-tier contrast.

The three cohort systems and their distinct inferential roles are documented in Supplementary Tables S54–S56. The central contribution is therefore a pretreatment structural response-quality axis that reduces uncertainty within pCR and RCB by resolving clinically consequential heterogeneity that pathology alone does not capture, rather than a new standalone clinical decision rule. Prospective, adequately powered outcome validation is required before clinical use.

## Extended Data Figures

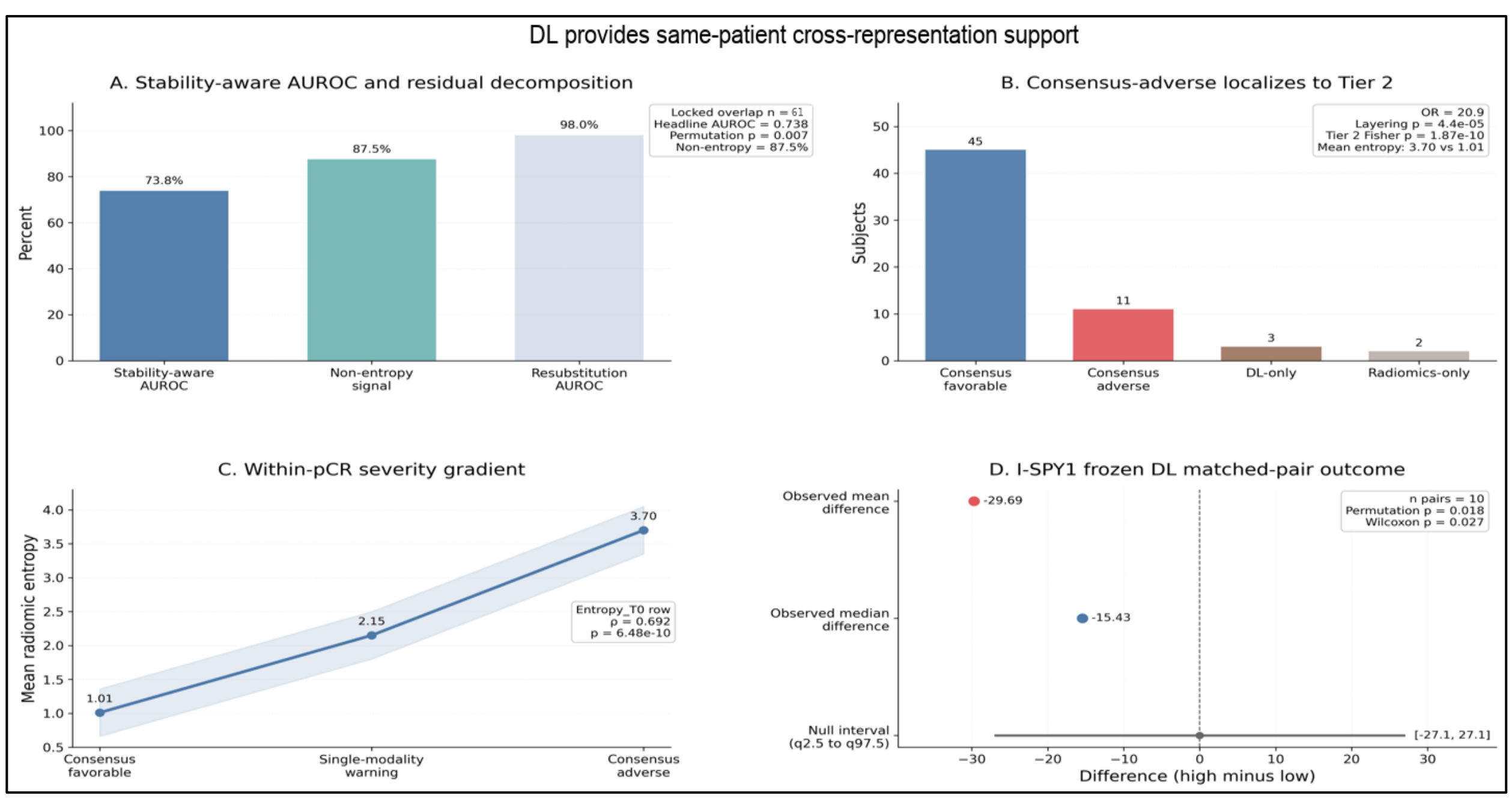

**Extended Data Figure 1 | Deep learning provides same-patient cross-representation support for the structural phenotype.** ***a,*** *In the locked overlap subset, the stability-aware DL estimate yields AUROC = 0.738 with permutation p = 0.007, while residual decomposition indicates that 87.5% of the DL signal is non-entropy. A lighter bar shows the in-sample resubstitution AUROC (0.980), included for contrast; the reported value is the conservative stability-aware generalisation estimate (0.738). Deep-learning convergence is support-bounded at full-cohort scale (κ = 0.069 at 220 patients; see Limitations).* ***b,*** *DL consensus states map strongly onto the structural framework: the four-cell decomposition yields 45 consensus-favourable, 11 consensus-adverse, 3 DL-only, and 2 radiomics-only cases; adverse layering is concentrated in Tier 2 (odds ratio = 20.9, layering $p = 4.4 \times 10^{-5}$, Tier 2 Fisher $p = 1.87 \times 10^{-10}$), with markedly higher entropy in the adverse group (3.70 vs 1.01).* ***c,*** *Within pCR achievers, DL severity tracks radiomic severity monotonically; using the Entropy_T0 row, the severity gradient is strong (Spearman ρ = 0.692, $p = 6.48 \times 10^{-10}$).* ***d,*** *In an external I-SPY1 frozen matched-pair analysis, DL-defined high-risk cases show worse paired outcomes than matched low-risk cases (observed mean difference = −29.69, Wilcoxon p = 0.027), with robustness under permutation (p = 0.018).*

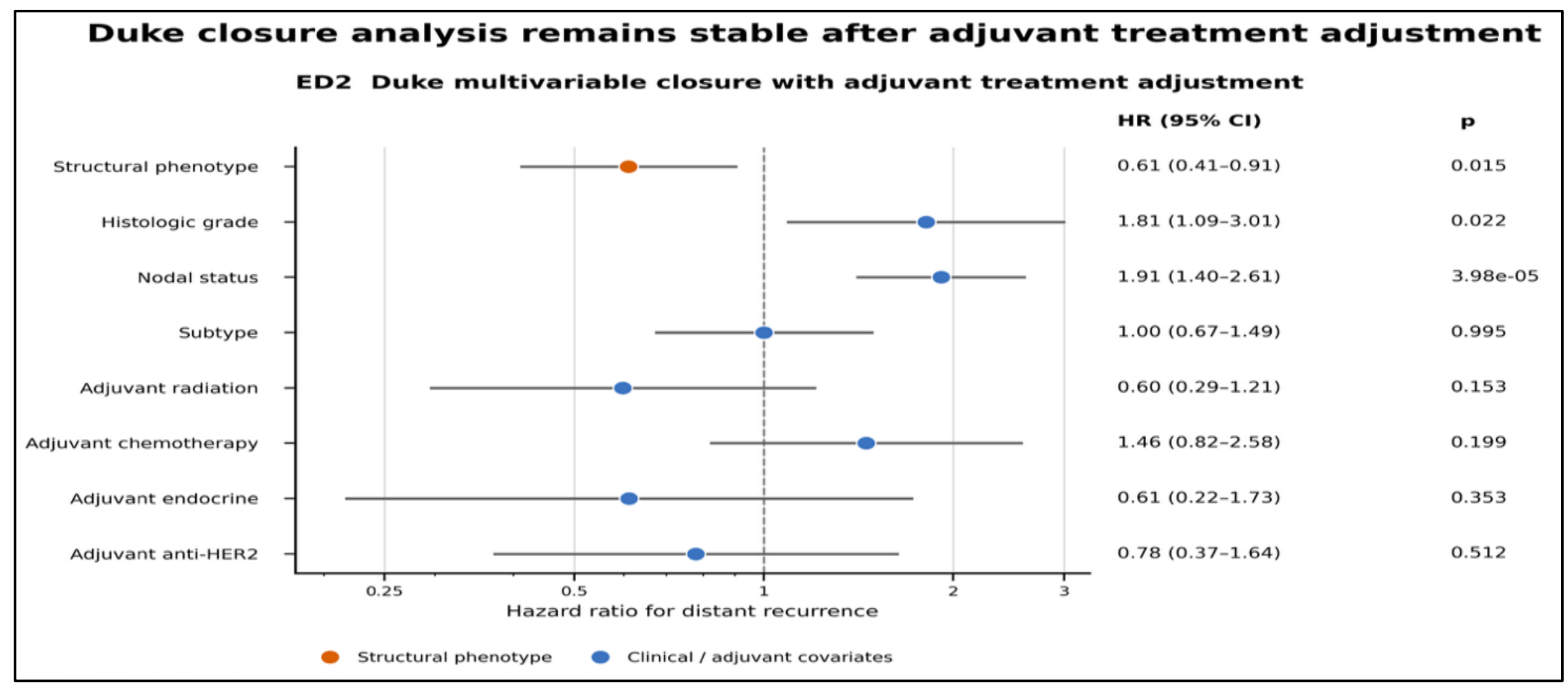

**Extended Data Figure 2 | *Duke multivariable closure analysis remains stable after adjuvant treatment adjustment.*** *Forest plot from the locked Duke eight-variable Cox proportional hazards model for distant recurrence-free survival (n = 908, 76 events). The model includes structural phenotype (baseline Shannon entropy, orange), histologic grade, nodal status, molecular subtype, and all four adjuvant treatment modalities: radiation, chemotherapy, endocrine therapy, and anti-HER2 therapy (blue). The structural phenotype remained independently associated with distant recurrence after full adjustment (HR = 0.61, 95% CI 0.41–0.91, p = 0.015), whereas no individual adjuvant treatment term reached significance in the fully adjusted model. Points denote hazard ratios and horizontal lines denote 95% confidence intervals; the dashed vertical line marks HR = 1. The structural phenotype term is highlighted separately from the clinical and adjuvant treatment covariates to distinguish the variable of interest from the adjustment set.*

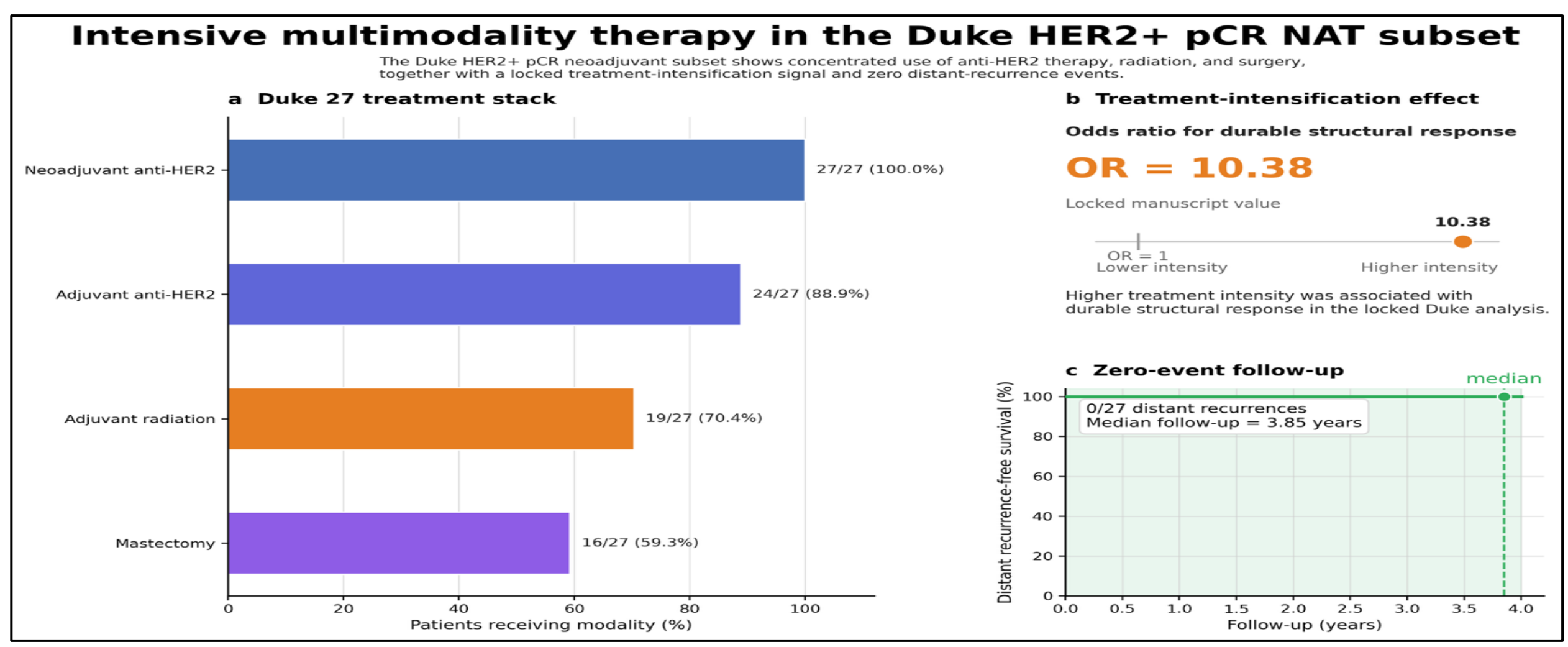

**Extended Data Figure 3 | *Intensive multimodality therapy and zero distant recurrence in the Duke HER2-positive pCR neoadjuvant subset. a,*** *Treatment-stack composition of the 27 HER2-positive patients who achieved strict pathologic complete response after neoadjuvant therapy in the Duke cohort. All 27 received neoadjuvant anti-HER2 therapy; 24/27 (88.9%) received adjuvant anti-HER2 therapy; 19/27 (70.4%) received adjuvant radiation; 16/27 (59.3%) underwent mastectomy.* ***b,*** *Treatment-intensity association with durable structural response. In the locked Duke logistic regression analysis, higher treatment-stack intensity was associated with membership in the durable zero-recurrence subgroup (OR = 10.38, p = 0.004). The odds ratio reflects the combined contribution of neoadjuvant and adjuvant treatment modalities, not baseline structural entropy, which was not a significant predictor in this model.* ***c,*** *Kaplan–Meier plot showing zero distant-recurrence events (0/27) over a median follow-up of 3.85 years, demonstrating that intensive multimodal therapy produced durable disease control in this subgroup despite the presence of structurally heterogeneous baseline profiles. These findings illustrate that treatment intensity is associated with reduced clinical expression of structural risk — the structural phenotype identifies risk that intensive therapy may overcome.*

**Driver / confounder causal architecture map**

Summary visualisation of Table 7 showing which variable classes drive, fail to explain, partially contribute to, or were not tested for each major response and recurrence outcome.

DRIVES NULL PARTIAL UNTESTED

| | pCR occurrence | pCR quality (Tier 1 vs 2) | Recurrence within pCR | Distant recurrence (Duke) | Local recurrence (Duke) |
|---|---|---|---|---|---|
| Structural phenotype | NULL | DRIVES | DRIVES | DRIVES | NULL |
| Immune programme | PARTIAL | DRIVES | PARTIAL | UNTESTED | UNTESTED |
| Drug mechanism | DRIVES | DRIVES | PARTIAL | UNTESTED | UNTESTED |
| Tumour burden | PARTIAL | NULL | NULL | PARTIAL | UNTESTED |
| Molecular subtype | DRIVES | NULL | NULL | NULL | UNTESTED |
| Genomic risk scores | PARTIAL | NULL | UNTESTED | UNTESTED | UNTESTED |
| Demographics | NULL | NULL | UNTESTED | PARTIAL | UNTESTED |
| Adjuvant therapy | UNTESTED | PARTIAL | UNTESTED | PARTIAL | UNTESTED |
| Surgery / menopause | UNTESTED | UNTESTED | UNTESTED | PARTIAL | UNTESTED |

**Extended Data Figure 4 |** ***Driver/confounder causal architecture map.*** *Summary visualisation of Table 6 showing how each variable class relates to five clinically relevant outcome layers: pCR occurrence, pCR quality (Tier 1 vs 2), recurrence within pCR, distant recurrence in Duke, and local recurrence in Duke. Green denotes variables that* ***drive*** *the outcome, grey denotes a tested* ***null*** *result, yellow denotes* ***partial*** *or context-dependent contribution, and red denotes* ***untested*** *relationships in the present analysis set. The map shows that structural phenotype is not a primary driver of pCR occurrence but does index pCR quality, recurrence stratification within pCR, and distant recurrence in Duke, while showing a null association with local recurrence in Duke, consistent with specificity for distant rather than local failure.*

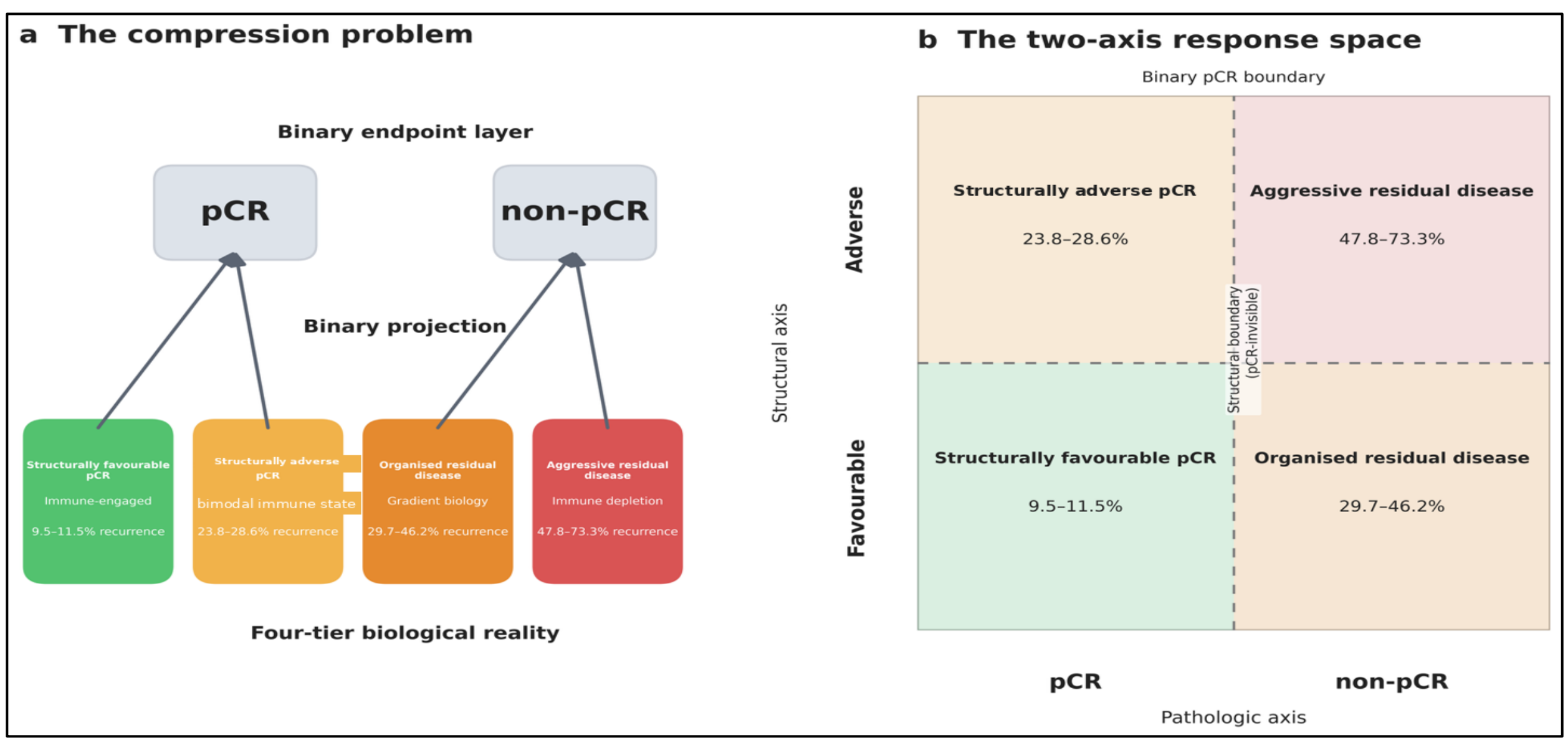

**Extended Data Figure 5 |** ***The two-axis response space and the four-tier framework. a,*** *Pathologic complete response (pCR) is a binary endpoint (top) that the four-tier framework reads as a simplified projection of a richer four-state response-quality space (bottom). Complete responders partition into a structurally favourable, immune-engaged subgroup (Tier 1) and a structurally adverse subgroup with bimodal substate distribution (Tier 2; Cluster 1 ~40% immune-depleted + Cluster 2 ~60% immune-engaged). Non-responders similarly partition into organised residual disease (Tier 3) and aggressive residual disease with an inflammatory/stromal-elevated state-defining programme (Tier 4). The binary projection (upward arrows) summarises this four-state architecture into two categories; the four-tier framework adds the within-category response-quality dimension.* ***b,*** *The two-axis response space. Combining the pathologic axis (pCR versus non-pCR, horizontal) with the structural axis (favourable versus adverse, vertical) resolves four tiers spanning a 7.7-fold recurrence gradient (9.5–73.3%) at response extremes. The pCR endpoint distinguishes the vertical boundary; the horizontal structural boundary — which separates Tier 1 from Tier 2 within pCR and Tier 3 from Tier 4 within residual disease — is the response-quality dimension that the structural axis adds. Within pathology-confirmed I-SPY1 pCR, Tier 2 carries elevated recurrence risk relative to Tier 1* (with concordant support from the pooled responder synthesis: HR = 2.87, 95% CI 1.38–5.96; see §3.3).

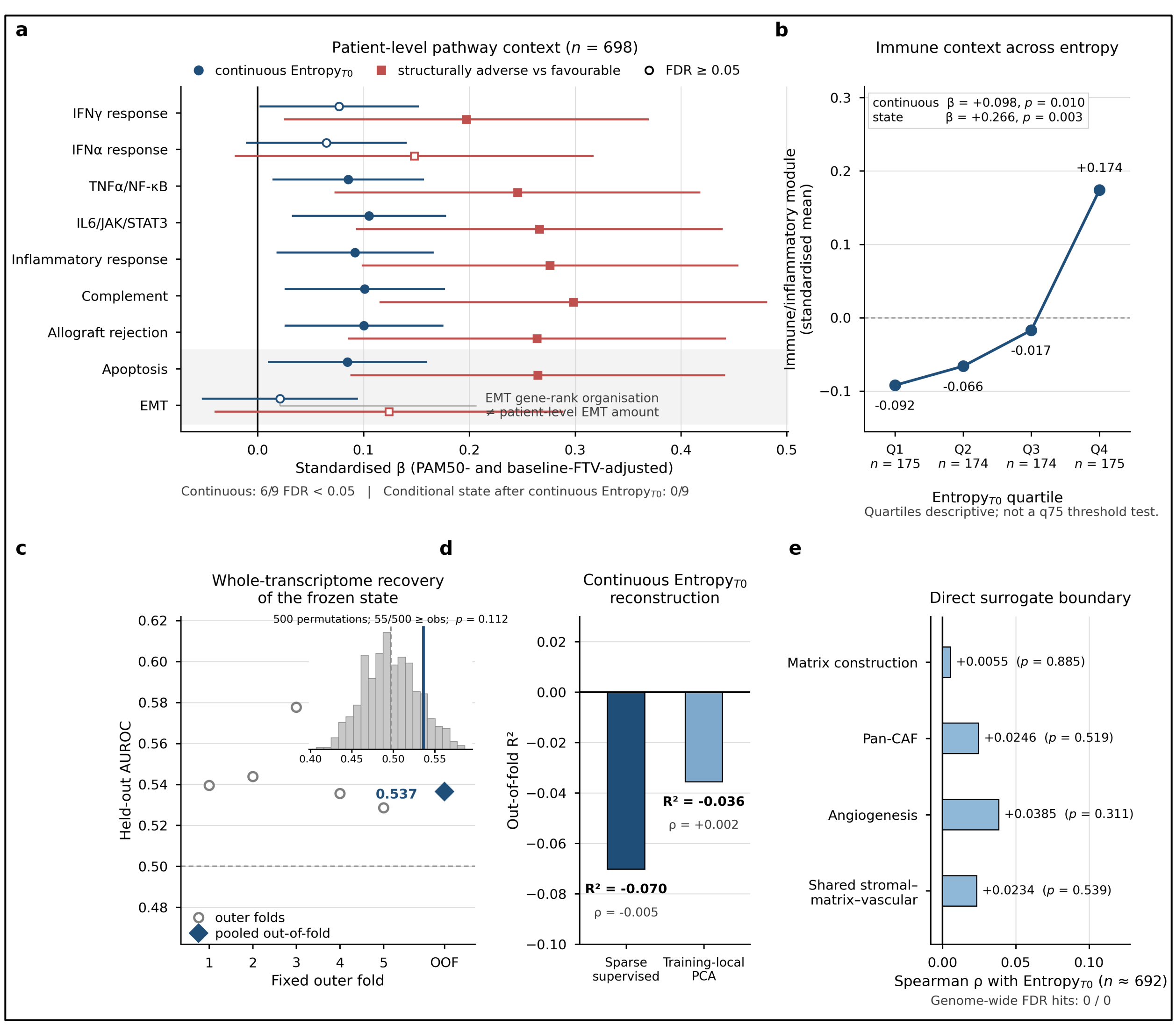

**Extended Data Figure 6 | *Molecular context without reconstruction of the MRI structural axis. a,*** *PAM50- and baseline-FTV-adjusted associations of continuous Entropy_T0 and the adverse structural state with the nine-pathway panel in 698 patients; lines show HC3 95% CIs and open markers denote FDR ≥ 0.05.* ***b,*** *Seven-pathway immune/inflammatory-module means across entropy quartiles; quartiles are descriptive and do not define a biological threshold.* ***c,*** *Corrected fold-local structural-state recovery (pooled OOF AUROC = 0.537); inset shows the empirical 500-permutation null (p = 0.112).* ***d,*** *Sparse and PCA-based bulk RNA did not reconstruct continuous Entropy_T0 (OOF R² = −0.070 and −0.036).* ***e,*** *Entropy_T0 was essentially unrelated to matrix-construction, pan-CAF, angiogenesis or shared stromal–matrix–vascular quantities. Together, bulk RNA provides biological context without materially reconstructing or directly substituting for the imaging phenotype.*

**Supplementary Information**

Supplementary Information includes four front-matter items (supplement roadmap, cohort-role orientation, master denominator audit, inference contract), six Extended Data Figures (ED1–ED6), seventy-eight Supplementary Tables (S1–S78), twenty-five supplementary methodological notes (D1–D25), and seventeen Supplementary Figures (S1–S17). Most items are cited at their first relevant appearance in the main text; additional supporting tables and figures are provided for completeness. A claim-to-evidence roadmap is provided as Supplementary Item A1. Supplementary materials and reproducibility documentation for this arXiv preprint are available at: [https://drive.google.com/file/d/1emZbVB-G2xqiMZtYj4Dc-UrNe3j18LrX/view?usp=drive_link](https://drive.google.com/file/d/1emZbVB-G2xqiMZtYj4Dc-UrNe3j18LrX/view?usp=drive_link)

**Acknowledgements**
The authors thank the I-SPY trial consortium, the ACRIN 6657 investigators, and the Duke University investigators for making their imaging and clinical data publicly available through The Cancer Imaging Archive. This work was supported by the Department of Biomedical Engineering and the Department of Computer Science at The George Washington University.

**Author Contributions**
D.K. conceived the study, designed and performed all analyses, developed the four-tier structural quality framework, and wrote the manuscript. M.H.L. supervised the research, provided critical intellectual guidance, and reviewed the manuscript. Both authors approved the final version.

**Competing Interests**
The authors declare no competing interests.

**Data Availability**
All imaging and clinical data used in this study are publicly available through The Cancer Imaging Archive (TCIA): I-SPY2 (https://doi.org/10.7937/TCIA.D8Z0-9T85), I-SPY1/ACRIN 6657 (https://doi.org/10.7937/K9/TCIA.2016.HdHpgJLK), and Duke Breast Cancer MRI (https://doi.org/10.7937/TCIA.e3sv-re93). UCSF imaging (the Breast-MRI-NACT-Pilot collection) is available through TCIA (https://doi.org/10.7937/K9/TCIA.2016.QHsyhJKy). The BreastDCEDL-ISPY2 dataset is available from TCIA (https://doi.org/10.7937/42WQ-TH78). The QIN-BREAST dataset is available from TCIA (https://doi.org/10.7937/K9/TCIA.2016.21JUEBH0). Transcriptomic datasets are available from the NCBI Gene Expression Omnibus: I-SPY2 baseline expression under accession GSE194040, and I-SPY1/ACRIN 6657 baseline expression under accession GSE22226. No new data were generated for this study.

**Code Availability**
Analysis code will be deposited in a public GitHub repository before peer-reviewed publication and is available from the corresponding author for reproducibility review. The repository will include radiomic feature extraction, four-tier framework construction, deep-learning convergence, statistical analyses, documented GSE194040 column mapping, and metagene-score reconstruction.